\documentclass[12pt,preprint]{aastex}

\usepackage{float}
\usepackage{times}
\usepackage{graphicx}
\usepackage{amssymb}
\usepackage{color}
\usepackage{url}

\newcommand{\msun}{\ensuremath{M_{\odot}}}
\newcommand{\lsun}{\ensuremath{L_{\odot}}}
\newcommand{\mbh}{\ensuremath{M_{\rm BH}}}
\newcommand{\mdef}{\ensuremath{M_{\lambda,{\rm def}}}}
\newcommand{\mvdef}{\ensuremath{M_{V,{\rm def}}}}
\newcommand{\massdef}{\ensuremath{M_{\star,{\rm def}}}}
\newcommand{\msig}{$M_{\rm BH}$--$\sigma$}
\newcommand{\mlum}{${M_{\rm BH}}$-$L$}
\newcommand{\ml}{\ensuremath{\Upsilon_{\star,{\rm dyn}}}}

\newcommand{\sersic}{S\'{e}rsic}
\newcommand{\cs}{core-S\'{e}rsic}
\newcommand{\kms}{\ensuremath{\mathrm{km~s}^{-1}}}

\def\aj{AJ}%
\def\araa{ARA\&A}%
\def\apj{ApJ}%
\def\apjl{ApJ}%
\def\apjs{ApJS}%
\def\apss{Ap\&SS}%
\def\aap{A\&A}%
\def\aaps{A\&AS}%
\def\mnras{MNRAS}%
\def\pasa{PASA}%
\def\pasp{PASP}%
\def\nat{Nature}%
\defcitealias{Kormendy-09a}{KFCB09}
\defcitealias{Kormendy-09b}{KB09}
\defcitealias{Rusli-13}{Paper~I}
\defcitealias{Lauer-05}{L05}

\shorttitle{Depleted Cores and \mbh}
\shortauthors{Rusli et al.}

\begin{document}

\title{Depleted Galaxy Cores and Dynamical Black Hole Masses}

\author{S.P. Rusli\altaffilmark{1,2}, P. Erwin\altaffilmark{1,2},
  R.P. Saglia\altaffilmark{1,2}, J. Thomas\altaffilmark{1,2},
  M. Fabricius\altaffilmark{1,2}, R. Bender\altaffilmark{1,2},
  N. Nowak\altaffilmark{3}}

\altaffiltext{1}{Max-Planck-Institut f\"{u}r extraterrestrische Physik, Giessenbachstrasse, D-85748 Garching, Germany}
\altaffiltext{2}{Universit\"{a}ts-Sternwarte M\"{u}nchen, Scheinerstrasse 1, D-81679 M\"{u}nchen, Germany}
\altaffiltext{3}{Max-Planck-Institut f\"{u}r Physik, F\"{o}hringer Ring 6, D-80805 M\"{u}nchen, Germany}

\begin{abstract} 

Shallow cores in bright, massive galaxies are commonly thought to be the
result of scouring of stars by mergers of binary supermassive black
holes. Past investigations have suggested correlations between the
central black hole mass and the stellar light or mass deficit in the
core, using proxy measurements of \mbh\ or stellar mass-to-light ratios
($\Upsilon$). Drawing on a wealth of dynamical models which provide both
\mbh\ and $\Upsilon$, we identify cores in 23 galaxies, of which 20 have
direct, reliable measurements of \mbh\ and dynamical stellar
mass-to-light ratios (\ml). These cores are identified and measured
using Core-S\'ersic model fits to surface brightness profiles which
extend out to large radii (typically more than the effective radius of
the galaxy); for approximately one fourth of the galaxies, the best fit
includes an outer (\sersic) envelope component. We find that the core
radius is most strongly correlated with the black hole mass and that it
correlates better with total galaxy luminosity than it does with
velocity dispersion. The strong core-size--\mbh\ correlation enables
estimation of black hole masses (in core galaxies) with an accuracy
comparable to the \mbh--$\sigma$ relation (rms scatter of 0.30 dex in
$\log \mbh$), without the need for spectroscopy. The light and mass
deficits correlate more strongly with galaxy velocity dispersion than
they do with black hole mass.  Stellar mass deficits span a range of
0.2--39 \mbh, with almost all (87\%) being $< 10 \, \mbh$; the median
value is 2.2 \mbh.

\end{abstract}


\maketitle

\section{Introduction}

In the present-day view, mergers are a common and important way of
building up massive elliptical galaxies. With the presence of
supermassive black holes in the centers of most galaxies
(\citealt{Kormendy-95}; \citealt{Magorrian-98}; \citealt{Richstone-98}),
merger processes are thought to have left signatures on the central
structure of the remnant galaxies. \citet{Begelman-80} suggest that a
binary black hole that is formed by the merger of two galaxies scours
the stars from the center of the newly created system as the binary
shrinks. The energy liberated by the hardening of the binary evacuates
the central part through the ejection of surrounding stars, causing less
light in the center.   

\textit{Hubble Space Telescope} (\textit{HST}) observations of galaxy
nuclei see two kinds of behavior in the central light profiles of
ellipticals, traditionally classified as ``power-law'' and ``core''
galaxies \citep{Kormendy-94, Ferrarese-94, Lauer-95, Lauer-05,
Byun-96, Lauer-07}. The power-law profiles took their name from their
approximation by a single power law at small radii ($r \lesssim 10$ or
20\arcsec). More modern interpretations have emphasized that these
profiles can be better understood as the inward continuation of the
galaxy's overall S\'ersic profile, usually modified by an additional,
nuclear-scale stellar component \citep{Graham-03, Trujillo-04,
Ferrarese-06, Kormendy-09a}. The core galaxy, in contrast, displays a
surface brightness profile with a distinct break from a steep outer
slope to a shallower inner cusp. Core profiles mainly occur in very
luminous elliptical galaxies and are considered the result of
dissipationless mergers of two galaxies that have central black holes.
 
Much numerical work has been done to explore the binary black hole idea.
The work of \citet{Ebisuzaki-91} suggests that the stellar mass ejected
from the core is comparable to the mass of the central black hole \mbh.
\citet{Makino-96} find that core radius roughly doubles after each major
merger. \citet{Nakano-99}, using isothermal King models, suggest
that the core size should scale with the final \mbh. Simulations
performed by \citet{Milosavljevic-01} show for the first time that
merging two galaxies with steep cusps can result in a merger remnant
with a shallow power-law cusp in the inner part (core).
\citet{Merritt-06} is the first to quantify the magnitude of the mass
deficit (or the mass ejected by the binary) with respect to the galaxy's
merger history. By following the binary evolution up to the stalling
radius, he formulates the mass deficit to be $0.5\,N$\mbh\ with $N$ is
the number of major merger events. An investigation into the
post-coalescence stage of the binary suggests that the mass deficit
could still be larger. This is because the coalescing binary emits
gravitational waves which impart a kick on the merged black hole and
eject it from the center. Due to dynamical friction, the resulting
oscillations of the black hole would gradually damp, returning it to the
center of the galaxy and ejecting more stars from the core. This process
could increase the mass deficit existing before the binary coalescence
by up to 5\mbh\ for a single merger \citep{Gualandris-08}.

Several publications in the past have performed function fitting to
galaxy profiles in order to quantify the core parameters such as the
core size, the deficit in light or mass, and to analyse the correlation
between these core properties and \mbh\ or the global properties of the
galaxies (\citealt{Graham-04}; \citealt{Ferrarese-06};
\citealt{Lauer-07}, \nocite{Kormendy-09a}Kormendy et al.\ (2009,
hereafter KFCB09), \nocite{Kormendy-09b}Kormendy \& Bender (2009,
hereafter KB09), \citealt{Richings-11}; \citealt{Dullo-12, Dullo-13}; see also
\citealt{Kormendy-13}). These studies are based on different samples,
datasets, approaches, parametrizations, and assumptions, which
apparently lead to different, partly systematic, core quantities. The
most important difference is the amount of central mass deficit with
respect to the black hole mass that can be expected in core galaxies.
\citet{Graham-04} and \citet{Ferrarese-06} find an average mass deficit
of around twice the black hole mass, \citet{Dullo-12} derive central
mass deficits of about 0.5--4 \mbh\ while \citetalias{Kormendy-09a} find
their mean mass deficit to be 10.5 \mbh. Aside from that, each of these
studies has a drawback in at least one of three aspects crucial to
achieve high accuracy.

The first is the radial range of the surface brightness profiles.
\citet{Lauer-07}, \citet{Richings-11}, and \citet{Dullo-12}  fit galaxy
profiles that extend typically to only 10--20\arcsec. Moreover, it
appears that some galaxy profiles in \citet{Dullo-12}, which are taken
from \nocite{Lauer-05}Lauer et~al.\ (2005, hereafter L05), suffer from
oversubtraction of the sky that biases the core-S\'ersic fits (see
Section \ref{previousfits}). A full galaxy profile is necessary to
properly measure the outer profile, to optimally remove the sky
background, and thus to provide better constraints on the light and mass
deficit.

The second is the determination of the stellar mass-to-light ratio
($\Upsilon_{\star}$) to convert the light deficit to the mass deficit.
In the past studies, either different values are assigned to the
galaxies according to their luminosities (\citealt{Ferrarese-06},
\citetalias{Kormendy-09a}) or a constant value is assumed for all the
galaxies in the sample (\citealt{Graham-04}, \citealt{Dullo-12}). This
of course introduces additional uncertainties and biases in deriving the
mass deficits.

The last important aspect lies in the black hole masses, which are
mostly derived from outdated black hole-bulge relations, especially the ones
between \mbh\ and the velocity dispersion of the host bulge (\msig\
relation). The majority of core galaxies are bright massive elliptical
galaxies. For this class of galaxies \citet{Lauer-07} find that the
\mlum\ relation predicts a higher density of massive black holes
($\sim10^{10}$ \msun) compared to the \msig\ relation. Furthermore, the
\msig\ relation used in these papers also fails to predict the existence of the largest black
holes ($\mbh > 10^{10}$\msun) that are found in distant quasars
\citep{Shields-06b} in the sample of local galaxies. The recent finding
of black holes of 10 billion solar masses
\citep{McConnell-11b,vandenBosch-12}, which might be the missing link
between the local black holes and the massive distant quasars, further
supports the indications that the previous \msig\ relations for massive bright
early type galaxies need to be reevaluated
\citep[e.g.,][]{Kormendy-13}. 
In addition, most previous studies have used \mbh\ measurements (or \msig\
relations based on \mbh\ measurements) which did not include dark-matter
halos in the modeling. \mbh\ values typically increase by $\sim 20$\% 
\citep{Schulze-11} or 30\% \citep[hereafter Paper I]{Rusli-13} when dark matter is
included in the modeling, but the increase can be as large as a factor
of 6 when the black hole's sphere of influence is not well resolved \citep{McConnell-11a}. 
In the particular sample that we use in this paper, which includes BCGs, we
find that the measured black hole mass is on average more than 3 times
higher than that predicted by the \msig\ relation (e.g.
\citealt{Tremaine-02}, \citealt{Gueltekin-09}, \citealt{Graham-11}.

In this paper, we combine improved core-property measurements with new
measurements and present the analysis of possible relations between the
core and black hole mass or the galaxy properties. We use only directly
measured black hole masses and consequently avoiding scatter and
possible biases inherent in using proxy measurements. We use
dynamically-determined, individual stellar mass-to-light ratios (\ml) to
estimate the mass deficit, instead of relying on a scaling function or a
constant value. We also use full galaxy profiles in the fitting and the
quantification of the core parameters. Our analysis is based on a sample
of 20 galaxies; seven of these galaxies are taken from
\citetalias{Rusli-13}, in which the black hole masses and 
\ml\ are derived taking into account the presence of dark matter halos.
For completeness, we add a further 3 galaxies to the sample; they are not
included in the main sample due to the lack of reliable literature
sources for the black hole masses. In order to derive the mass deficit,
we assume that the original (pre-scouring) profile is the extrapolated
\sersic\ profile that fits the light distribution at large radii. In
Section~\ref{themethod}, we describe the method that we use to identify
the core galaxies and to fit the observed surface brightness profiles.
In Section~\ref{corefitting}, we present the best-fitting models for
each galaxy; cases where galaxies are best fit with the addition of an
outer envelope are discussed in Section~\ref{multi-component-fits}. We
compare the fitting results with previous studies in
Section~\ref{previousfits}. Section~\ref{coredeficit} presents the
details of how the luminosity and mass deficit in the core are computed
based on the best-fitting models. The core-related quantities derived in
Section~\ref{corefitting} and \ref{coredeficit} are compared with the
black hole mass, velocity dispersion and galaxy luminosity in Section
\ref{bhcore}. The last Section discusses and summarizes the results.

\section{The method}
\label{themethod}

\subsection{Identifying cores}
\label{identifycore}

All of the photometric profiles examined in this work are quite
extended, reaching out to at least 9 kpc. Therefore, we base our
analysis on a function that is designed to fit not just the inner or the
outer part of galaxies, but rather the galaxy as a whole. To identify
core galaxies, we follow the criteria in \citet{Graham-03} and \citet{Trujillo-04}, who define
the core as ``a downward deviation from the inward extrapolation of the
outer (S\'{e}rsic) profile''. This involves
fitting the galaxy in question with both the \sersic\ and the \cs\
functions.

The \sersic\ profile \citep{Sersic-63, Sersic-68} is written as:
\begin{equation}
I(r) = I_e \, {\rm exp} \left\{-b_n \left[ \left( \frac{r}{r_e} \right)^{1/n}-1 \right] \right\}.
\label{sersiceq}
\end{equation}
$I_e$ is the intensity at $r_e$, the projected half-light radius. $n$ is
called the \sersic\ index which describes the shape or curvature of the
light profile. The quantity $b_n$ is a function of $n$, defined in such
a way that $r_e$ encloses half of the total luminosity. We approximate
$b_n\sim 2n-1/3+4/405n+...$ by using the asymptotic expansion of
\citet{Ciotti-99}, Eq. 18 for $n > 0.36$. Many other approximation
formulas are available and are summarized in \citet{Graham-05}. For
$n=1$, the \sersic\ function reduces to an exponential function and for
$n=4$, it becomes the de Vaucouleurs profile \citep{deVaucouleurs-48}.

The \cs\ function introduced by \citet{Graham-03} and \citet{Trujillo-04} is expressed as:
\begin{equation}
I(r) = I^{\prime} \left[1+\left(\frac{r_b}{r}\right)^\alpha\right]^{\gamma/\alpha}\,{\rm exp}\left[-b_n\left(\frac{r^\alpha+r_b^\alpha}{r_e^\alpha}\right)^{1/n\alpha}\right]
\label{coresersiceq1}
\end{equation}
with
\begin{equation}
I^{\prime} = I_b2^{-\gamma/\alpha}\,{\rm exp} \left[b_n\left(2^{1/\alpha}r_b/r_e\right)^{1/n}\right].
\label{coresersiceq2}
\end{equation}
This profile consists of a \sersic\ profile in the outer part, specified
by the projected half-light radius $r_e$ and the \sersic\ index $n$, and
a power-law profile in the inner part with a slope of $\gamma$. The
change from one to another regime occurs at the break radius $r_b$ and
the sharpness of the transition is specified by the parameter $\alpha$.
Higher values of $\alpha$ mean sharper transitions. $b_n$ is
approximated in the same way as for the \sersic\ function.

We summarize four criteria in \citet{Trujillo-04} that we use to
identify core galaxies below. In Section \ref{corefitting} we examine
each criterion.
\begin{enumerate}
\item A characteristic pattern should be visible in the residuals when
fitting a \sersic\ profile to an idealized core galaxy. This serves as a
qualitative evidence.
\item A \cs\ function gives a significantly better fit than a \sersic\
model. Quantitatively, it is expected that the reduced $\chi^2$ value
for the \sersic\ fit $\chi^2_{S}$ is larger than twice that of the \cs\
fit $\chi^2_{CS}$.
\item To avoid an ambiguous core detection, the potential core must be
well-resolved by the data. Quantitatively, $r_b$ should be $> r_2$ where
$r_2$ is the second innermost datapoint in the observed profile.
\item The power-law slope has to be consistently less than the slope of
the \sersic\ fit inside $r_b$.
\end{enumerate}

We decided to adopt an additional criterion based on resolution limits.
Trujillo et al.\ did not include such a limit, partly because many of
their profiles did not probe radii smaller than the \textit{HST} resolution limit
due to dust, etc.; however, they did note that one of their ``possible
core'' galaxies (NGC 3613) had a break radius smaller than the
resolution limit  of 0.16\arcsec\ originally suggested by
\citet{Faber-97}. The latter limit is $\sim 2$ times the FWHM of most of
the PC data used by \citet{Faber-97}; generalizing this, we suggest the
following additional criterion:
\begin{enumerate}
\setcounter{enumi}{4}
\item The potential core should have $r_b
\gtrsim 2$ the FWHM of the imaging data used for the innermost part of
the profile. For most of the galaxies considered in this paper, which
use optical WFPC2 or ACS/WFC data, this corresponds to $r_b \gtrsim
0.15\arcsec$; for two galaxies using NICMOS2 data (NGC 4261 and NGC
4374), the limit would be $\approx 0.26\arcsec$, and for NGC 1550, where
we use SINFONI data with a FWHM of 0.17\arcsec, the limit would be
0.34\arcsec.
\end{enumerate}

\subsection{The fitting procedure}

We construct a one-dimensional light profile along the circularized
radius (mean profile), instead of along the semi major radius as is
commonly done. The circularized radius is defined as $r=\sqrt{ab}$ where
$a$ and $b$ are the semi major and semi minor axes radius of the
isophote. This way, we take into account the variation of ellipticity
with radius, giving a fair representation of the isophotes. The use of
this mean profile is consistent with the implementation of the PSF
(point spread function) correction (Section \ref{psfconv}). 

The fitting is done using a one-dimensional profile fitting software
written by P. Erwin. An initial model is generated based on user-given
starting values. It is compared to the observed profile and the $\chi^2$
is calculated. The best-fitting solution is achieved  by minimizing the
$\chi^2$ of the fit, using the Levenberg-Marquardt algorithm
\citep{Levenberg-44,Marquardt-63}. The user provides a starting value and
its range of allowed values for each free parameter of the model. There
are three free parameters in the \sersic\ model, i.e. $n$, $r_e$, $I_e$
and there are six free parameters in the \cs\ model, i.e. $n$, $r_b$,
$r_e$, $\mu_b$, $\alpha$, $\gamma$.

Since we are mainly interested in the inner region of the galaxy, the
blurring of the central light profile by the PSF becomes important and
has to be taken into account. For several galaxies, the photometric
profiles are derived from deconvolved images (see
Table~\ref{tab:coresample}). We fit these profiles as described in the
previous paragraph. For the other galaxies we incorporate a PSF
correction routine, described in the next Section, into the fitting
program such that the initial model is convolved first before being
compared to the observed light profile. In this case, the minimised
$\chi^2$ represents the difference between the convolved model and the
data. The PSF convolution is done at every iteration.

\subsection{The implementation of the PSF convolution}
\label{psfconv}

\begin{figure}
\centering
  \includegraphics[scale=0.70]{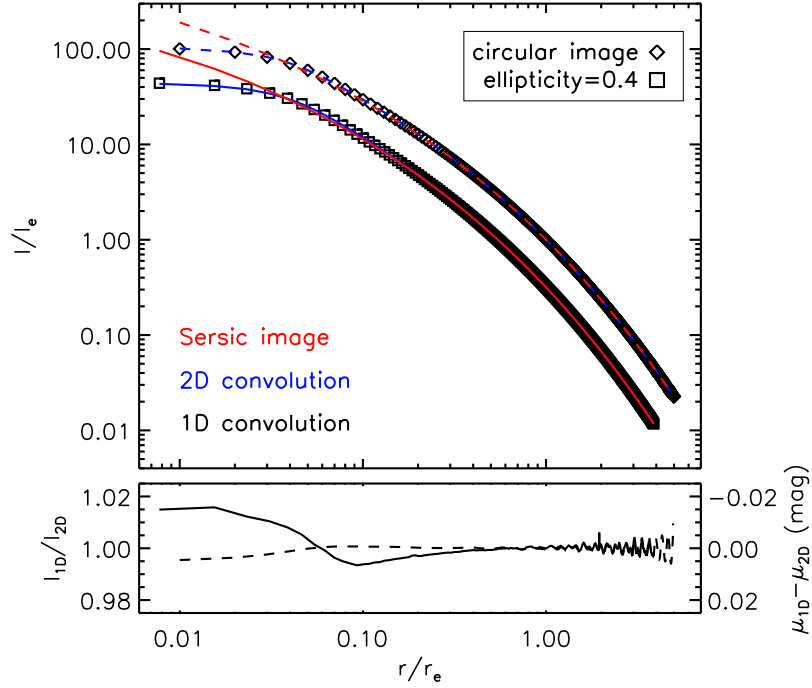}
  \caption[]{{\bf Top panel:} radial profiles of \sersic\ images and
    their convolution with a point-symmetric Gaussian image with
    $\sigma$ equal to 1/40 of the half-light radius of the \sersic\
    model. Dashed lines and diamonds show the case of a circular
    \sersic\ image; the solid lines and rectangles are for an
    elliptical S\'ersic image (ellipticity = $1-b/a = 0.4$). The red
    lines are the unconvolved \sersic\ models. The black symbols are
    the results of one-dimensional convolution (equation
    \ref{equ5}-\ref{equ7}) and the blue lines are the radial profiles
    of the convolved images, derived using ellipse-fitting. {\bf
      Bottom panel:} the flux ratio or the difference in surface
    brightness ($\mu$) between the one-dimensionally convolved
    \sersic\ profile (${\rm I_{1D}}$ or $\mu_{1D}$) and the radial
    profile of the convolved \sersic\ image (${\rm I_{2D}}$ or
    $\mu_{2D}$) for both the circular and elliptical cases.\\}
\label{testconv}
\end{figure}

For galaxies whose light profiles are not seeing-corrected, we take the
PSF into account in the fitting by means of convolution. We use the PSF
corresponding to the galaxy image used to derive the innermost galaxy
isophotes. For the galaxies from \citetalias{Rusli-13}, the PSF image is
the same one which was used in the deprojection. For the other galaxies,
the PSF image is generated using TinyTim by supplying the appropriate
filter and camera. From the image, we generate a circularized profile of
the PSF by averaging its radial profile at different azimuthal angles
(done using MIDAS). We further normalize the PSF profile such that the
total flux is one. The implementation of the convolution follows
\citet{Saglia-93} (their Appendix A). Below, we review the method.

The PSF convolution is carried out by multiplying the galaxy surface
brightness profile and the PSF in the Fourier space. On the sky, the
galaxy surface brightness is a function of a two-dimensional position
{\bf r} and so is the PSF. If $I({\bf r})$ and $P({\bf r})$ denote the
surface brightness and the PSF distribution, then the convolved
surface brightness profile becomes:
\begin{equation}
  I_{\rm conv}({\bf r}) = \int{\rm exp}\,(i{\bf k\cdot r})\,\hat{P}({\bf k})\,\hat{I}({\rm \bf k})\frac{d^2k}{(2\pi)^2}
\label{eqq}
\end{equation}
where 
\begin{eqnarray}
\hat{I}({\bf k}) &=& \int{\rm exp}\,(i{\bf k\cdot r})\,I({\bf r})\,d^2r \nonumber \\
\hat{P}({\bf k}) &=& \int{{\rm exp}\,(i{\bf k\cdot r})\,P({\bf r})\,d^2r}. \nonumber \\
\end{eqnarray}

Since we use a one-dimensional, circularized surface brightness
profile, I or P are a function of only the radius $r$. In this case,
$\int^{2\pi}_0{{\rm exp}\,(ikr\,cos\,\theta)\,d\theta} = 2\pi J_0(kr)$
where $J_0(kr)$ is the Bessel function of order zero. Then, equation
\ref{eqq} becomes:
\begin{equation}
I_{\rm conv}(r)=\int^\infty_0{\frac{1}{2\pi} J_0(kr)\, \hat{P}(k)\, \hat{I}(k)\, k \,dk}
\label{equ5}
\end{equation}
with 
\begin{equation}
\hat{P}(k)=\int^\infty_0{2\pi P(r) J_0(kr) r dr}
\label{equ6}
\end{equation}
\begin{equation}
\hat{I}(k)=\int^\infty_0{2\pi I(r) J_0(kr) r dr}
\label{equ7}
\end{equation}

In this way, although the surface-brightness and the PSF profiles are
one dimensional, the convolution is implemented such that the light
coming from different angles is also taken into account. This is
important because the light is scattered, due to the seeing, not only
in the radial but also in the azimuthal direction. Equations
\ref{equ5}-\ref{equ7} are also called the Hankel transform. We compute
the convolved surface brightness profile numerically, making use of
integration routines in the GNU Scientific Library.

Fig.~\ref{testconv} shows how well this convolution works in the case of
circular (upper profiles) and non-circular (lower profiles) images. For
the former, we generate a galaxy image which follows a \sersic\ law with
$n=4$ and zero ellipticity everywhere using the \texttt{makeimage}
program from the two-dimensional image-fitting package \textsc{imfit}
(Erwin, in
preparation)\footnote{\url{http://www.mpe.mpg.de/~erwin/code/imfit}}.
For the PSF, we create a circular Gaussian image with $\sigma$ = 1/40 of
the half-light radius of the galaxy model. We extract the radial profile
of both images and use these to perform the one-dimensional convolution
as described above. As an independent check, we separately convolve the
galaxy image with the Gaussian image in \textsc{imfit}. The radial
profile of this convolved image (${\rm I_{2D}}$) is compared to the
result of the 1D convolution (${\rm I_{1D}}$). The ratio is shown in the
lower panel of the figure by the dashed line. This process is then
repeated for a S\'ersic image with an ellipticity of 0.4. The
circularized profile of the convolved elliptical image is shown by the
solid blue line and this is compared to the 1D convolution result (the
rectangles); the ratio of both is displayed as the solid line in the
lower panel. The method works well in the case of circular galaxy and
shows a deviation of less than 0.02 mag arcsec$^{-2}$ for an elliptical
image; we have repeated this exercise for different S\'ersic indices,
resulting in similar conclusions. Since the ellipticity of 0.4 is larger
than the ones measured in the majority of galaxies in our sample, we
expect smaller errors in the actual galaxy case.

\section{Core-S\'{e}rsic vs. S\'{e}rsic fit}
\label{corefitting}

The list of galaxies examined in this Section is given in
Table~\ref{tab:coresample}, along with the data sources. We compiled
this list taking into account all galaxies which had both dynamically
measured central black hole mass and evidence for cores, but in the end
retained only those with clear evidence for a core \textit{and} good
enough data for measuring the core properties. The recent compilation of
\citet{McConnell-13} lists 9 more galaxies which we do not include:
A1836-BCG, A3565-BCG (IC 4296), NGC 524 NGC\,3607, NGC\,4473, NGC\,5077,
NGC\,5128, NGC\,5576, NGC\,7052. Strong nuclear dust in A1836-BCG, IC
4296, NGC\,3607, NGC 5128 and NGC\,7052 makes it very difficult to
robustly compute the core properties, so these galaxies are not
included. NGC~4473 has been shown not to be a core galaxy by
\citet{Ferrarese-06}, \citetalias{Kormendy-09a}, and
\citet{Dullo-12,Dullo-13}, while the same is true for NGC~5576
\citep{Trujillo-04}; see also \citet{Dullo-13}. In addition,
\citet{Trujillo-04} were only able to classify NGC~5077 as a ``possible
core'' galaxy, so we exclude it as well.  NGC~524 is a particularly
problematic case, since there is evidence for very complex morphology --
e.g., \citet{laurikainen10} fit a ground-based image using a S\'ersic
bulge, an exponential disk, and two ``lens'' components. This means that
a simple core-S\'ersic fit -- or even core-S\'ersic + outer envelope, as
we use for some galaxies -- is probably not appropriate. In addition,
the best available high-resolution data is \textit{HST}-NICMOS2, and the
fit to this data by \citet{Richings-11} suggests a core break radius
$r_b \sim 0.18\arcsec$, which is smaller than our adopted resolution
limit for NICMOS2 data (see Section \ref{identifycore}). (WFPC2 images
are not usable due to the extensive circumnuclear dust lanes.) We
reluctantly exclude this galaxy as well.

NGC\,1374 represents a case where the core detection is ambiguous, so it
is discussed independently in a separate section. This galaxy was
classified as a non-core galaxy in \citet{Dullo-12} using a limited
dataset from \citetalias{Lauer-05}, and also by \citet{Dullo-13}
using a more extended ACS-WFC profile from \citet{Turner-12}; in
Section~\ref{n1374} we show the fit using our surface brightness profile
which is slightly more extended in radius and which does not suffer
the possible sky-subtraction uncertainties of the ACS-only Turner et
al.\ data \citepalias[see the discussion in][]{Rusli-13}.
Table~\ref{tab:csparams} summarizes the list of galaxies for which the
presence of core is confirmed. The best-fitting \cs\ parameters are also
given in the same table. The best-fit \sersic\ and \cs\ functions of
these galaxies are shown in Fig.~\ref{fittingplots} and the criteria
which lead to the core identification are discussed below.

\subsection{The core galaxies}
\label{coregals}

When possible, we use all the available datapoints in the fitting. For
all galaxies, the \cs\ function gives a much better fit to the whole
profile than the \sersic. In several galaxies, however, we find that the
fit residuals become rather large at several outermost datapoints. Since
these datapoints do not seem to be well-described by a \sersic\
function, they are excluded from the fit. We keep only the datapoints
where the fit residual is less than 0.15 mag. The deviation from the
S\'ersic model at large radii by more than 0.15 mag may be caused by
imperfect sky subtraction or it may represent a real departure from the
\sersic\ profile, for example when where an outer halo is present;
see Section~\ref{multi-component-fits} for details on fits using extra
components. The fit is then repeated using only the remaining
datapoints, after excluding the outliers. The fit residuals, their rms
and the $\chi^2$ values are presented in Fig.~\ref{fittingplots}. In
this figure, the surface brightness profiles shown are not corrected for
Galactic extinction.

In all the galaxies in Fig.~\ref{fittingplots} we see a pattern in the
residuals of the \sersic\ fit that qualitatively suggests the presence
of the core -- compare to Fig.~3 in \citet{Trujillo-04}. This residual
pattern disappears when the galaxy is fitted with the \cs\ profile. So
the first criterion in Section \ref{identifycore} is secured. In each
case, the rms of the residuals for the best \sersic\ fit is much larger
than that for the \cs. The $\chi^2$ values for each galaxy are
calculated by adopting the residual rms of the best \cs\ fit as the
error on the surface brightness measurements. As discussed in \citetalias{Kormendy-09a},
this estimate for the real error is reasonable as long as the rms is of
order of the profile measurement error, i.e., a few hundredths of a mag
arcsec$^{-2}$, which is indeed the case for all the galaxies that
we examine. Using this error approximation, the $\chi^2$ of the \cs\ fit
is artificially set to 1.0. Note that the rms and $\chi^2$ values are
dependent on the radial extent and the radial sampling (logarithmic
sampling is used here). We find that the (reduced) $\chi^2$ value for
each \sersic\ fit is much more than twice that of the \cs. The average
ratio of $\chi^2_S/\chi^2_{CS}$ is around 49, with a minimum of 6.8,
easily fulfilling the second criterion. To assess the third criterion,
one can compare the position of the second innermost datapoint to the
position of the dotted vertical line which marks the break radius. It is
thus clear that we have cases of resolved cores. The fourth point is
also satisfied, since inside the break radius the \cs\ function shows a
consistent deficit in light compared to the extrapolated \sersic\
profile (shown by the blue line). For the cases where we exclude some
datapoints, the core is already well identified when fitting all the
datapoints, i.e. fulfilling the four criteria simultaneously. Limiting
the fitting range only improves the distinction. Lastly, the break
radius for each galaxy is larger than twice the FWHM of the imaging data
used to construct the innermost part of the surface brightness profile,
described more quantitatively in Section \ref{identifycore}.

Among the 23 galaxies, several exhibit more complex structures. Inside
$\sim0.7\arcsec$, the surface-brightness profile of IC\,1459 seems to
show a small excess of light. We model this nuclear excess as an extra
Gaussian component in the center on top of the \cs\ function, which
after PSF convolution provides a good fit to the data.  (In the cases of
NGC\,1407 and NGC\,4552, it is not clear that their inner excesses can
be modeled with a Gaussian, so we exclude the data at $r < 0.16\arcsec$
for NGC\,1407 and $r < 0.04\arcsec$ for NGC\,4552.)  For six galaxies
(including NGC\,1407), we find that significantly better fits are
achieved by including one or two extra \sersic\ or exponential
components, representing outer halos. This approach is discussed in more
detail in Section~\ref{multi-component-fits}.

The parameter values of the best-fit \cs\ model for each galaxy are
provided in Table~\ref{tab:csparams}. The uncertainties of the \cs\
parameters are determined through a Monte Carlo simulation (see next
Section). The \sersic\ indices for these galaxies are almost all larger
than 4, something which is common for giant ellipticals (e.g.,
\citealt{Caon-93}, \citetalias{Kormendy-09a}). The exceptions are four
of the galaxies which we model with multiple components (see
Section~\ref{multi-component-fits}). The best-fit $\alpha$ parameter for
NGC\,7768 is found to be 16.6. This large value indicates a very sharp
transition between the \sersic\ and the power-law profile. When the fit
is repeated with $\alpha=100$, which is a good approximation for
$\alpha=\infty$ \citep{Graham-03}, the best-fit values for the other
five parameters hardly change. Since for large $\alpha$ ($>$ 10) the
profile becomes insensitive to the exact value of $\alpha$, we prefer to
use the latter fit (fixing $\alpha$ to 100) to the light profile of
NGC\,7768. This choice prevents wildly varying values of $\alpha$ that
are not physically meaningful in the Monte Carlo simulations.

For six galaxies whose surface brightness profiles are taken from
\citetalias{Kormendy-09a} (i.e., NGC\,4261, NGC\,4374, NGC\,4472,
NGC\,4486, NGC\,4552, and NGC\,4649), the best-fitting \sersic\ model of
\citetalias{Kormendy-09a} is also overplotted on the right-hand panels
of Fig.~\ref{fittingplots}. Our inward-extrapolated \sersic\ components
and their \sersic\ fits do not coincide, reflecting the differences in
the way the core is defined. Note that \citetalias{Kormendy-09a} use
major-axis profiles while we use circularized profiles, resulting in a
seemingly poorer fit of KFCB's \sersic\ models in
Fig.~\ref{fittingplots}. For a more thorough comparison between our and
their fitting procedure, see Appendix \ref{comparisonkfcb09}.

NGC\,3091 and NGC\,7619 show small values in $\alpha$, i.e. $\alpha <
2$. \citet{Dullo-12} produce a best-fitting \cs\ model of NGC\,7619 by
keeping $\alpha$ constant at 5, which enforces a rather sharp transition
from \sersic\ to power-law profiles. Doing this exercise using our
circularized profile of NGC\,7619 results in a smaller $n$ (7.71),
larger $r_b$ (0.73\arcsec) and $\gamma$ (0.47) and a larger rms (0.033).
It is worth emphasizing that \citet{Dullo-12}'s profile for NGC\,7619 is
not as extended as ours and stops short of $r \approx 20\arcsec$.

\begin{figure*}
\centering
  \includegraphics[scale=0.5]{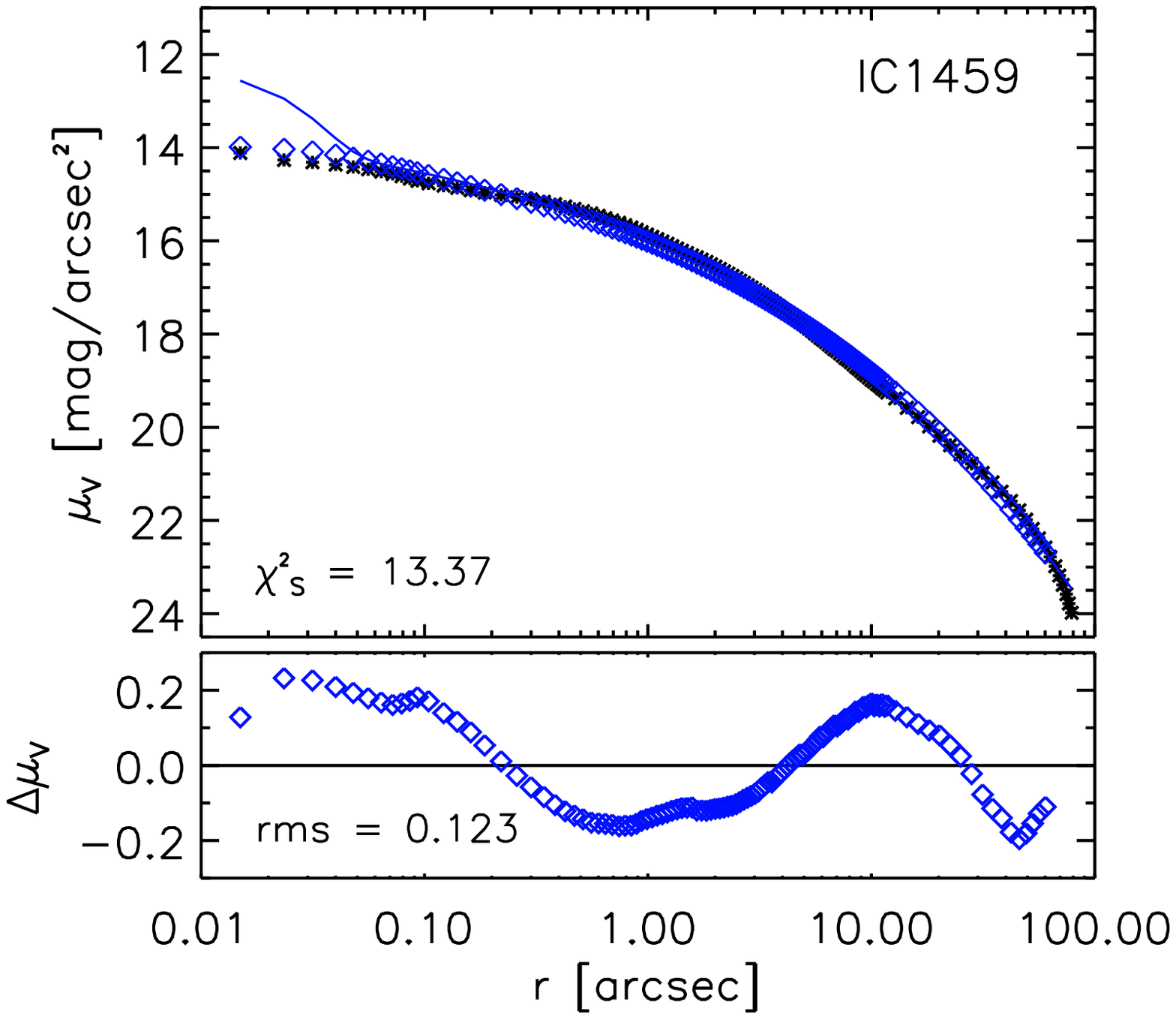}\includegraphics[scale=0.5]{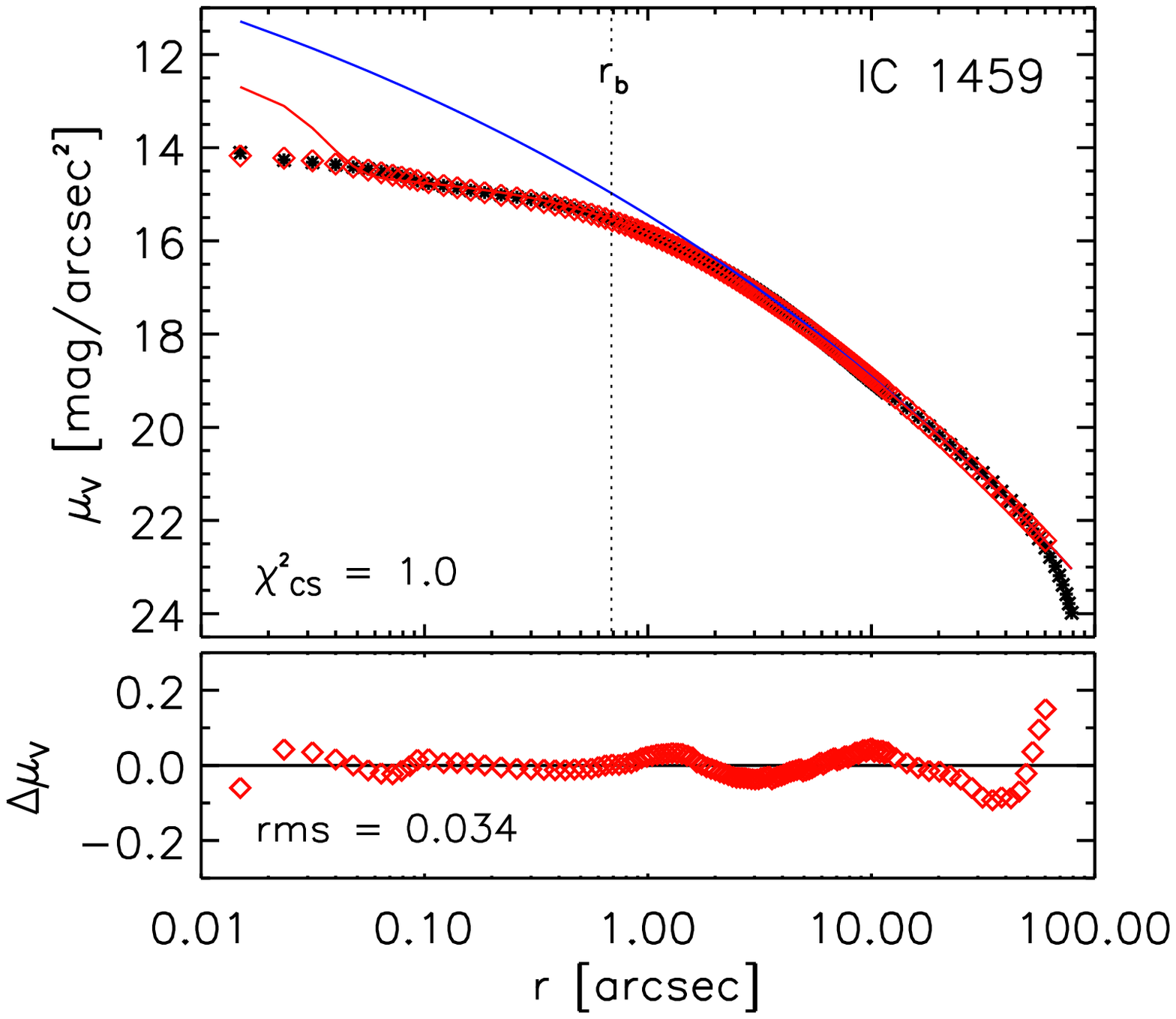}
  \includegraphics[scale=0.5]{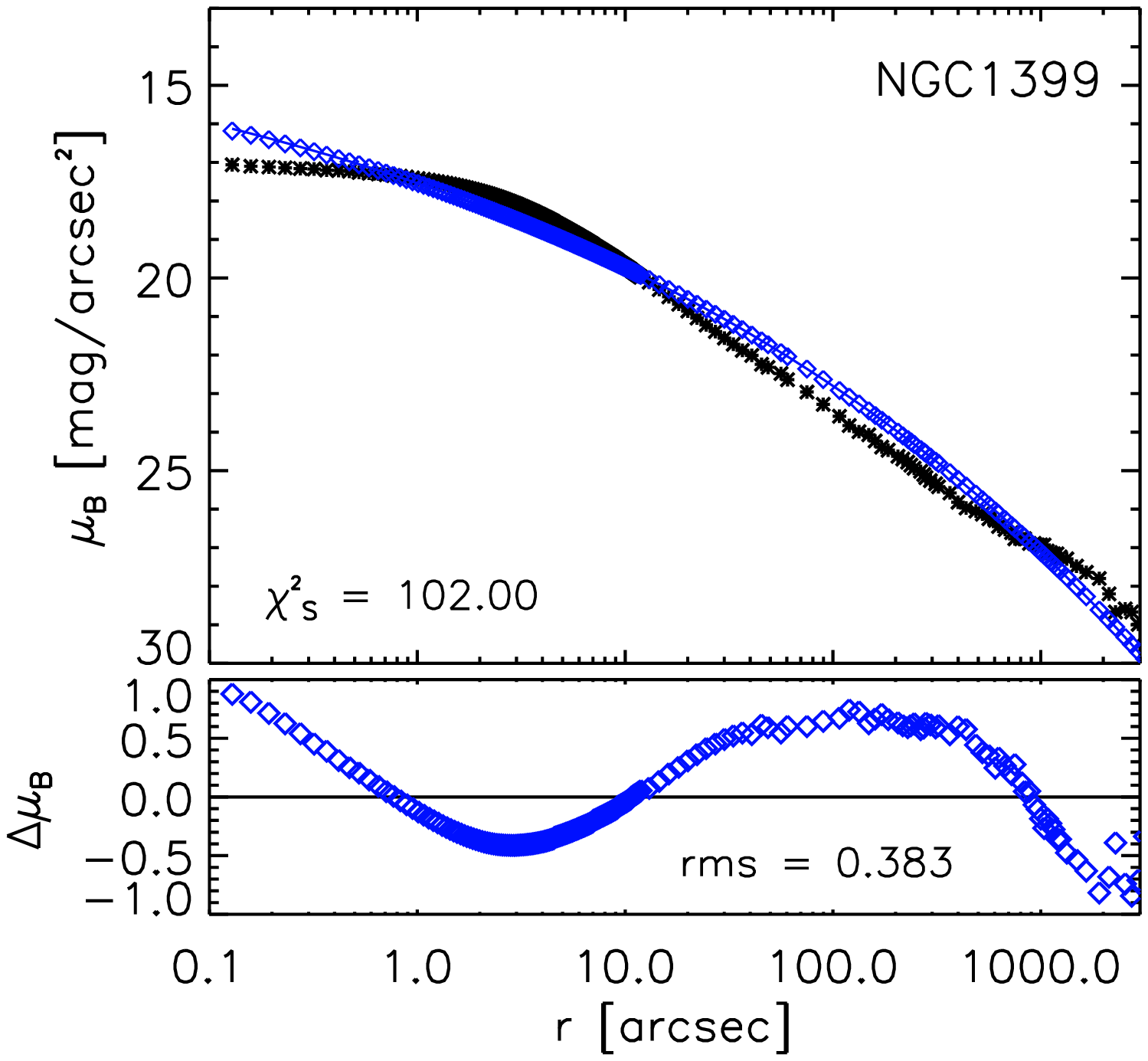}\includegraphics[scale=0.5]{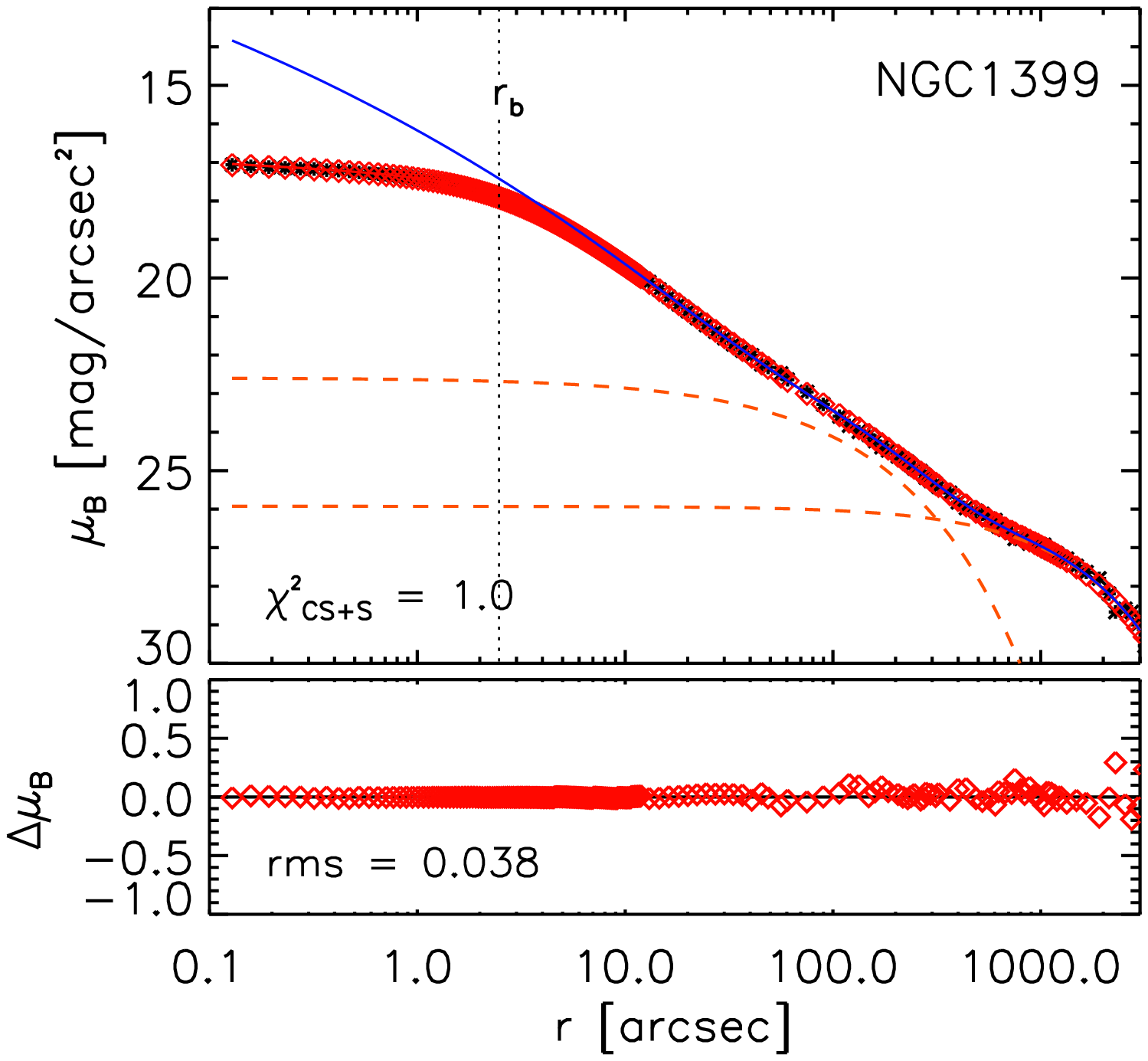} 
  \caption[]{\small The model fits to the surface-brightness profiles
    of all the galaxies in Table~\ref{tab:coresample}, except for
    NGC\,1374 (each row for each galaxy). The small lower panel of
    each plot shows the residual profile, i.e. the difference between
    the convolved best-fit model and the observed profile. In the left
    column, we show the \sersic\ fit. The black asterisks show the
    observed profile, the blue diamonds in the upper panels are
    the convolved best-fit \sersic\ function and the blue line is the
    intrinsic \sersic\ function (before PSF convolution). In the right
    column, we show the \cs\ (or \cs\ + envelope) fit. The red diamonds in the upper
      panels are the convolved best-fit \cs\ (+ envelope) function, the solid red line
    shows the intrinsic function (before PSF convolution)
    and the blue line is the inward extrapolation of the outer, \sersic\ part of 
    the \cs\ function. For fits with an outer envelope, the dashed red
    lines indicate the envelope components by themselves. The diamonds coincide exactly with the
    lines of the same color for galaxies whose light profile is
    derived from a deconvolved image, since there is no PSF
    convolution in the fit. The vertical dotted lines mark
    the break radius ($r_b$). The rms of each
    \sersic\ and \cs\ (+ envelope) fit is shown and the latter is taken as the
    uncertainty in the surface brightness profile, thereby setting the
    $\chi^2$ value of the \cs\ (+ envelope) fit to 1.0. The fit range is limited to
    a certain radius for several galaxies (see text). For galaxies
    with profiles taken from \citetalias{Kormendy-09a}, a green dashed line is plotted
    based on the \sersic\ function that best fits the outer part of
    the galaxy according to \citetalias{Kormendy-09a}. For IC\,1459, an additional
      Gaussian component is included in the fits to account for excess
      nuclear light.}
\label{fittingplots}
\end{figure*}

\addtocounter{figure}{-1}

\begin{figure*}
\centering
  \includegraphics[scale=0.5]{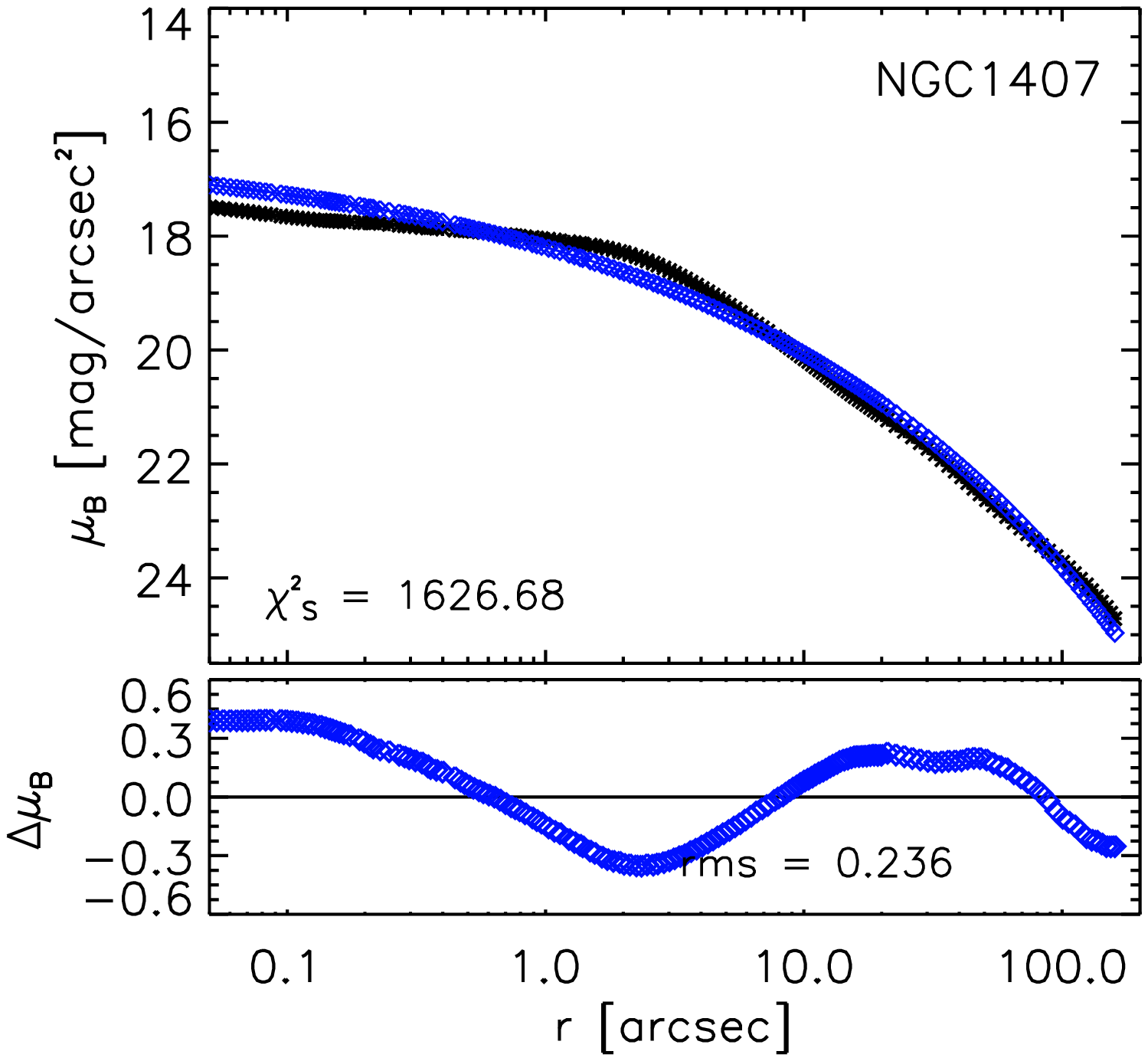}\includegraphics[scale=0.5]{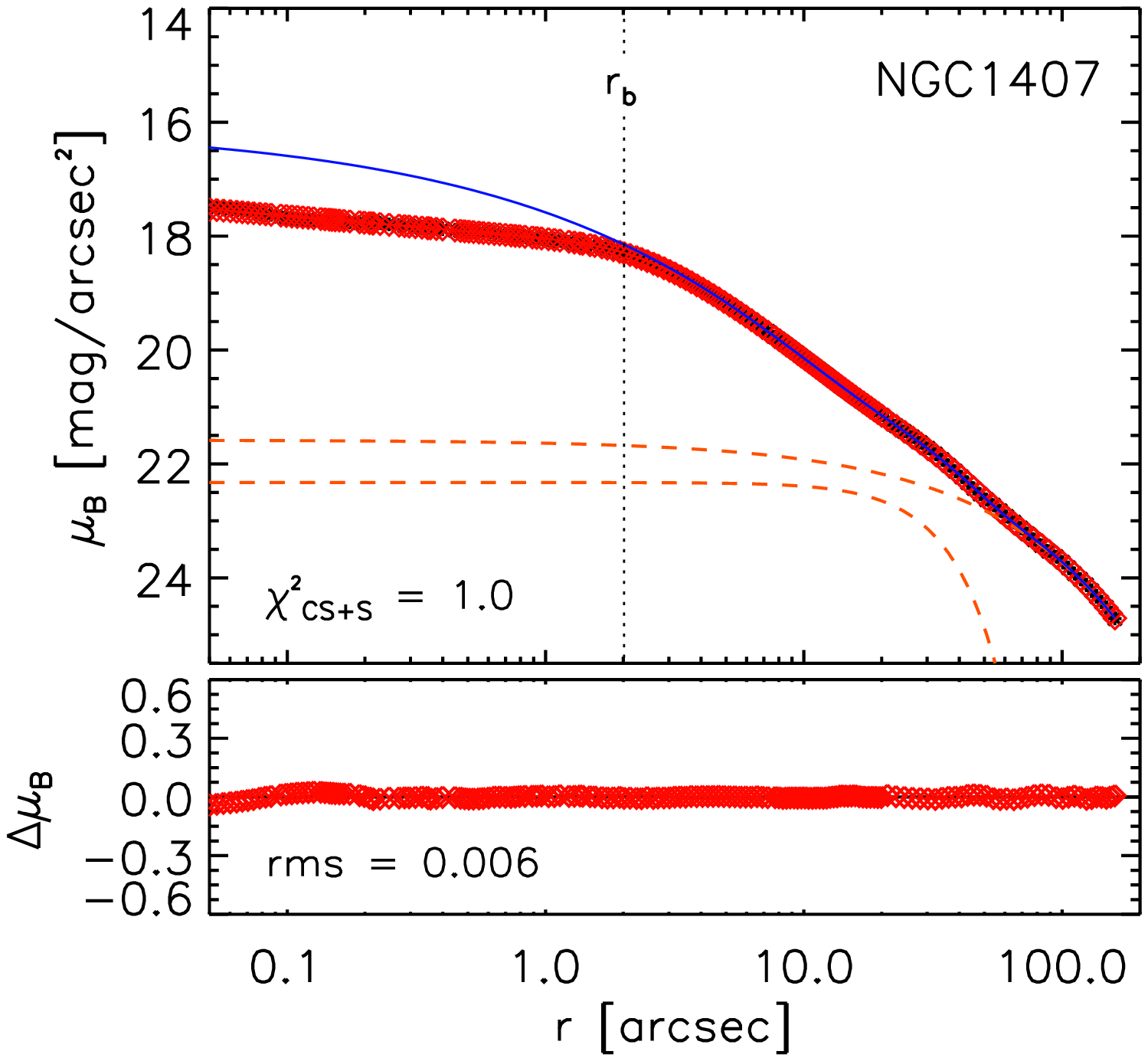}
  \includegraphics[scale=0.5]{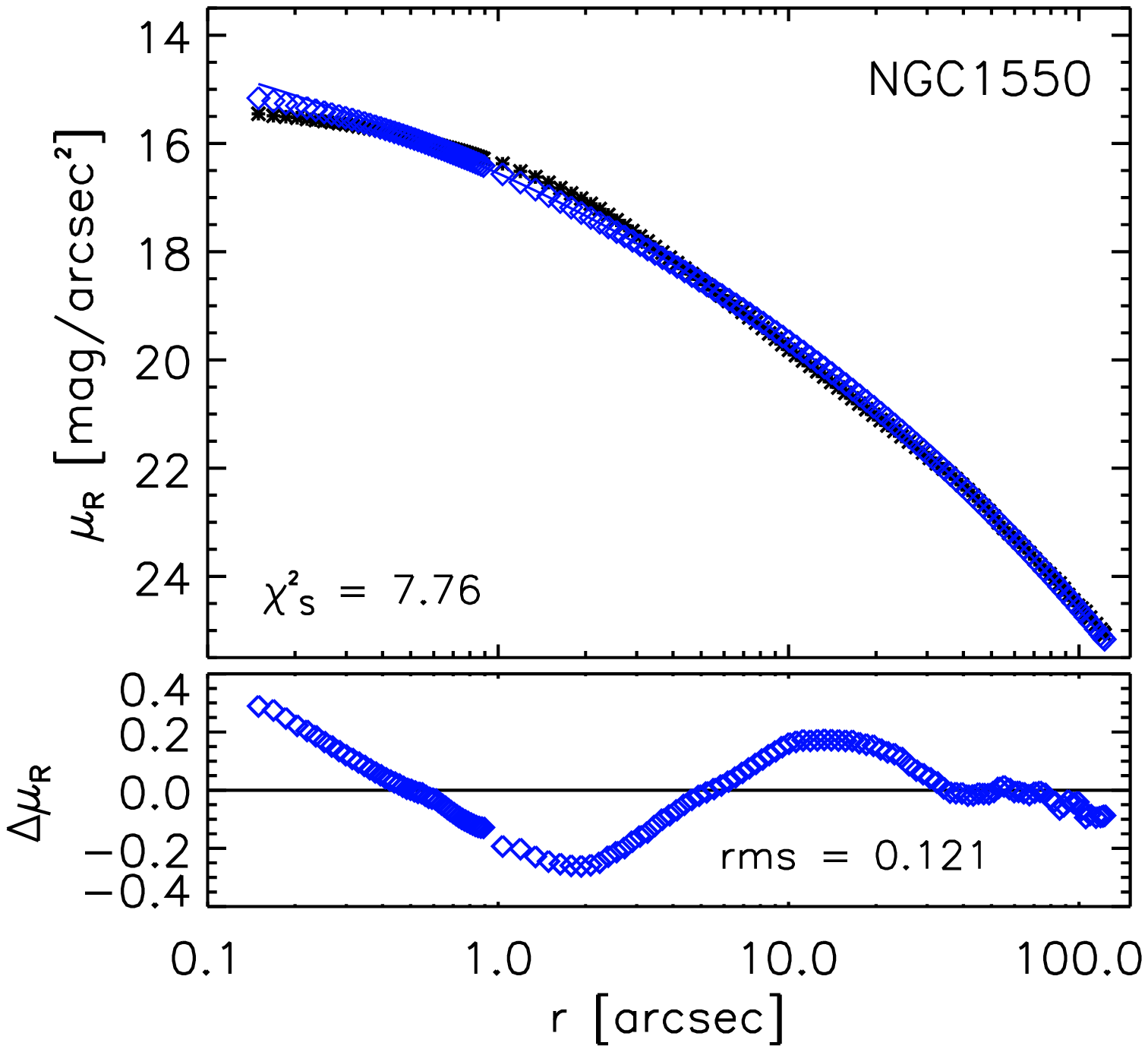}\includegraphics[scale=0.5]{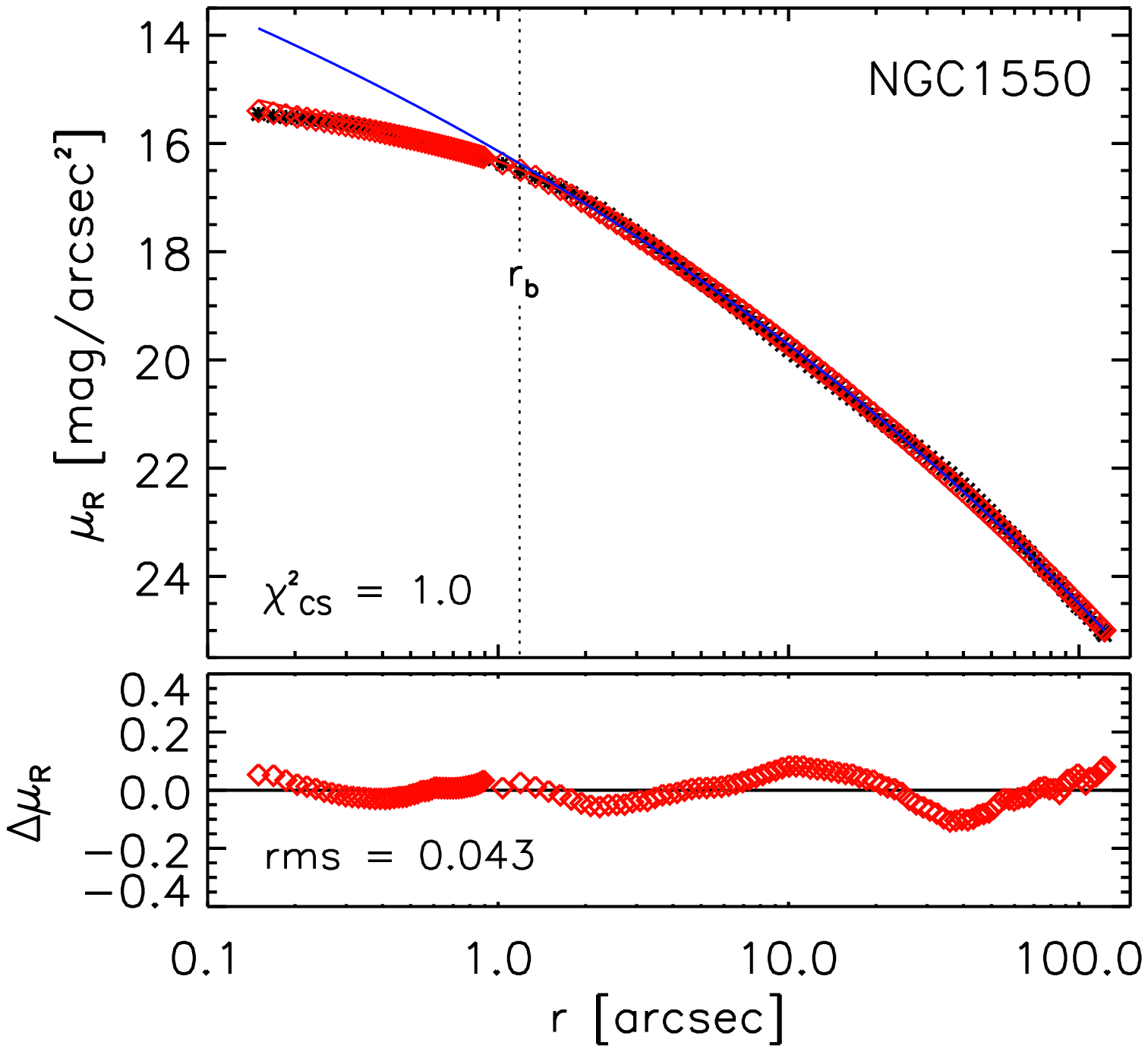}
  \includegraphics[scale=0.5]{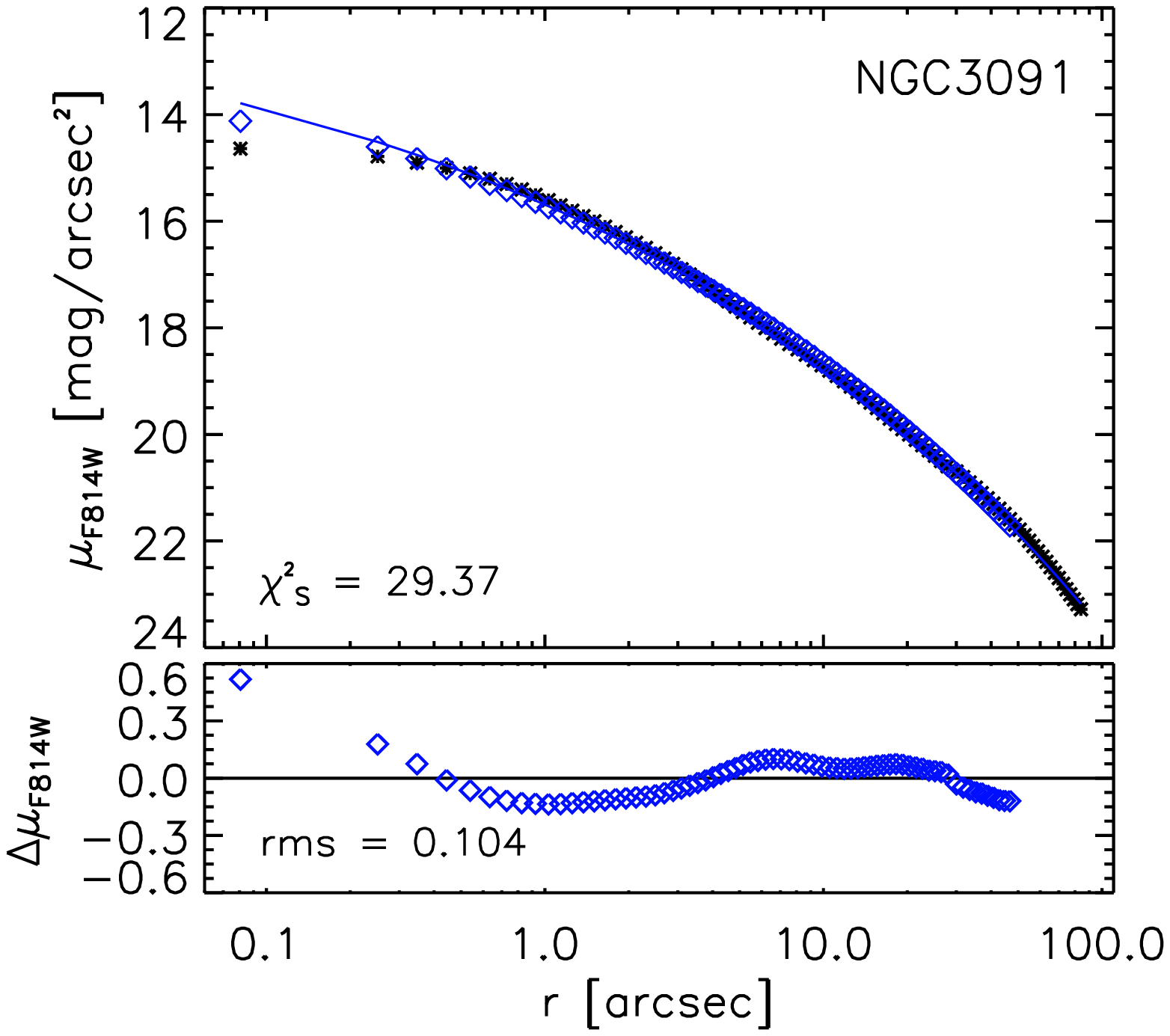}\includegraphics[scale=0.5]{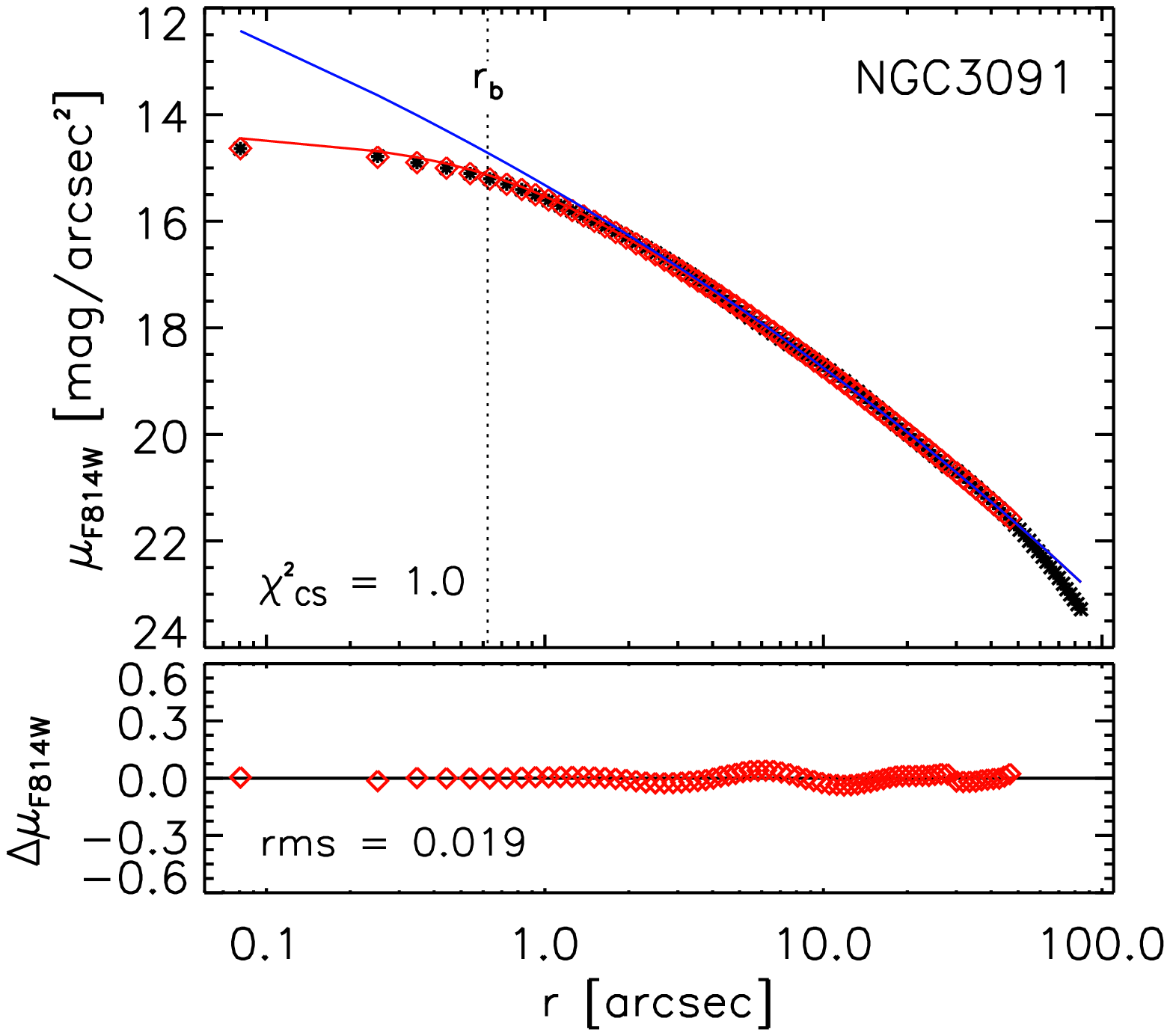}
\caption[]{continued.}
\end{figure*}

\addtocounter{figure}{-1}

\begin{figure*}
\centering
  \includegraphics[scale=0.5]{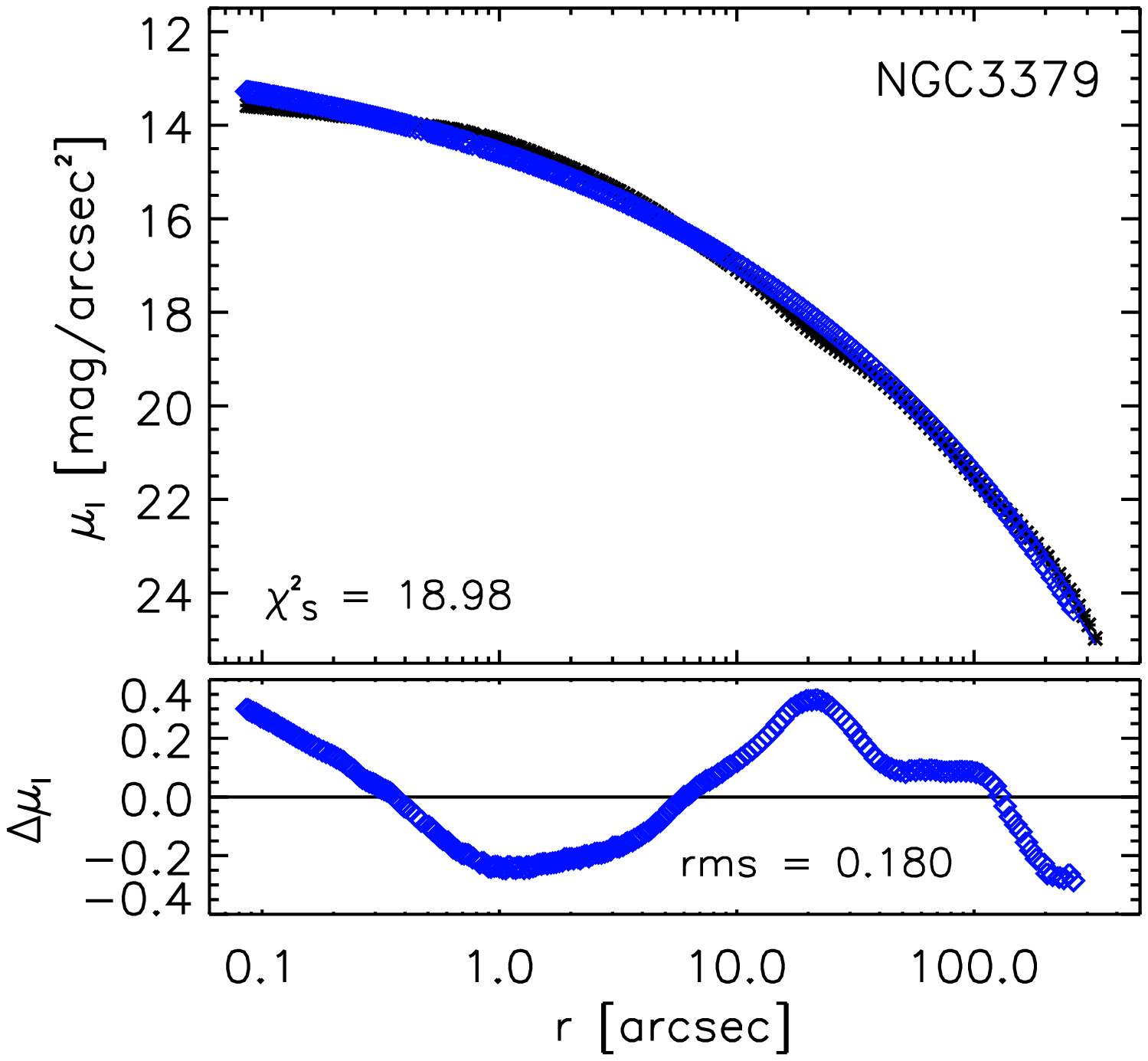}\includegraphics[scale=0.5]{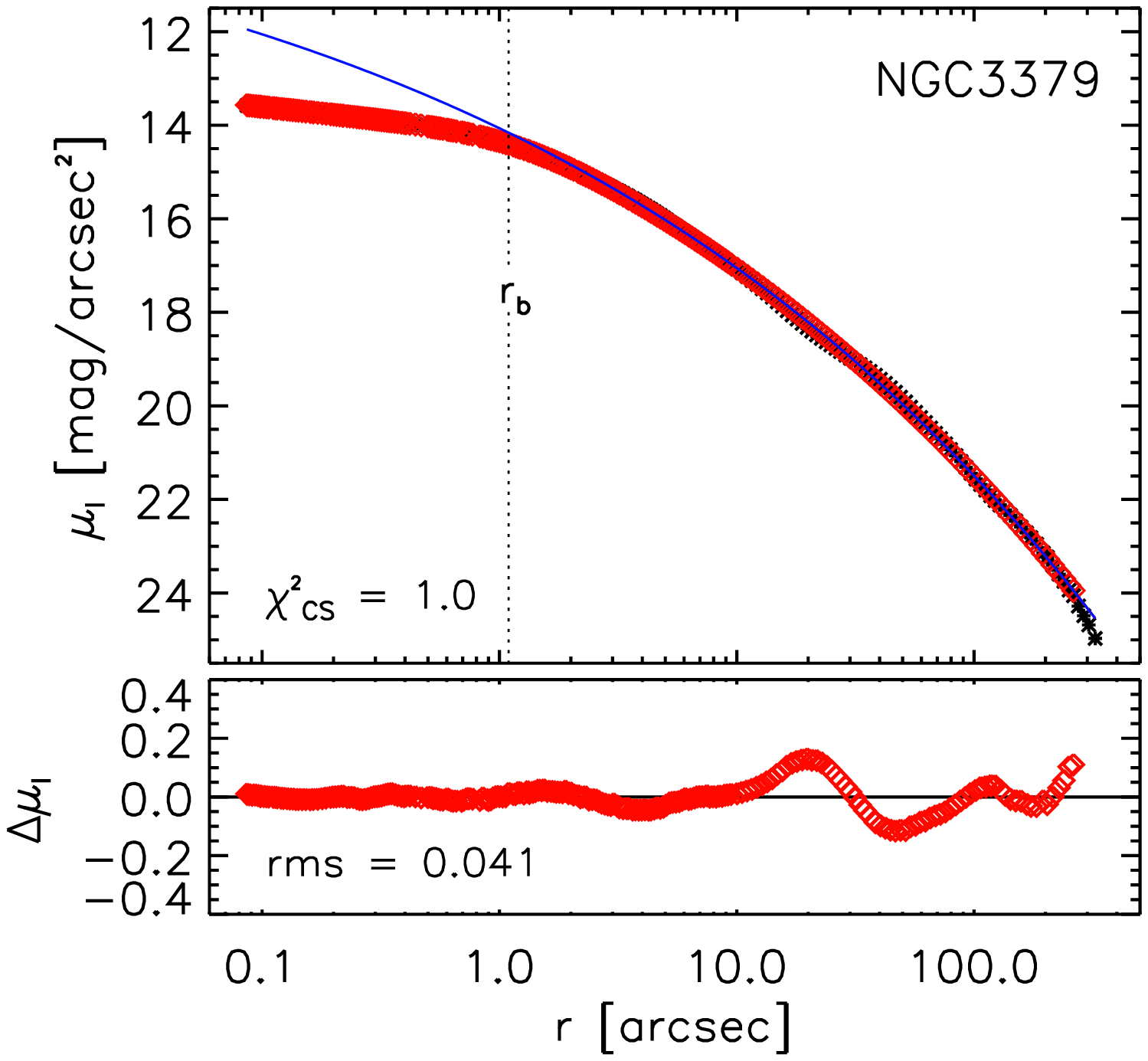}
  \includegraphics[scale=0.5]{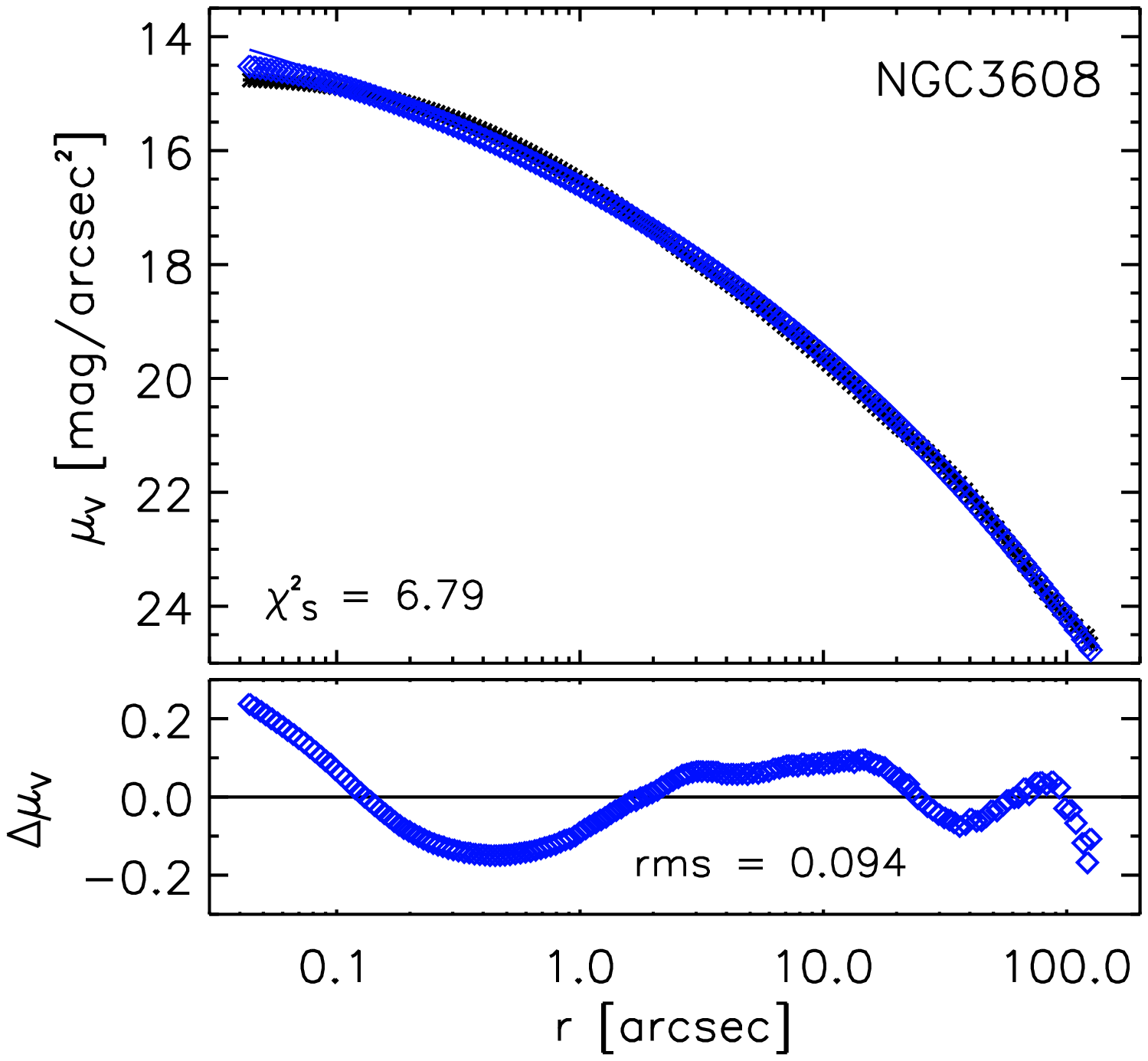}\includegraphics[scale=0.5]{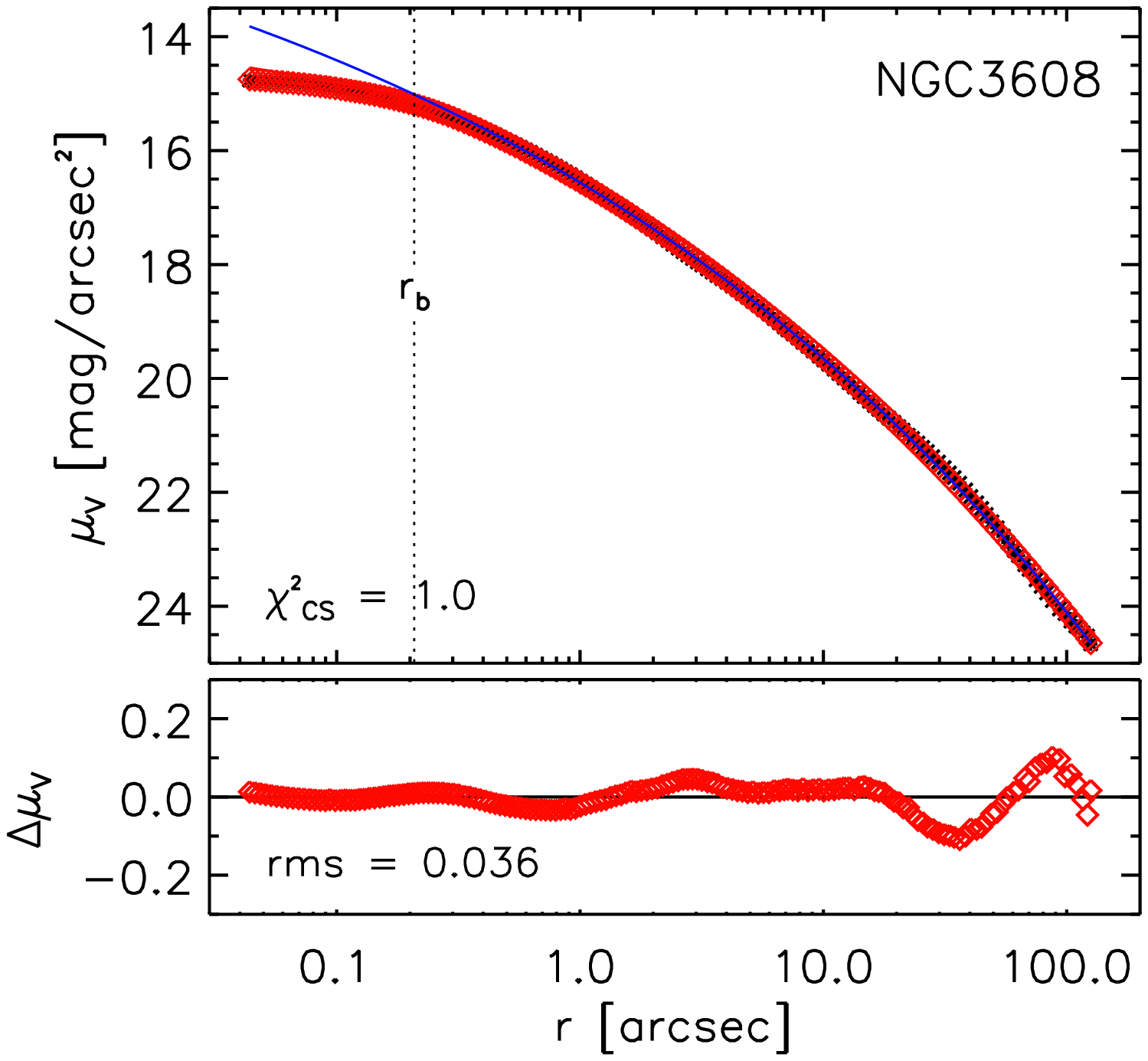}
  \includegraphics[scale=0.5]{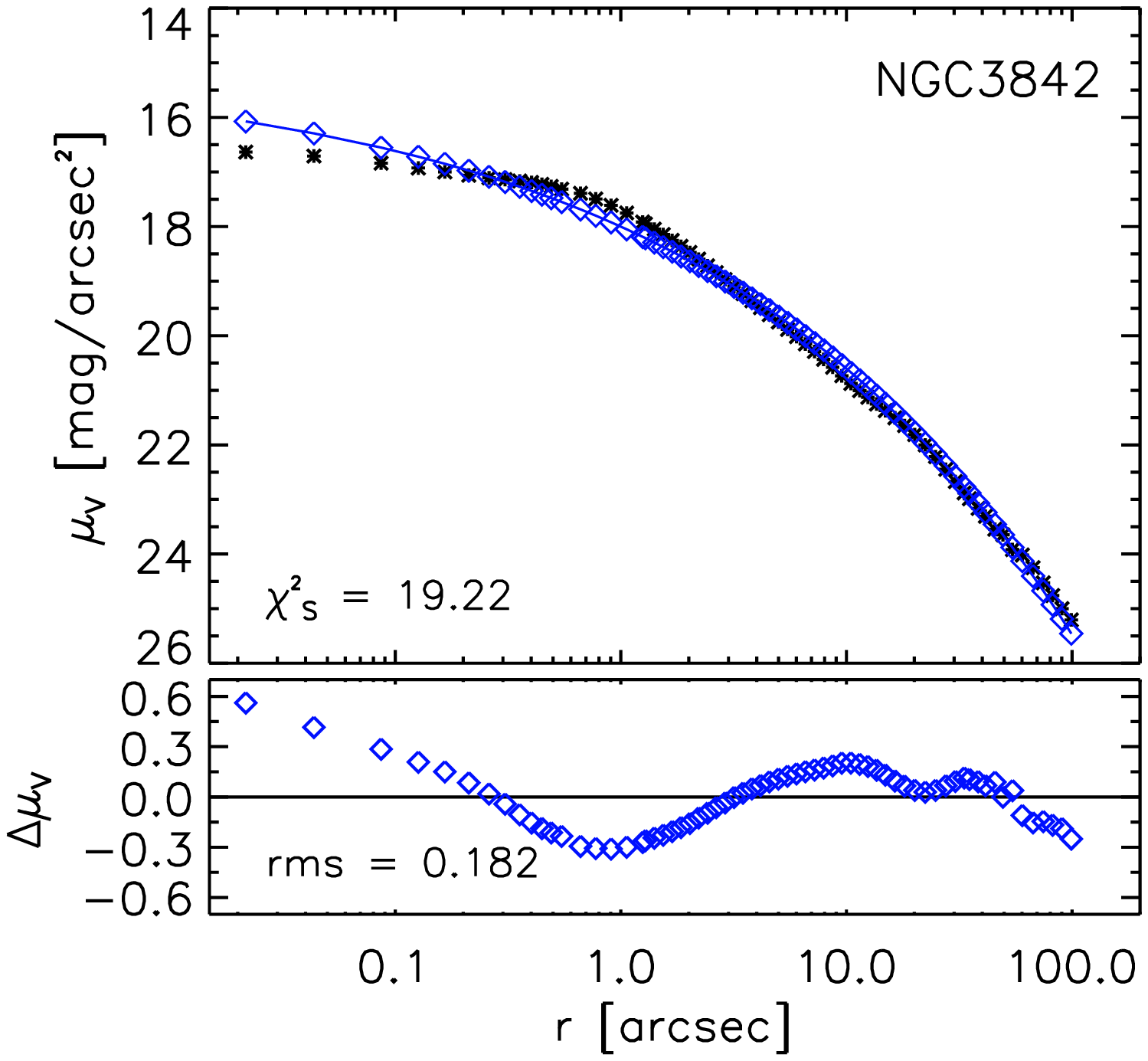}\includegraphics[scale=0.5]{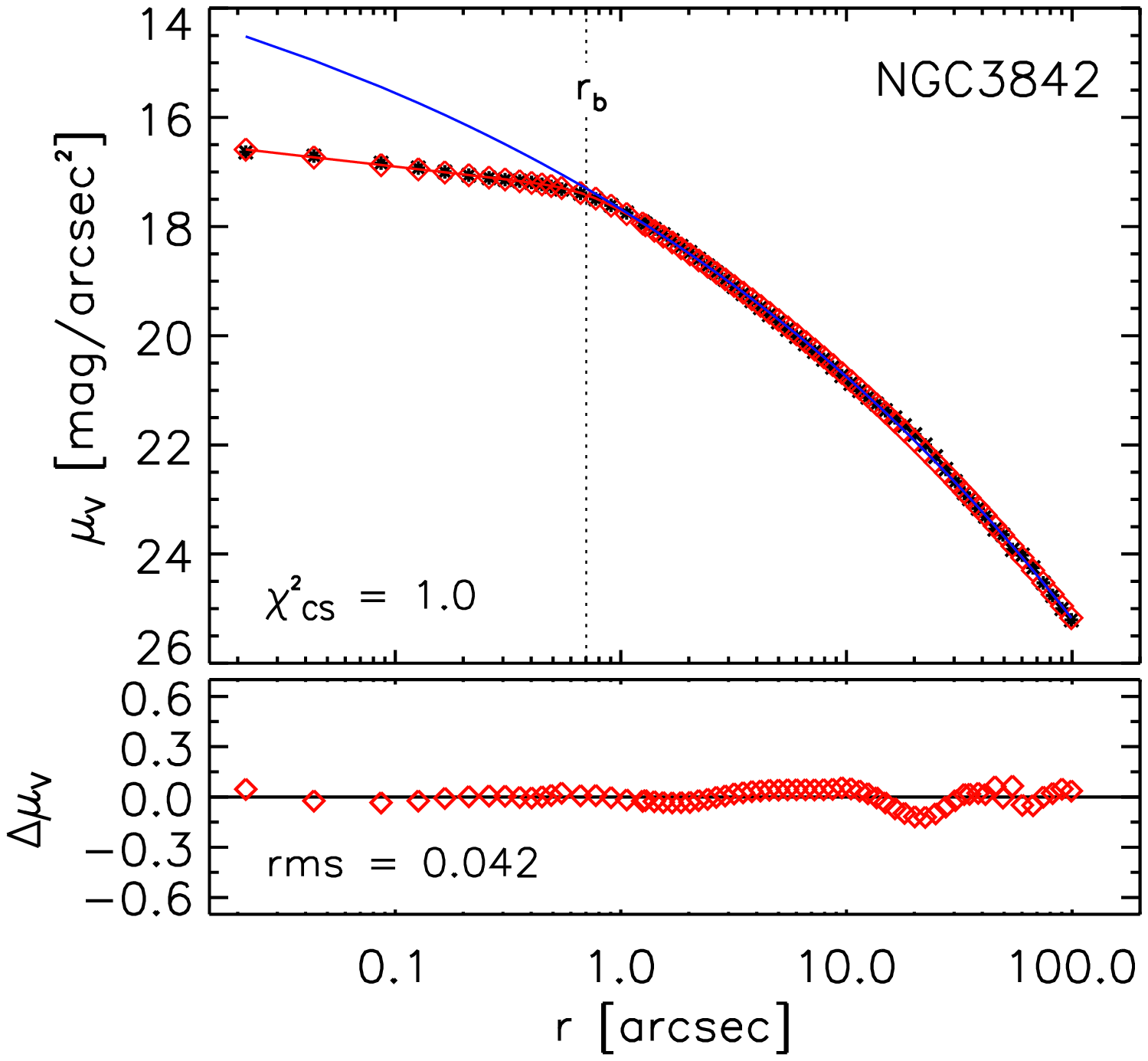}
\caption[]{continued.}
\end{figure*}

\addtocounter{figure}{-1}

\begin{figure*}
\centering
  \includegraphics[scale=0.5]{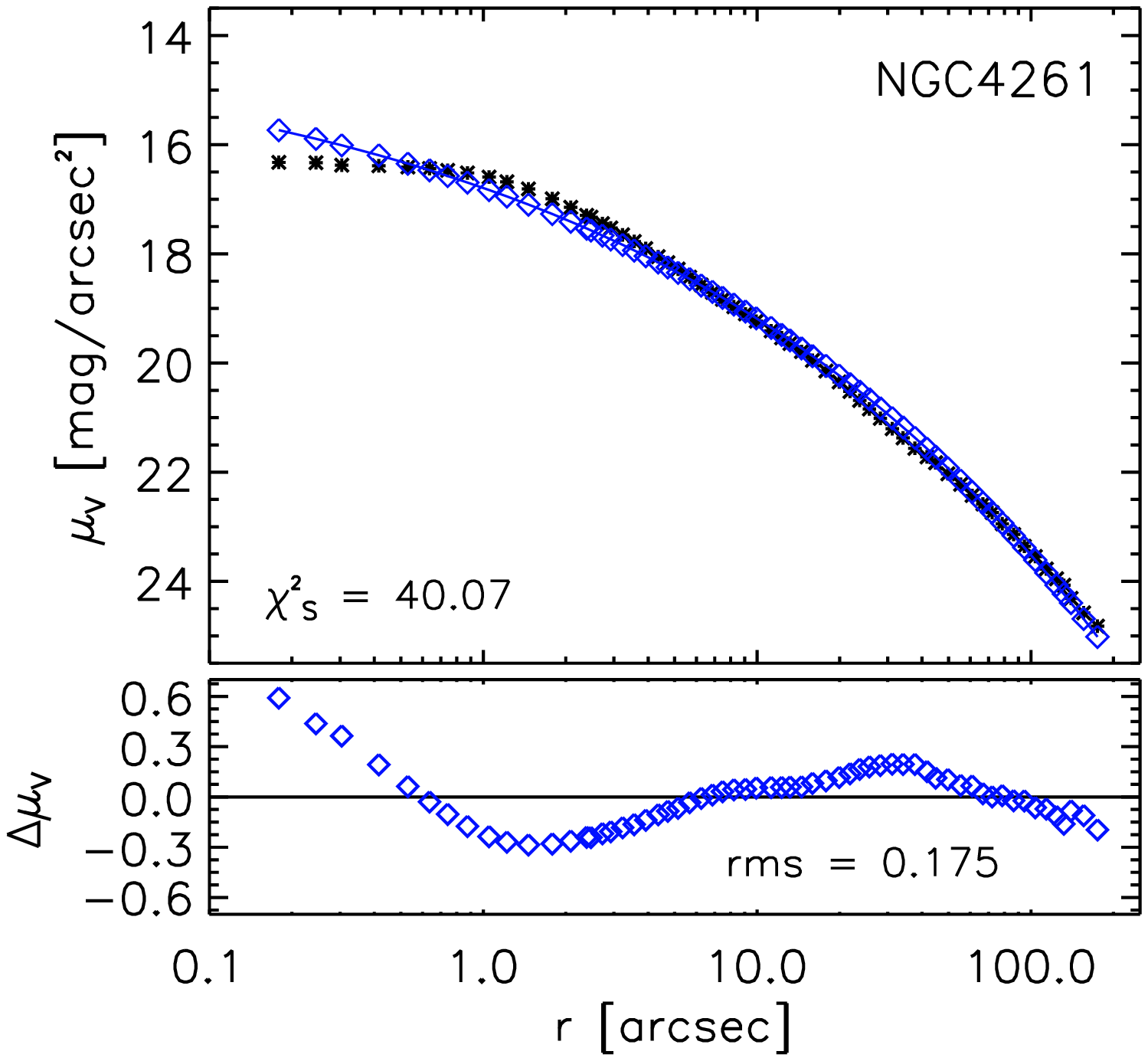}\includegraphics[scale=0.5]{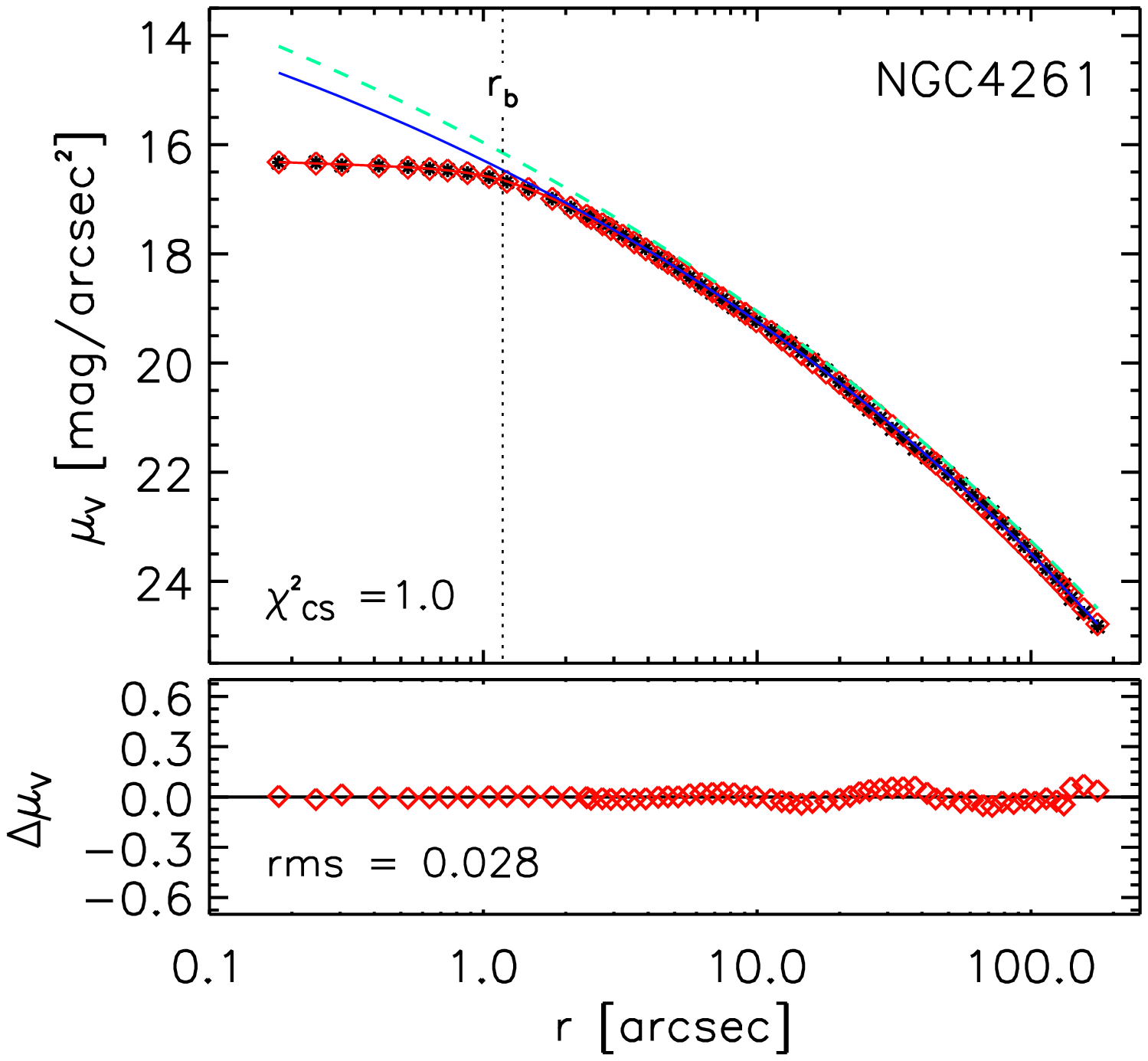}
  \includegraphics[scale=0.5]{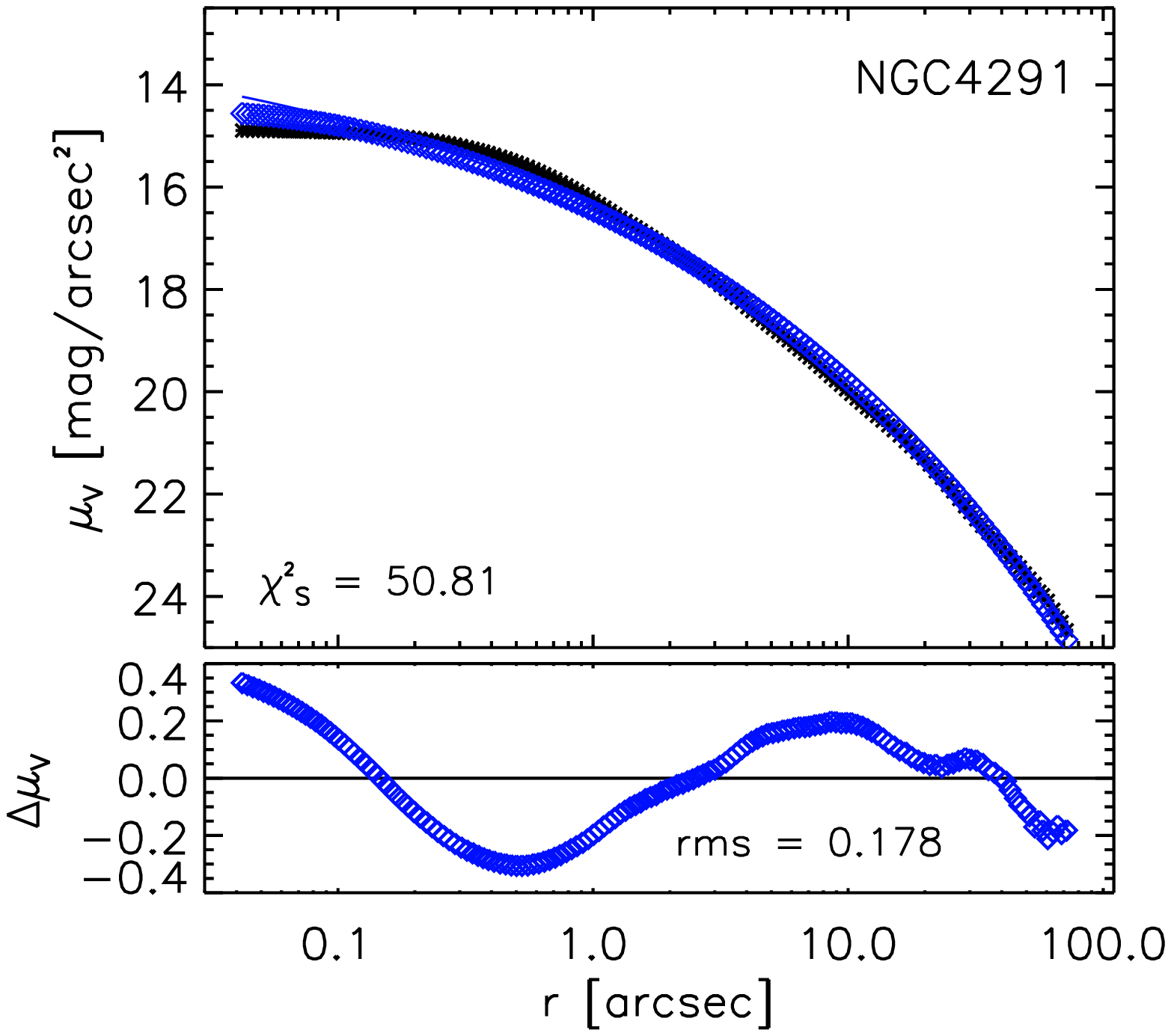}\includegraphics[scale=0.5]{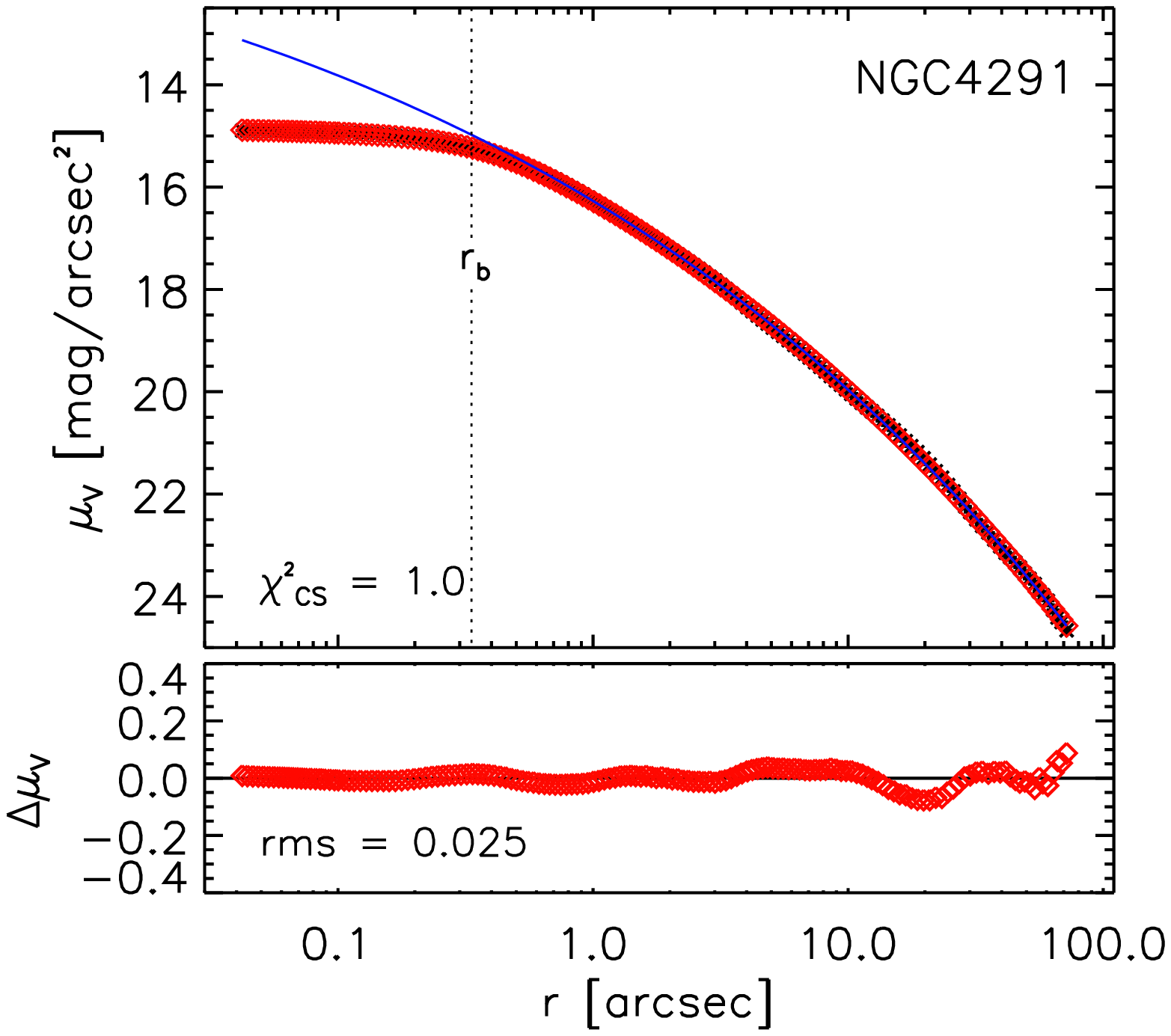}
  \includegraphics[scale=0.5]{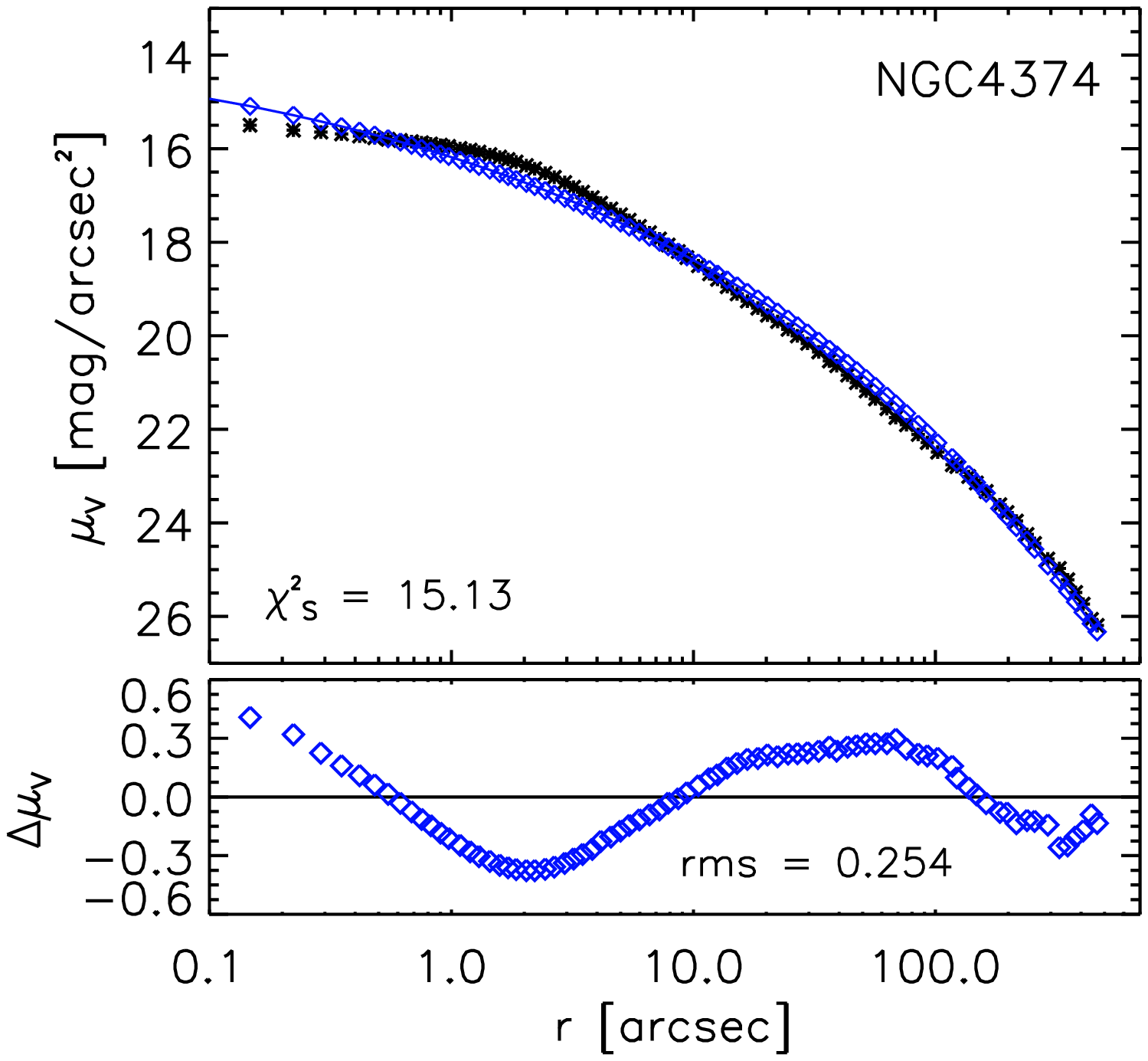}\includegraphics[scale=0.5]{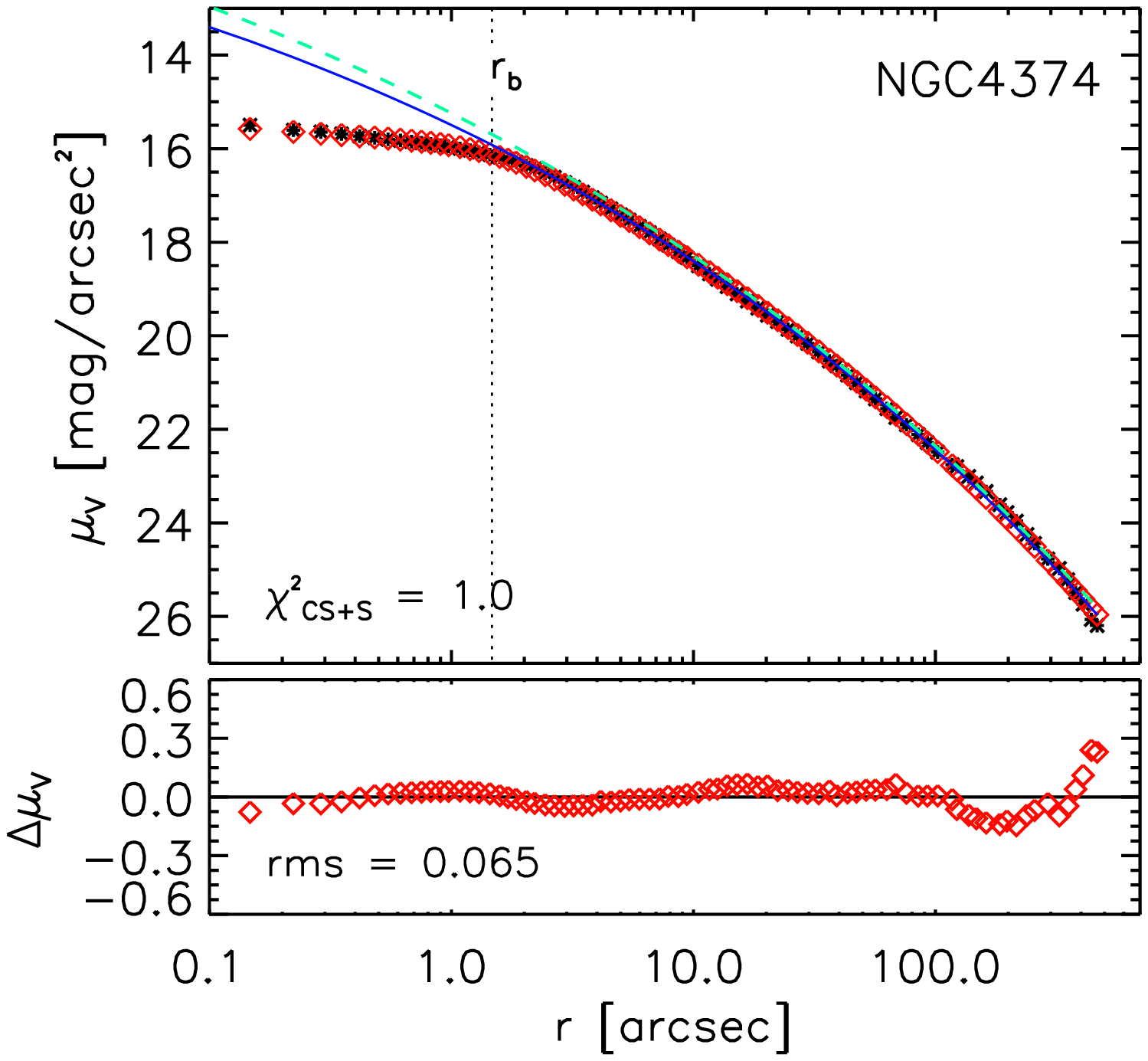}
\caption[]{continued.}
\end{figure*}

\addtocounter{figure}{-1}

\begin{figure*}
\centering
  \includegraphics[scale=0.50]{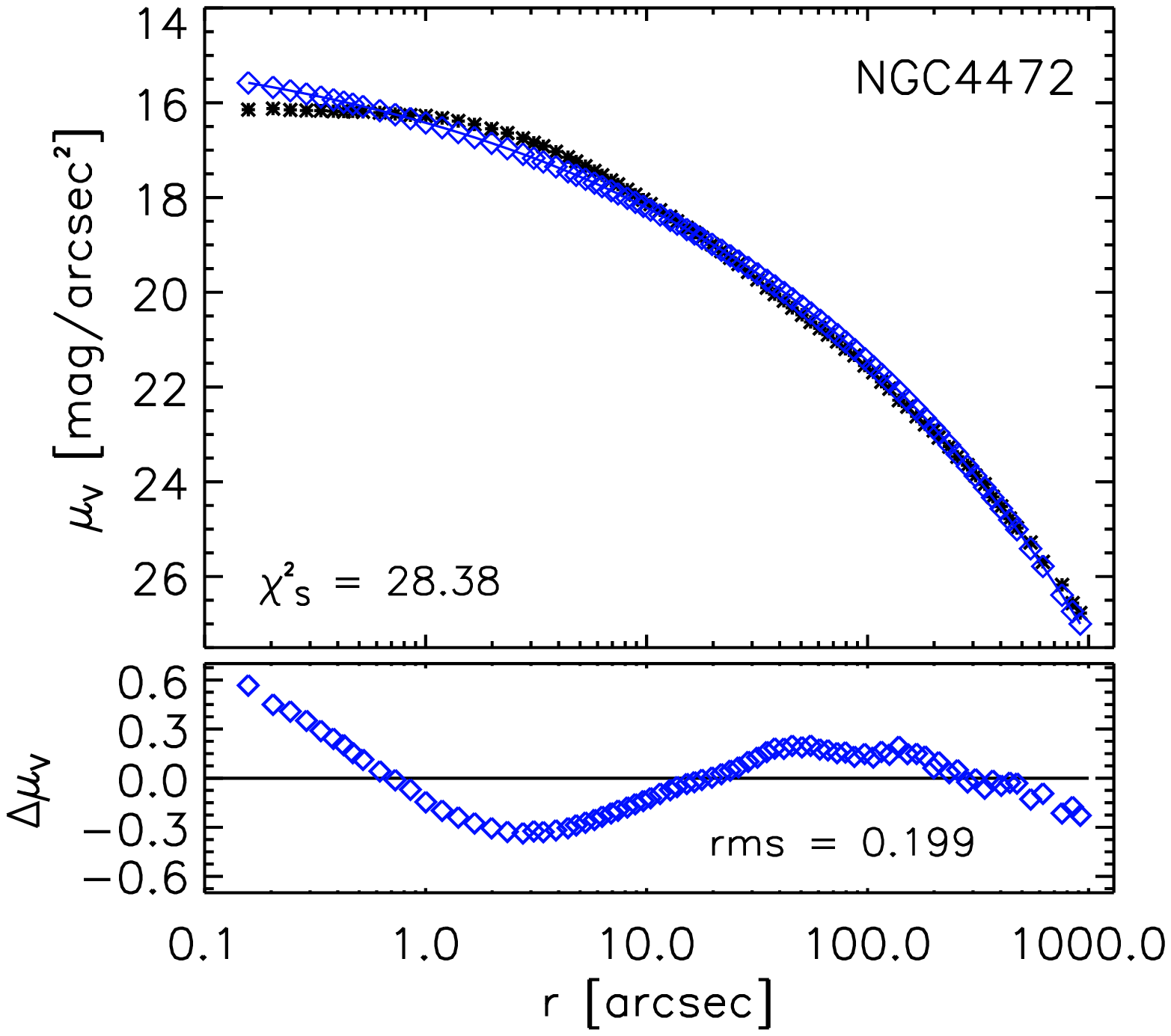}\includegraphics[scale=0.5]{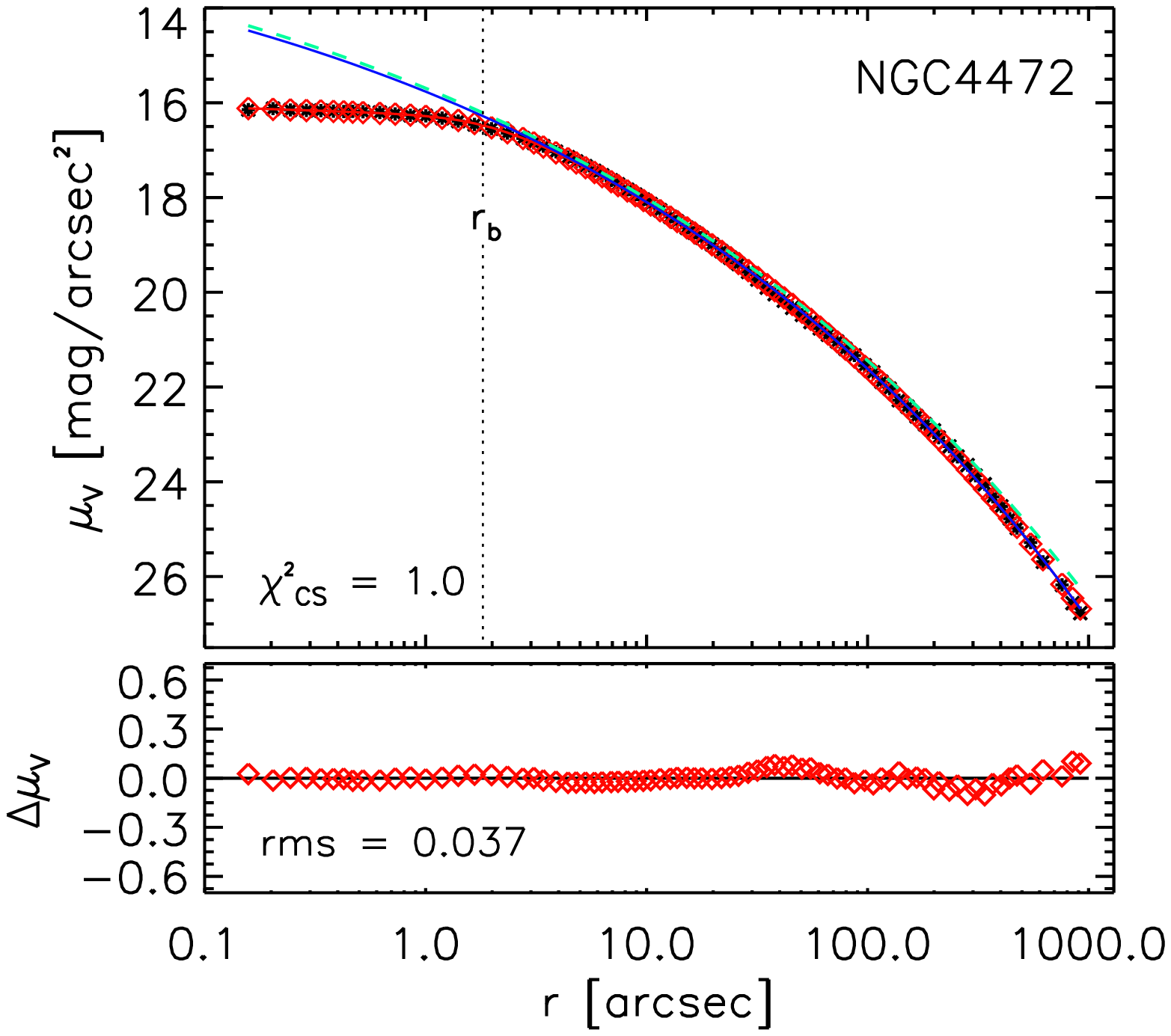}
  \includegraphics[scale=0.50]{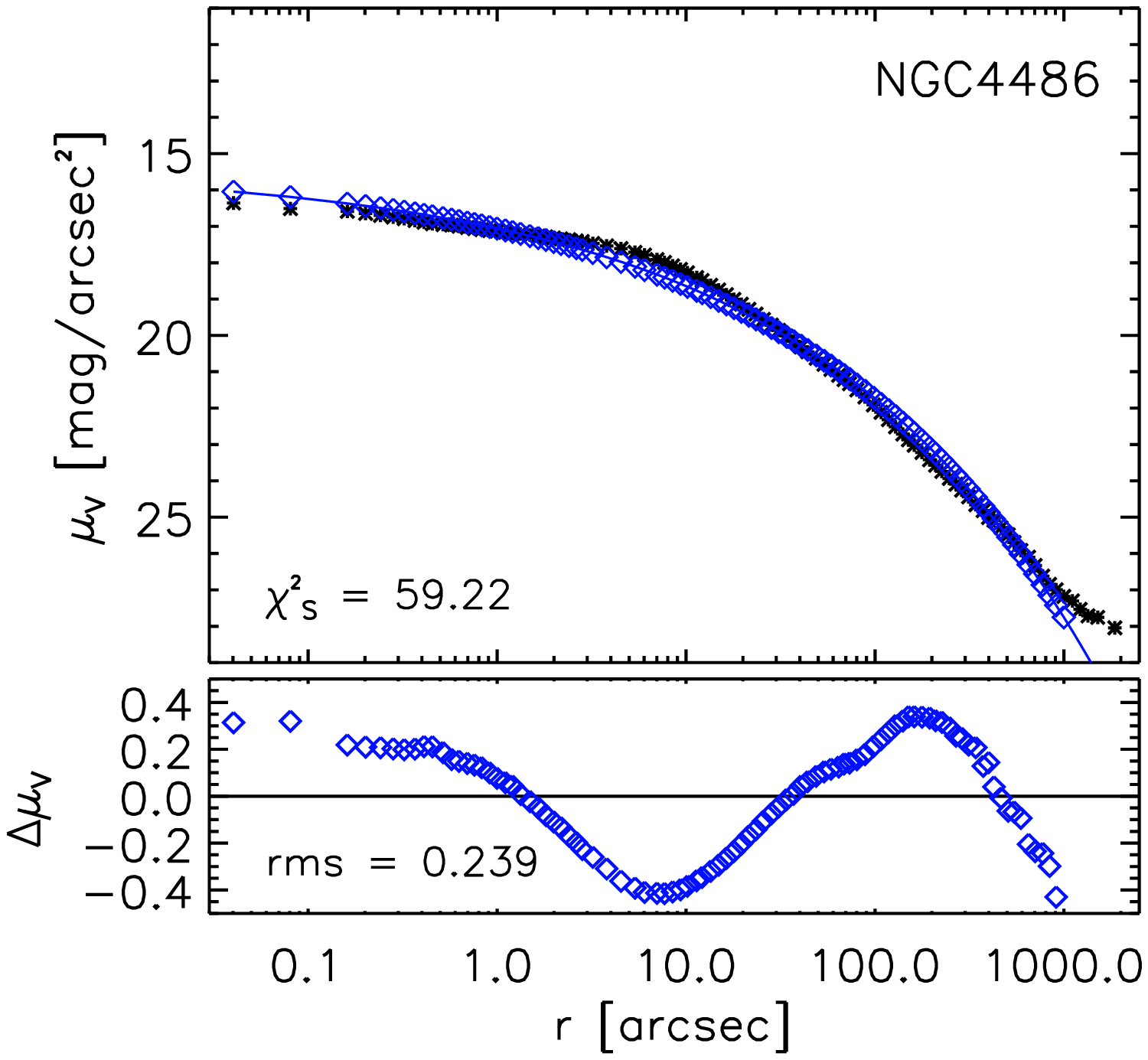}\includegraphics[scale=0.50]{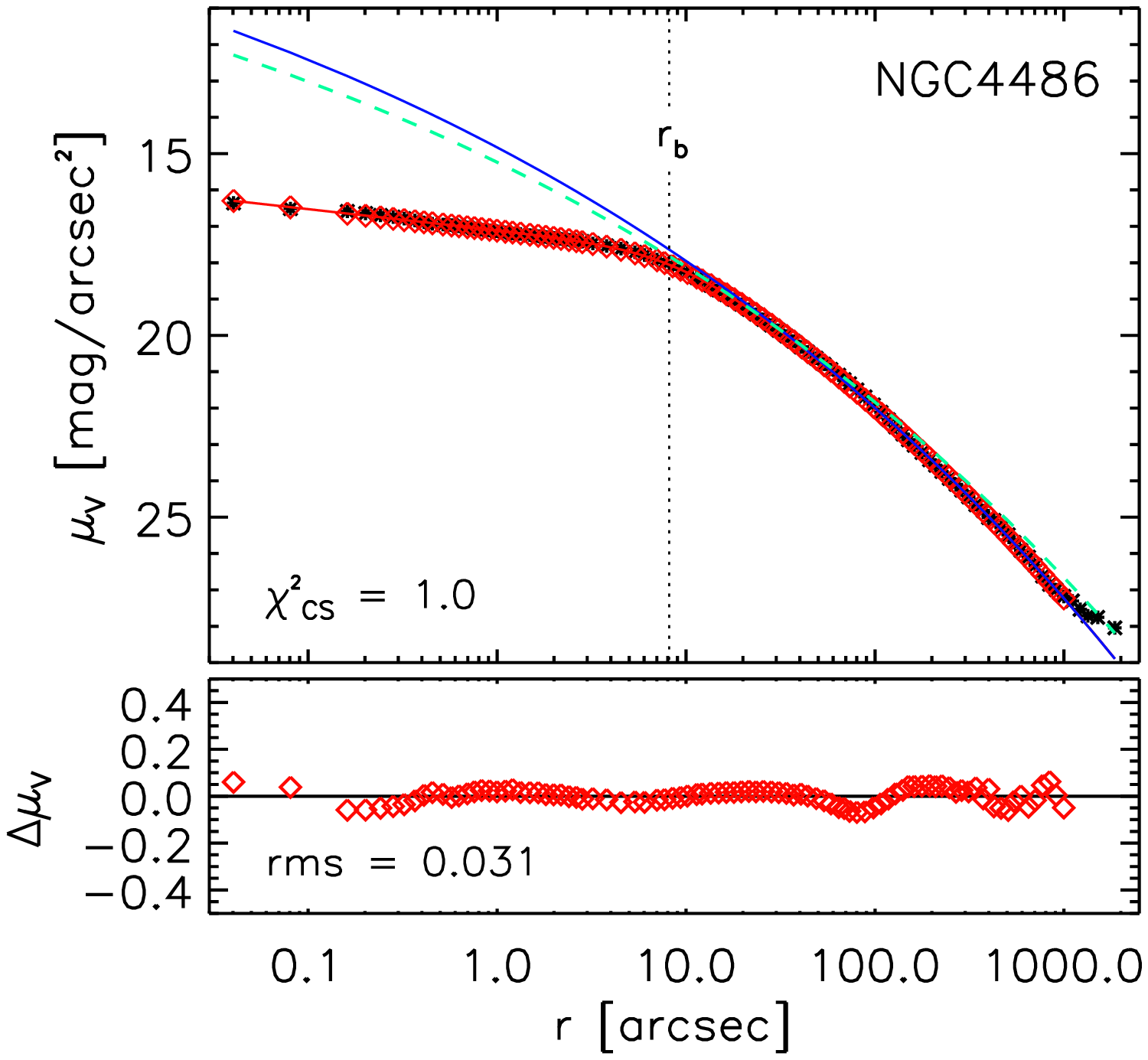}
  \includegraphics[scale=0.50]{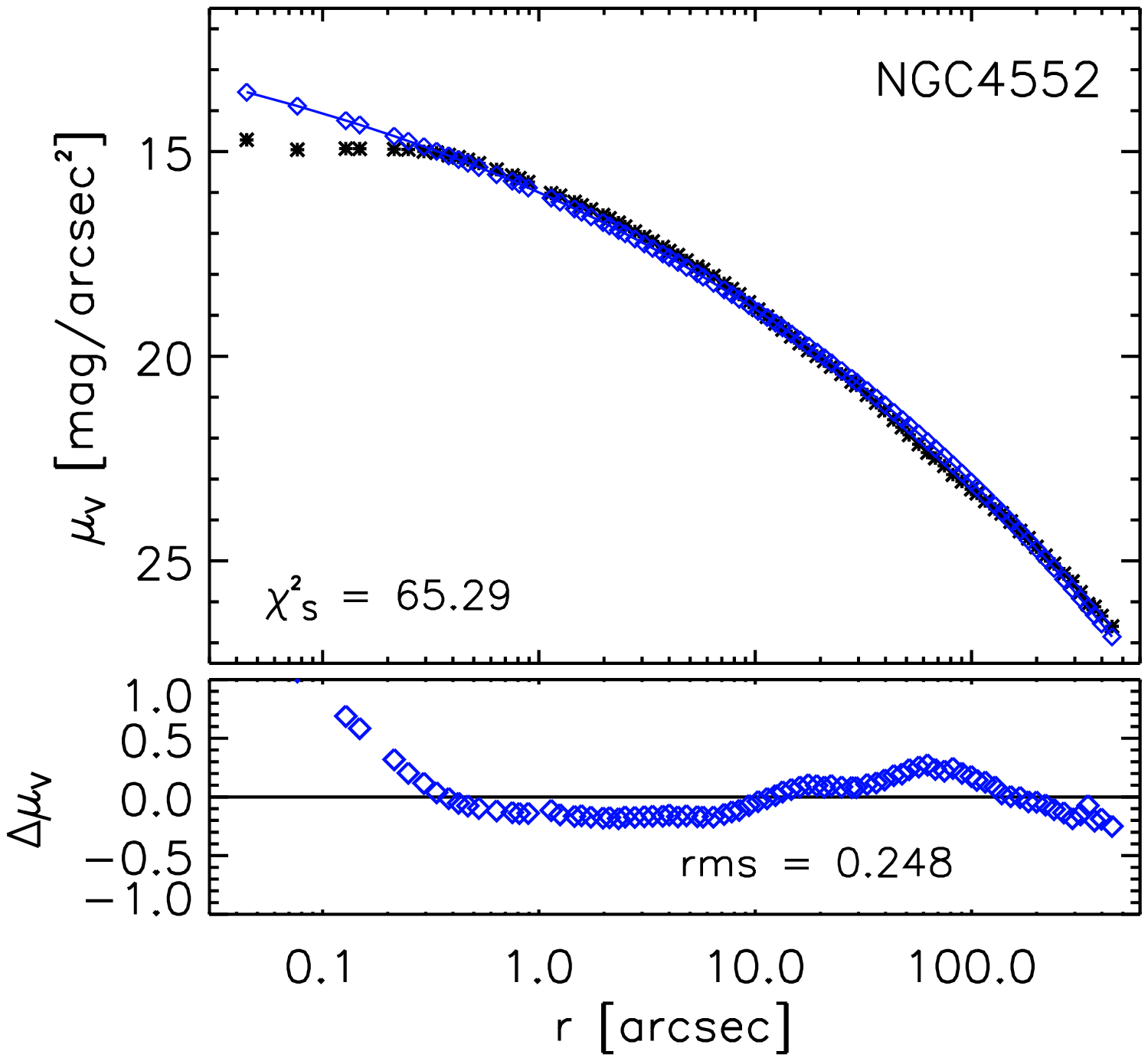}\includegraphics[scale=0.50]{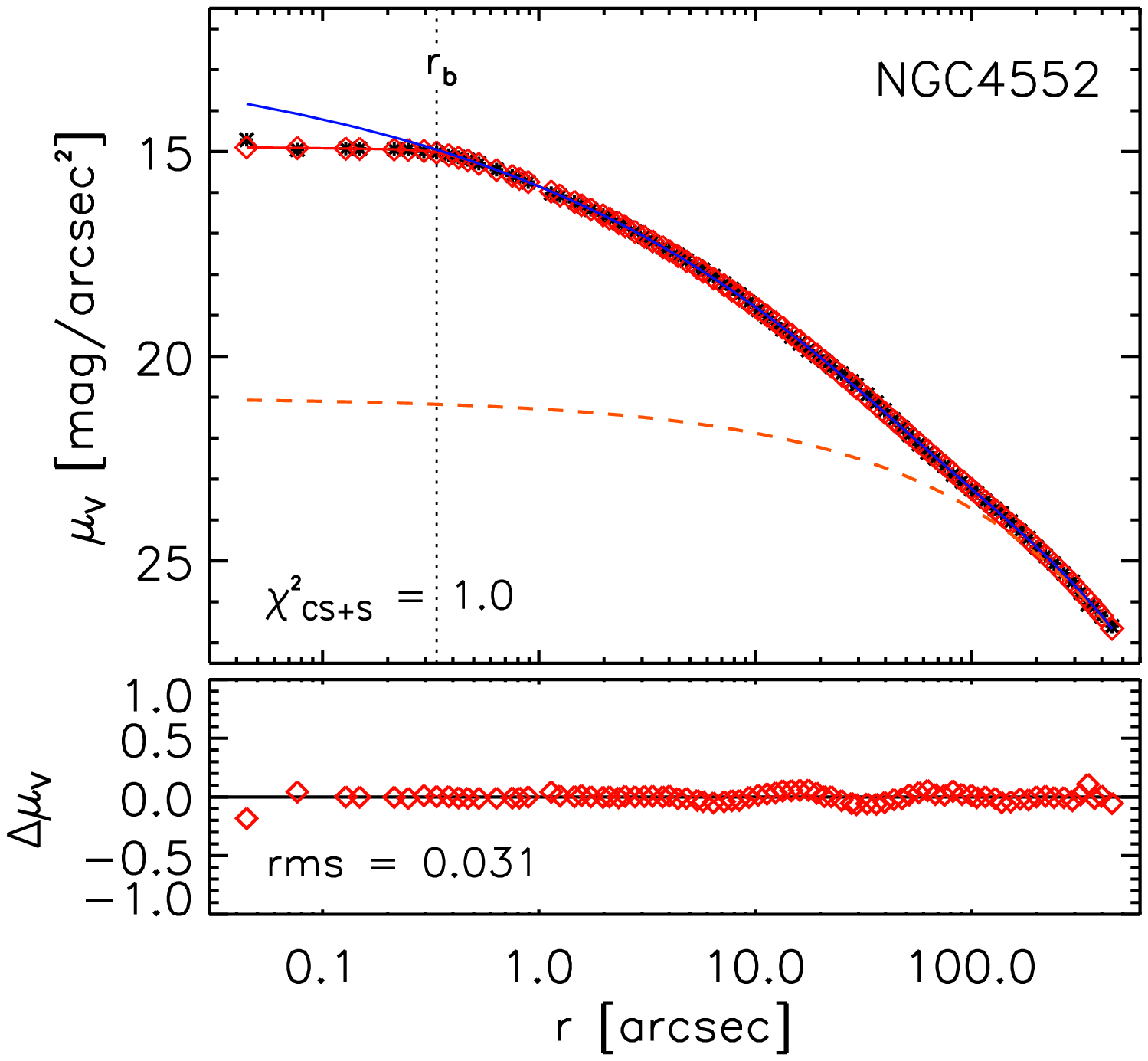}
\caption[]{continued.}
\end{figure*}

\addtocounter{figure}{-1}

\begin{figure*}
\centering
  \includegraphics[scale=0.50]{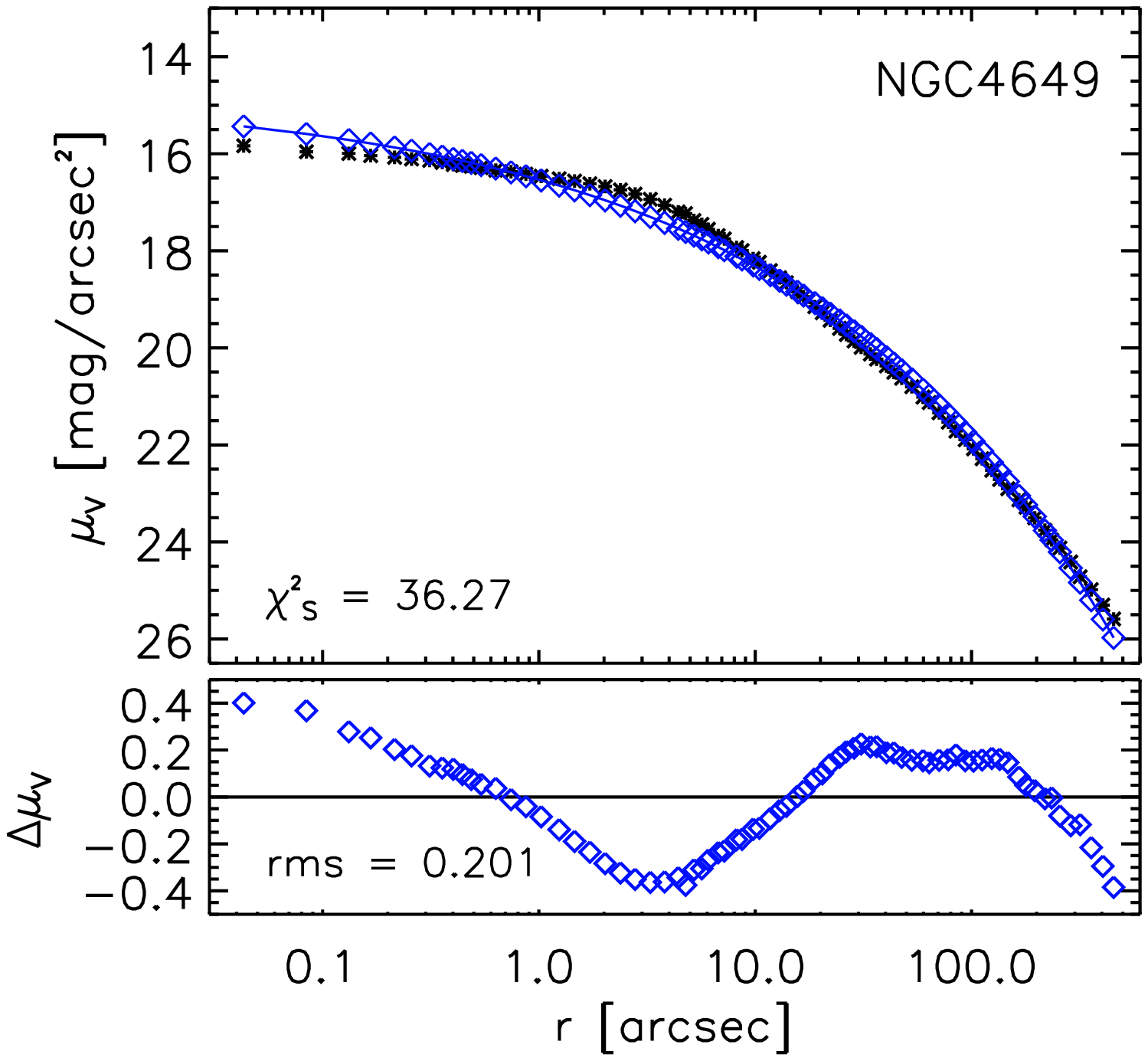}\includegraphics[scale=0.50]{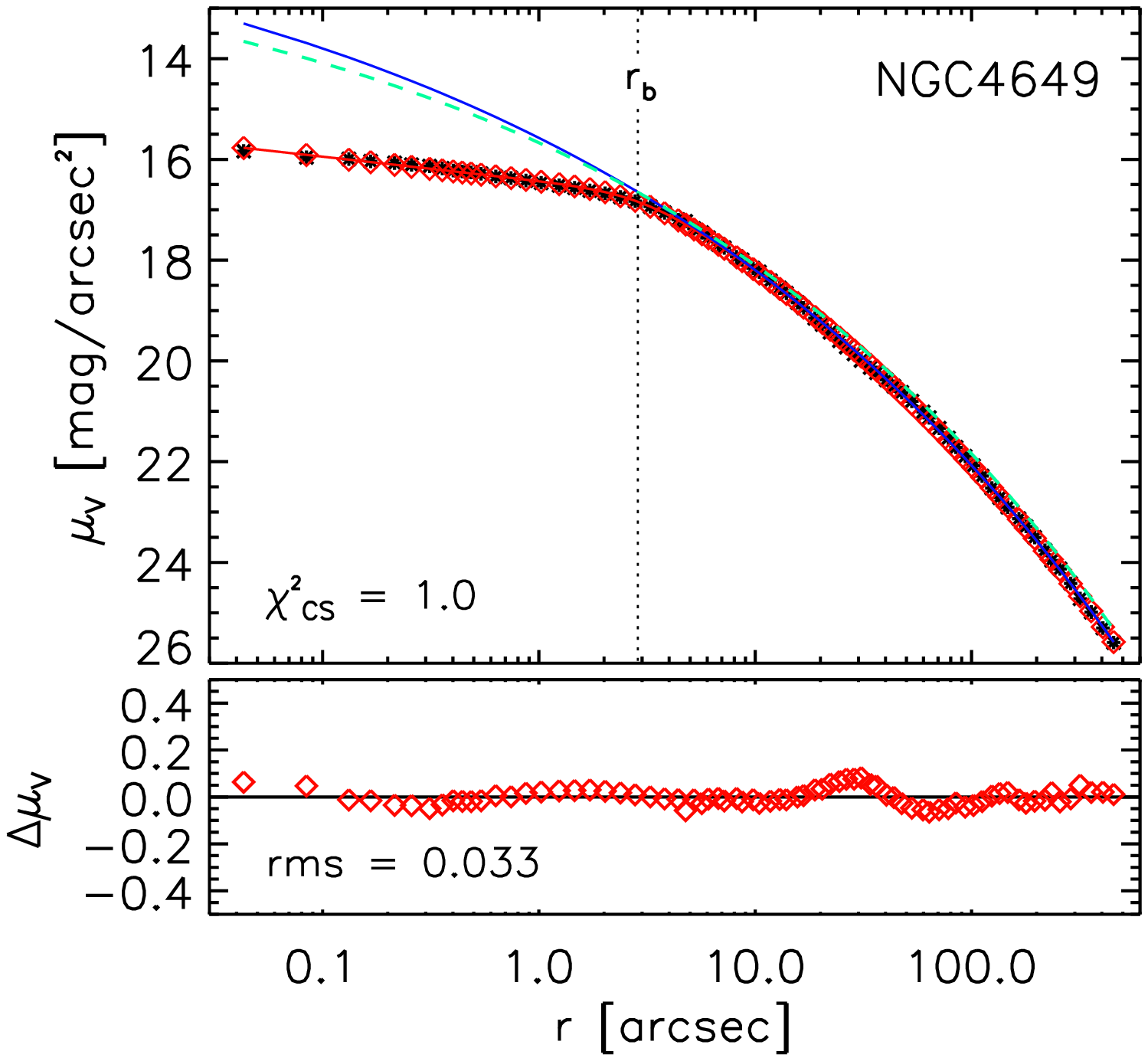}
  \includegraphics[scale=0.50]{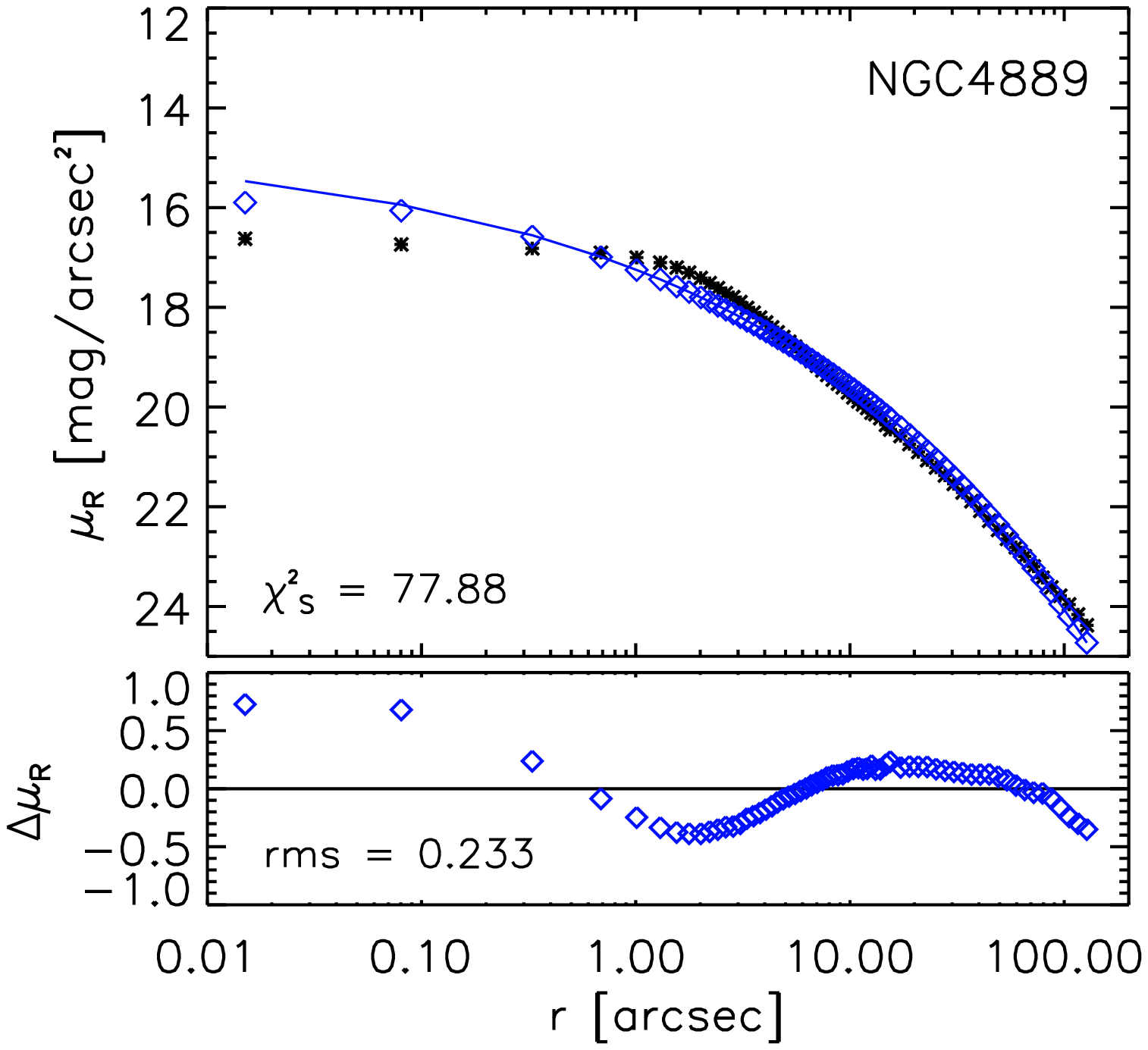}\includegraphics[scale=0.50]{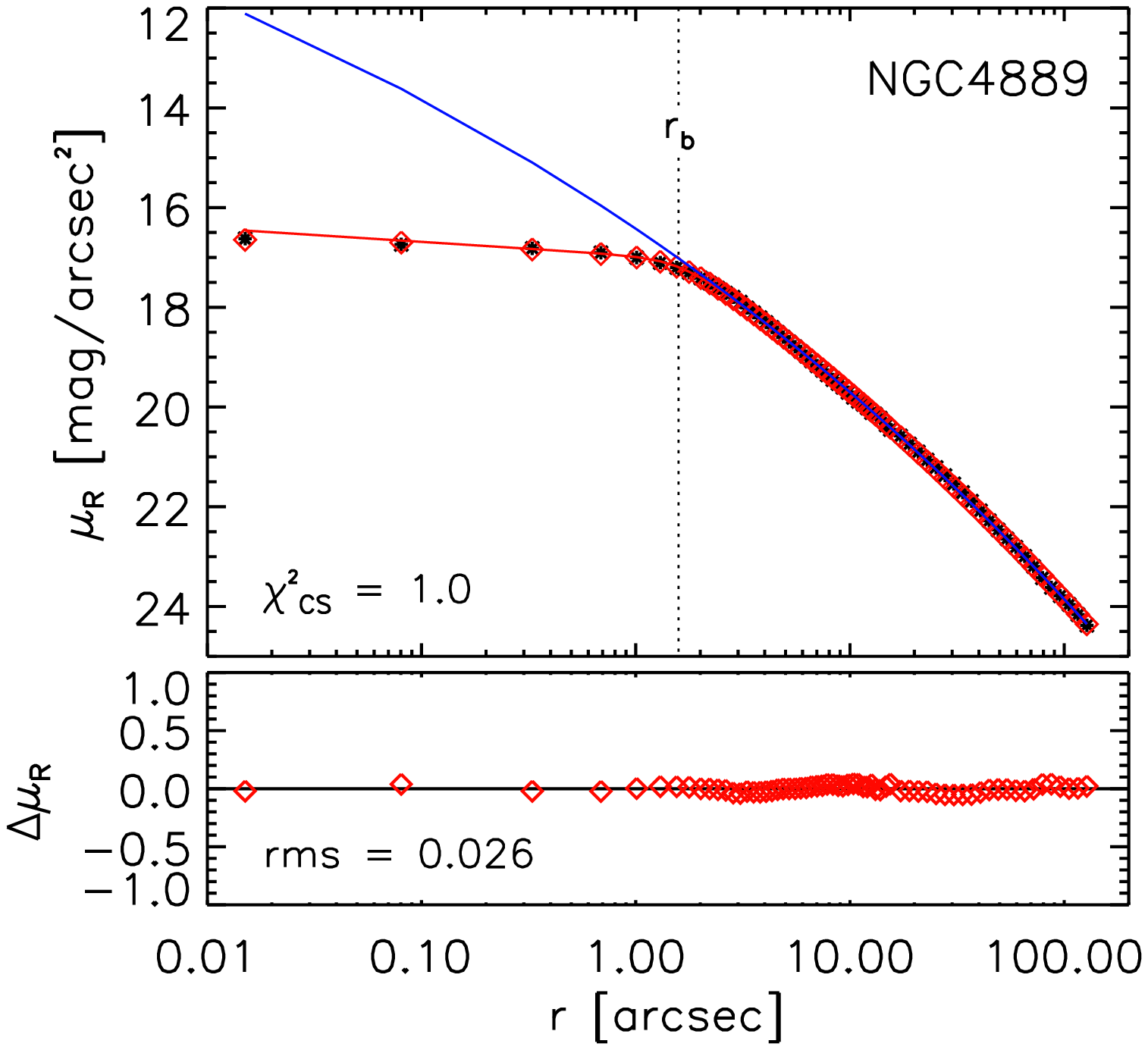}
  \includegraphics[scale=0.50]{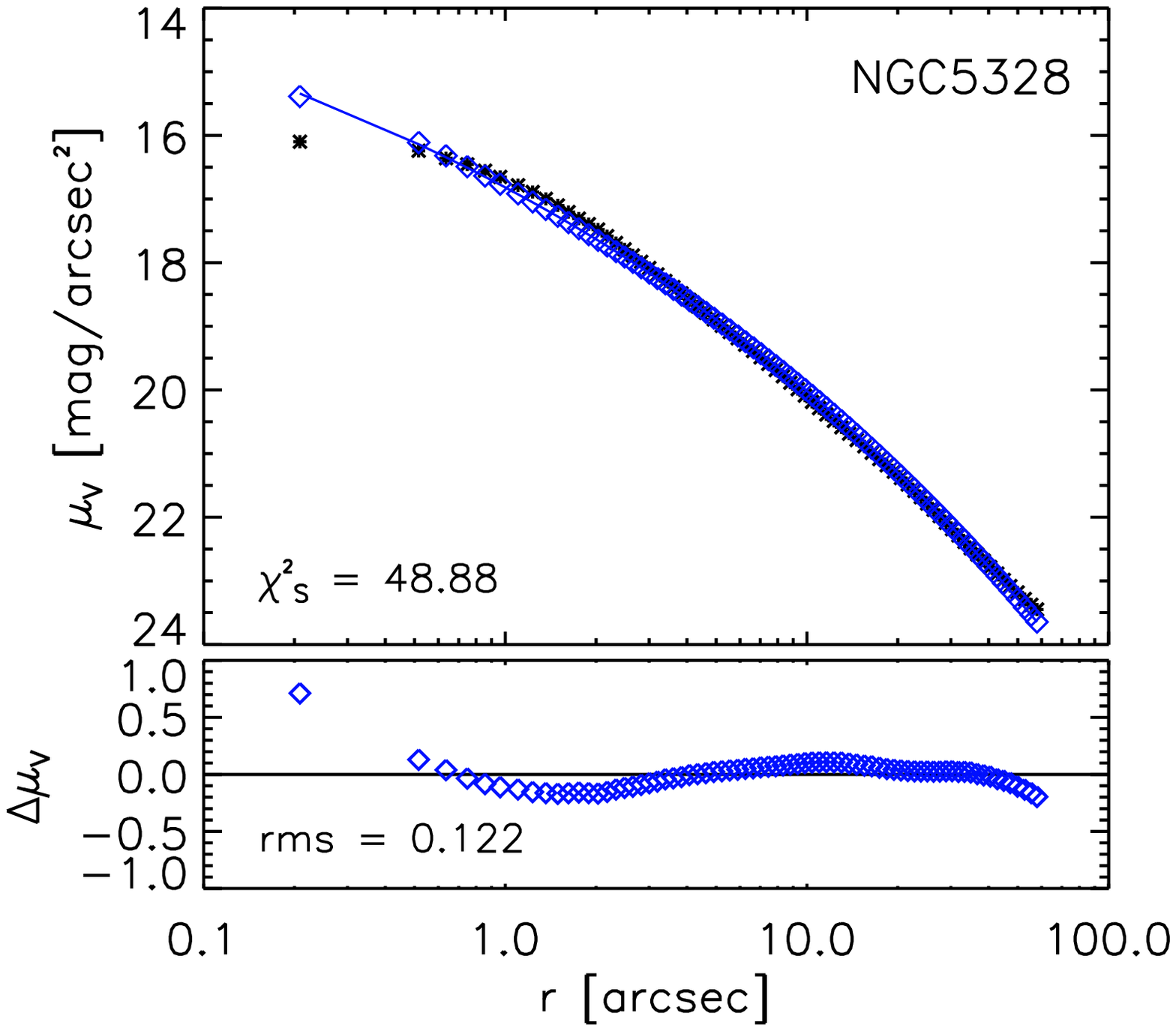}\includegraphics[scale=0.50]{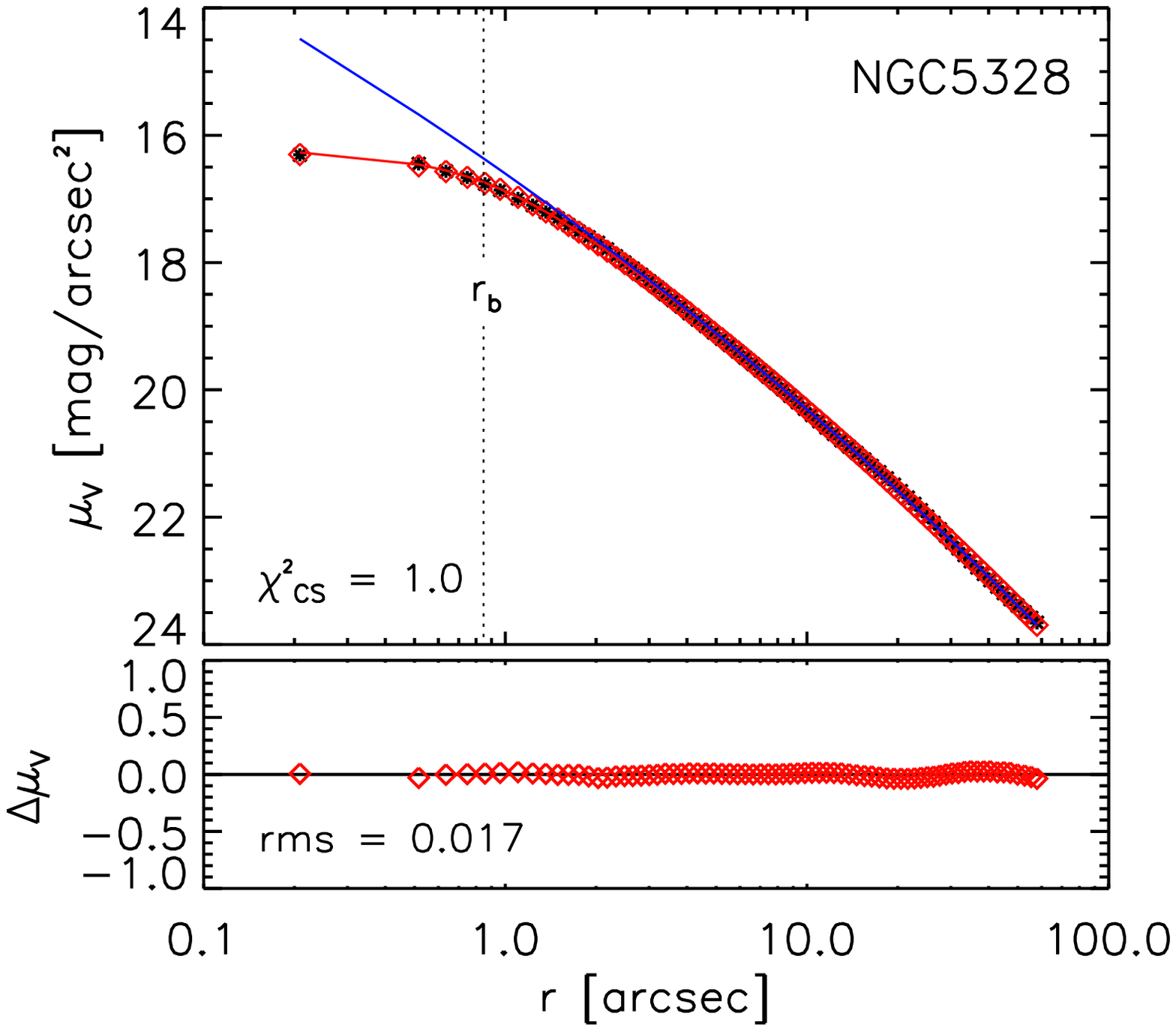}
\caption[]{continued.}
\end{figure*}

\addtocounter{figure}{-1}

\begin{figure*}
\centering
  \includegraphics[scale=0.50]{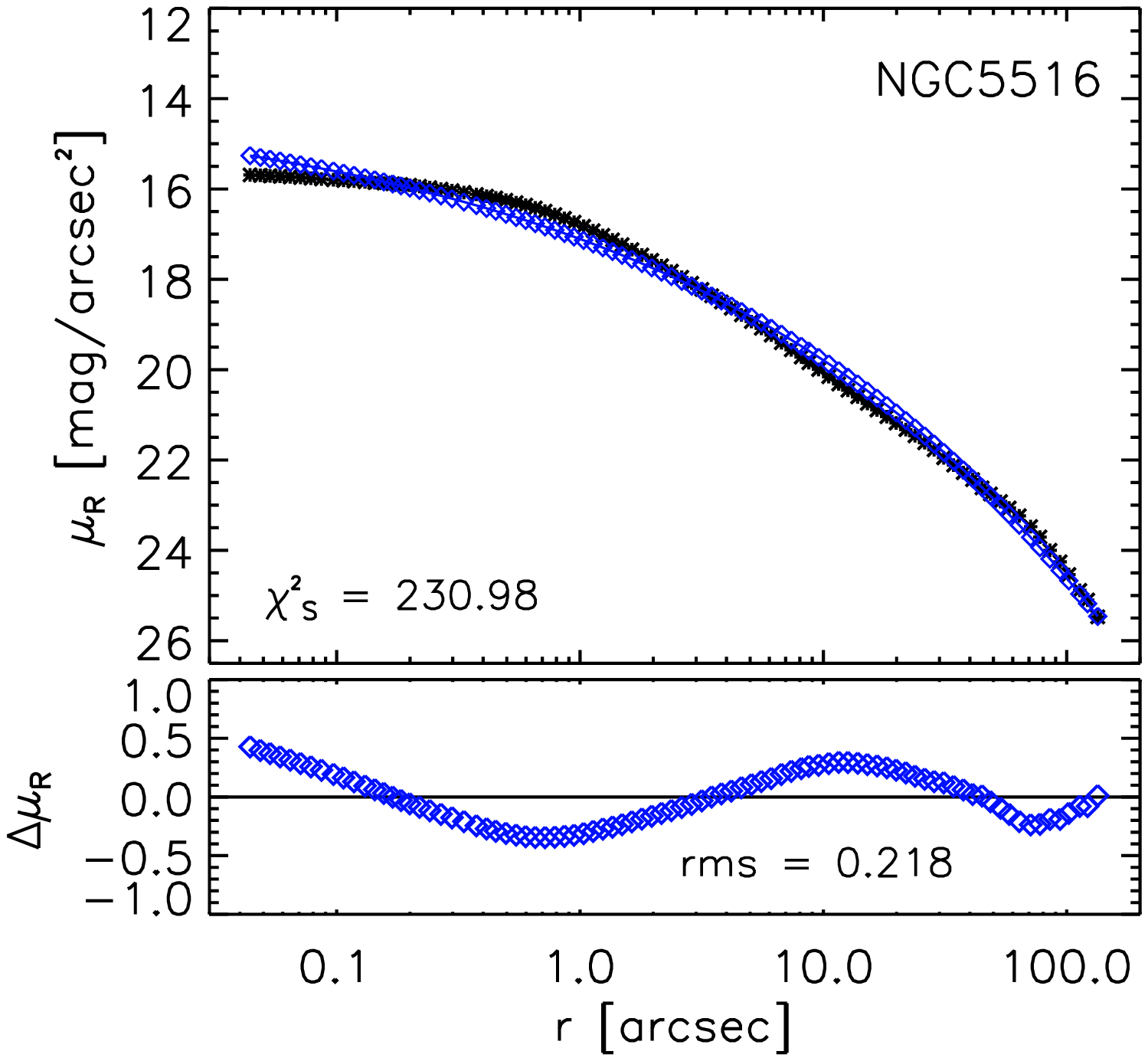}\includegraphics[scale=0.50]{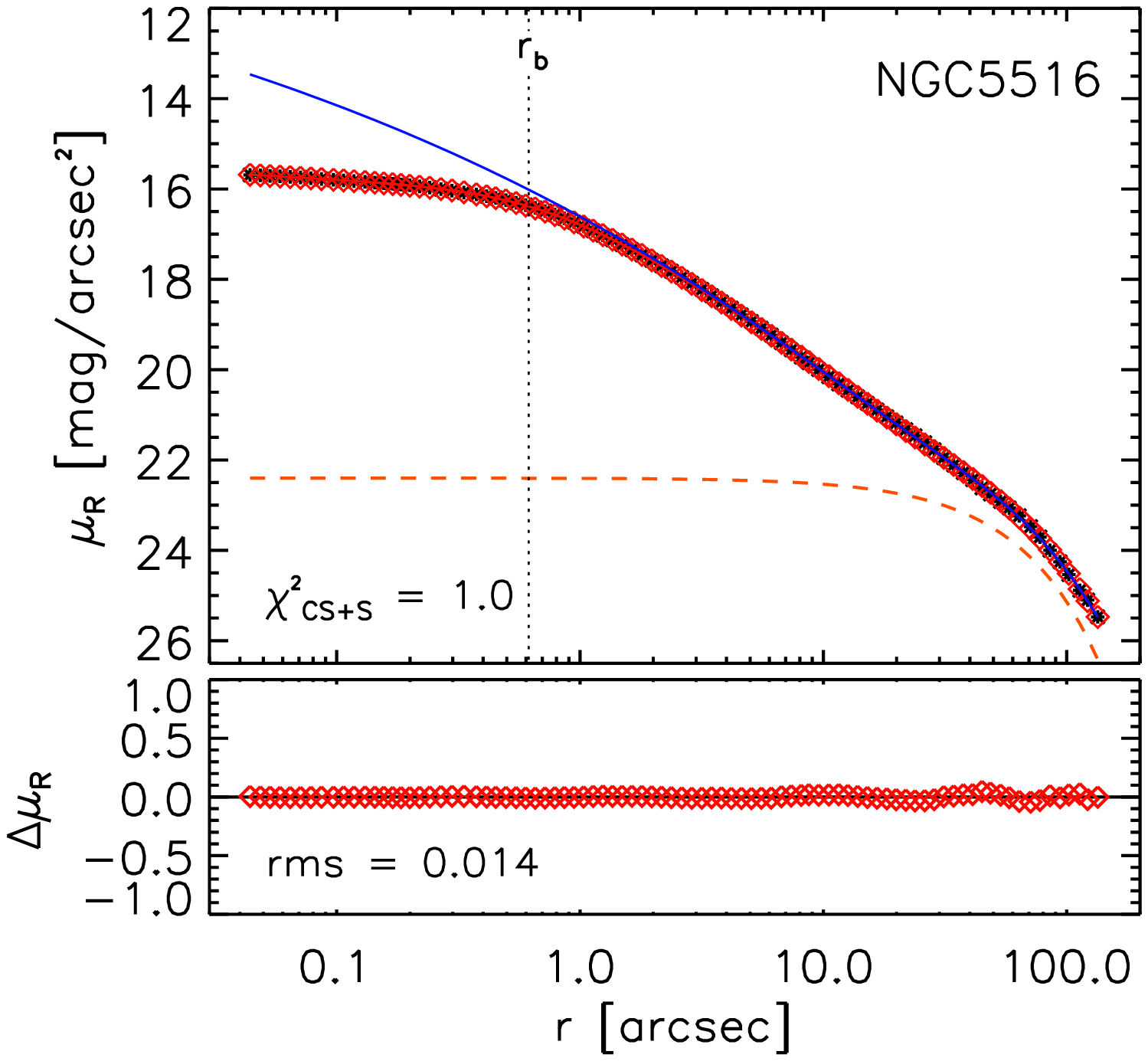}
  \includegraphics[scale=0.50]{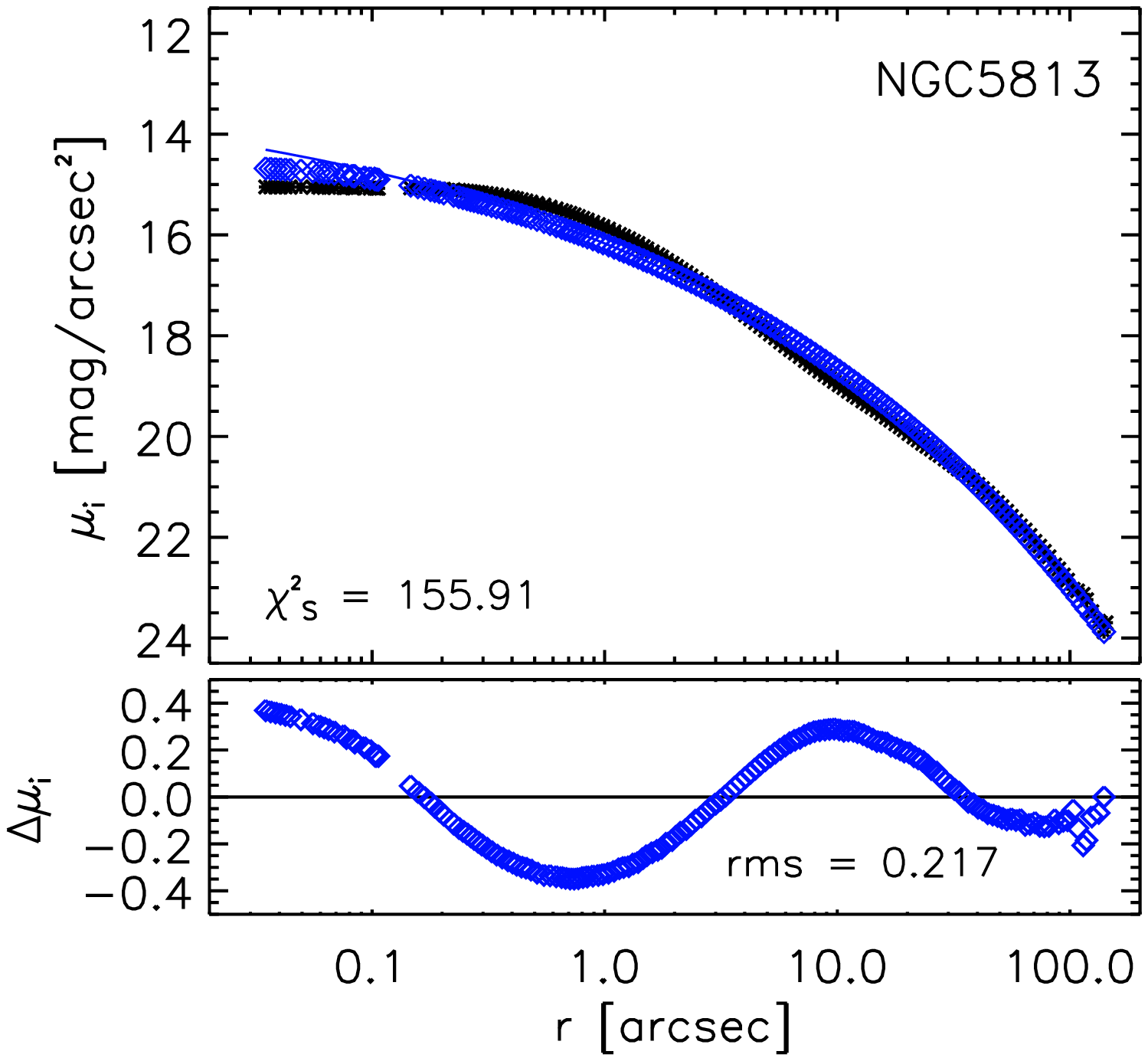}\includegraphics[scale=0.50]{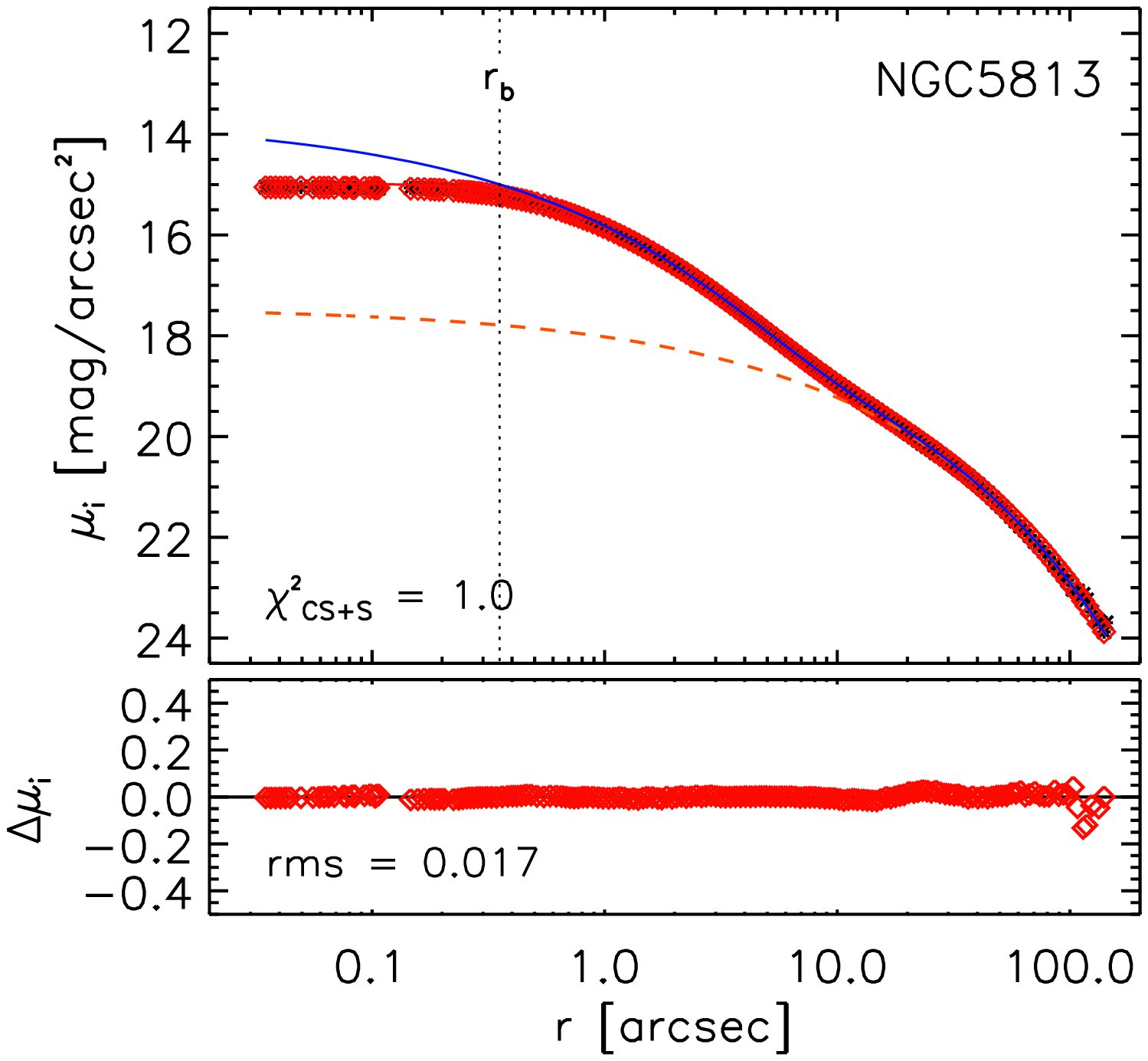}
  \includegraphics[scale=0.50]{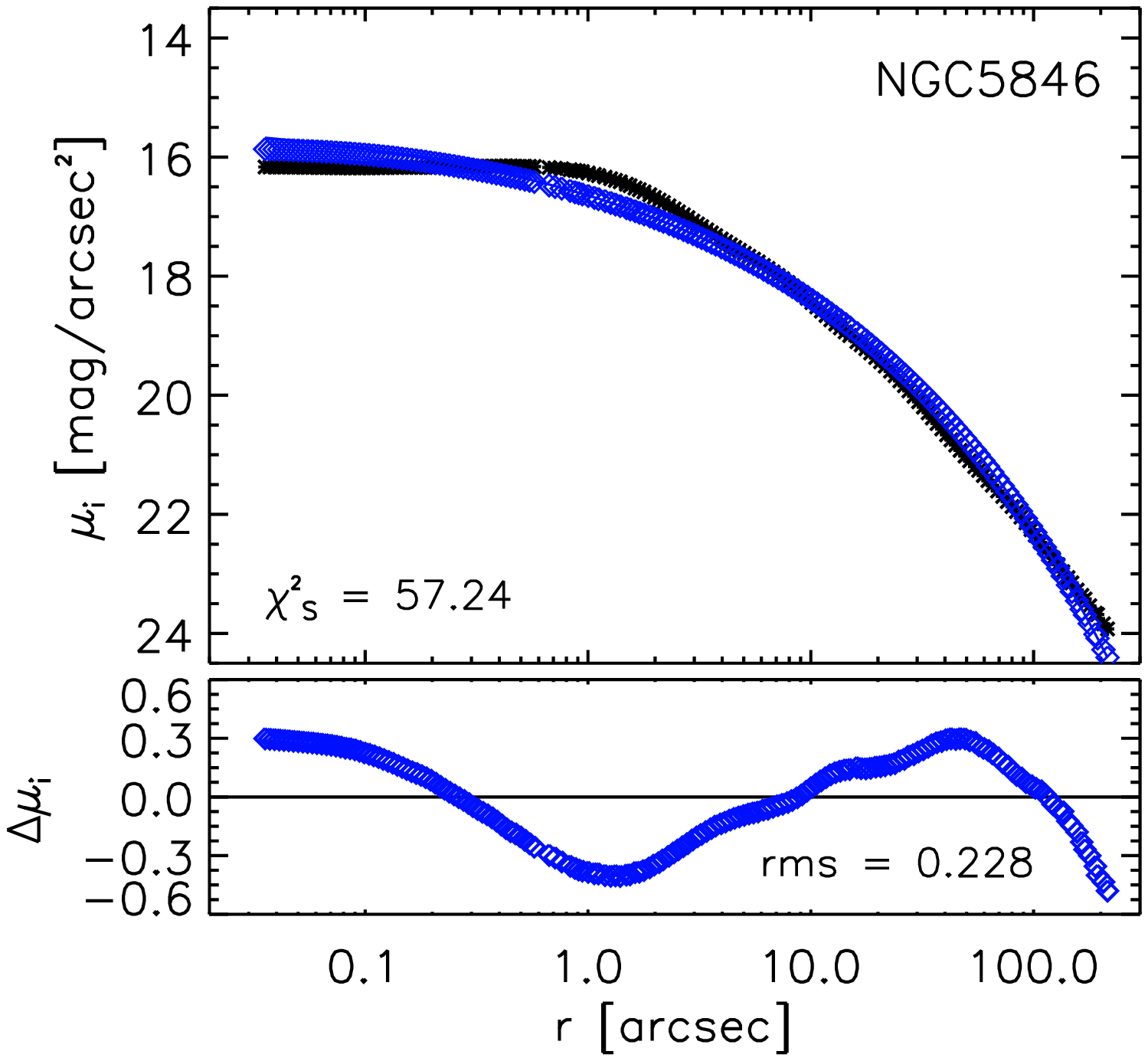}\includegraphics[scale=0.50]{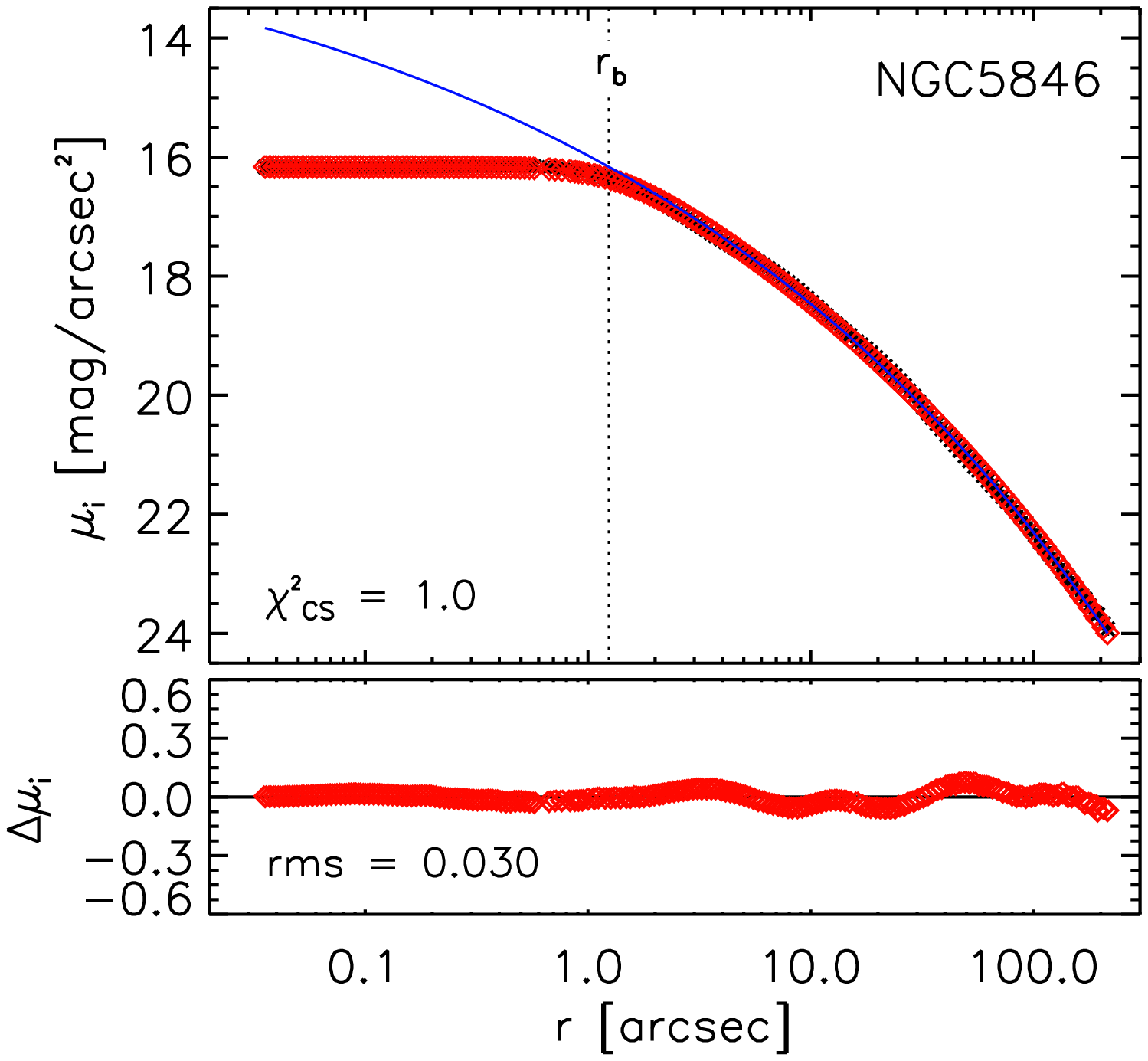}
\caption[]{continued.}
\end{figure*}

\addtocounter{figure}{-1}

\begin{figure*}
\centering
  \includegraphics[scale=0.50]{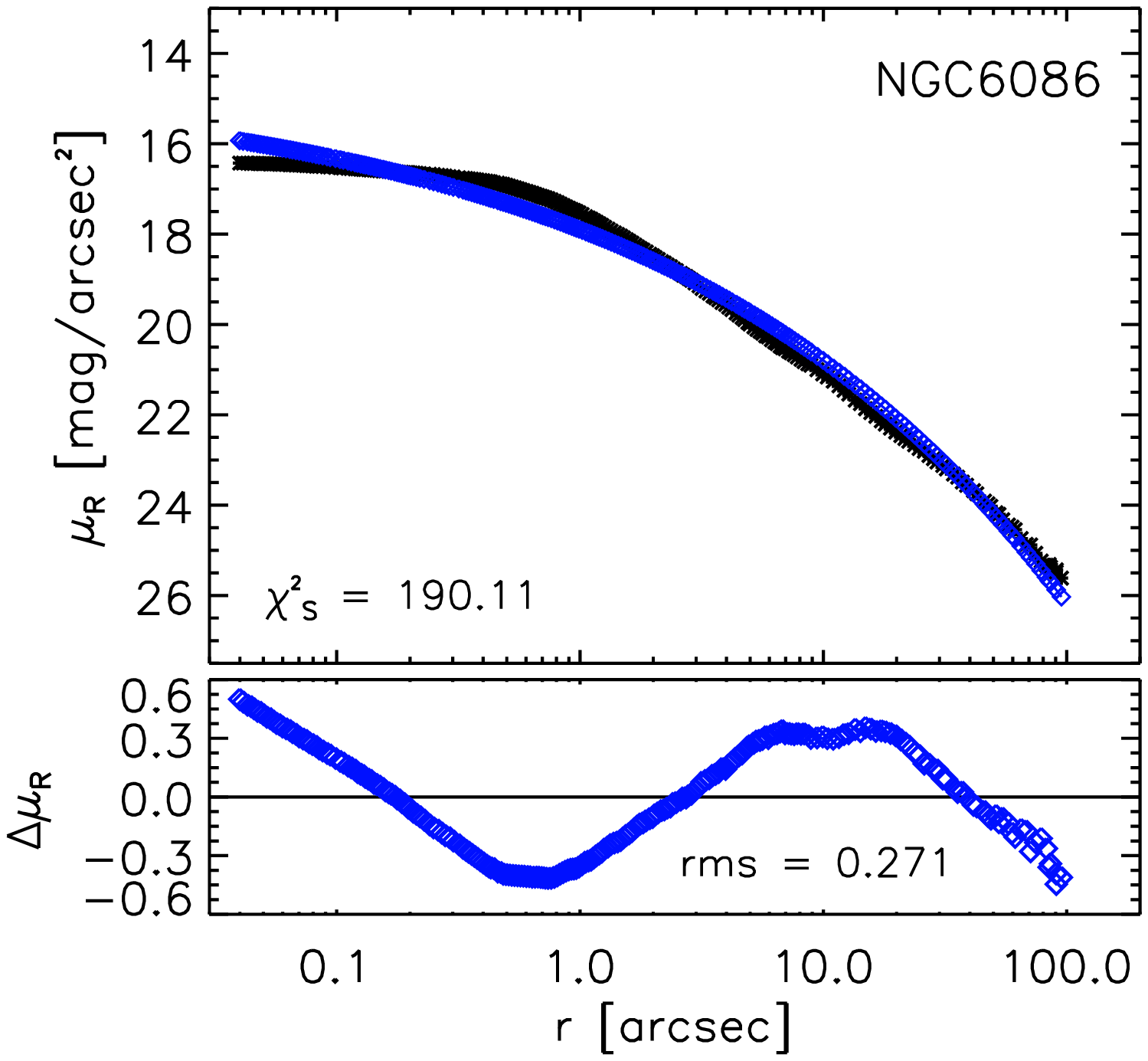}\includegraphics[scale=0.50]{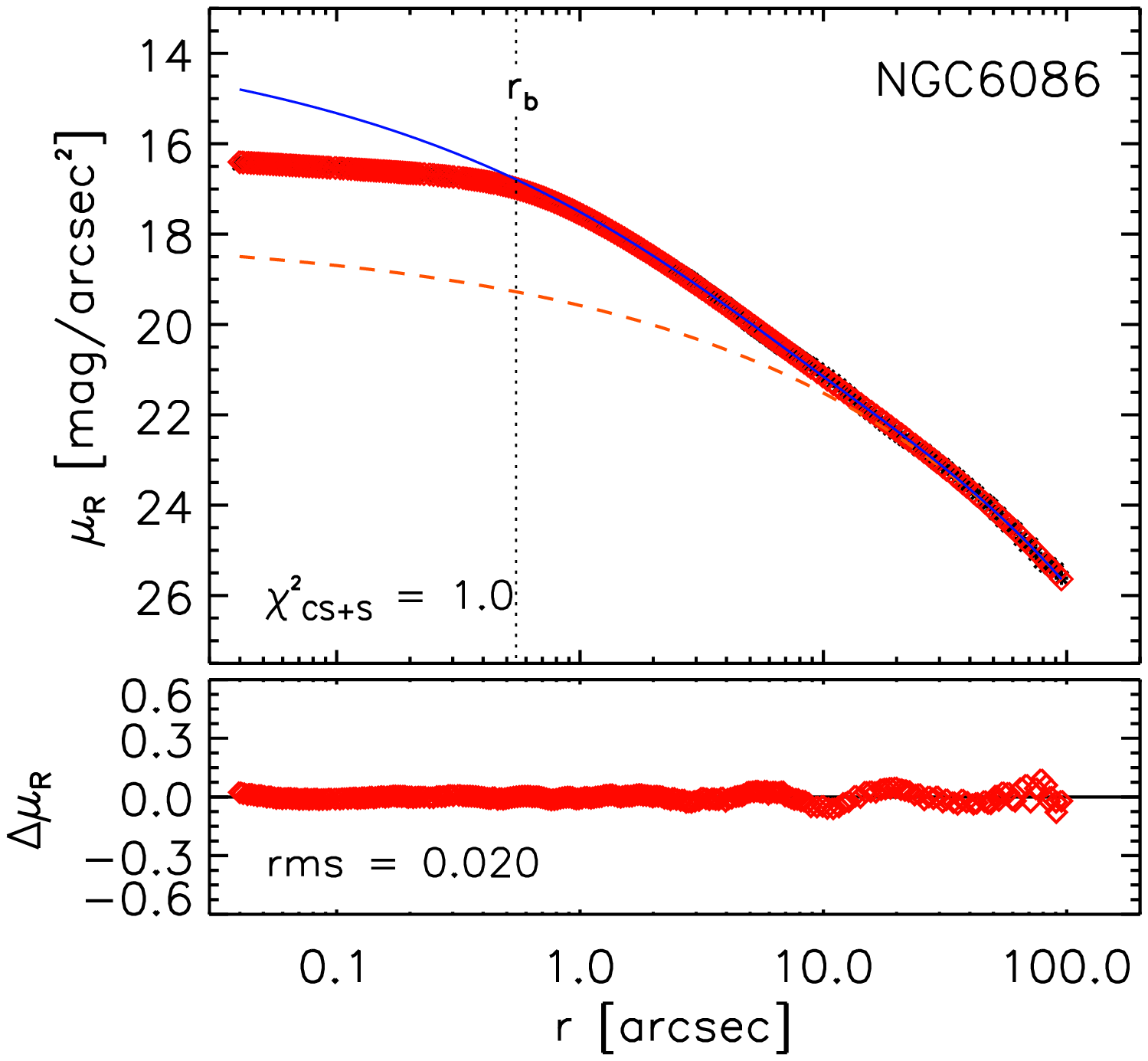}
  \includegraphics[scale=0.50]{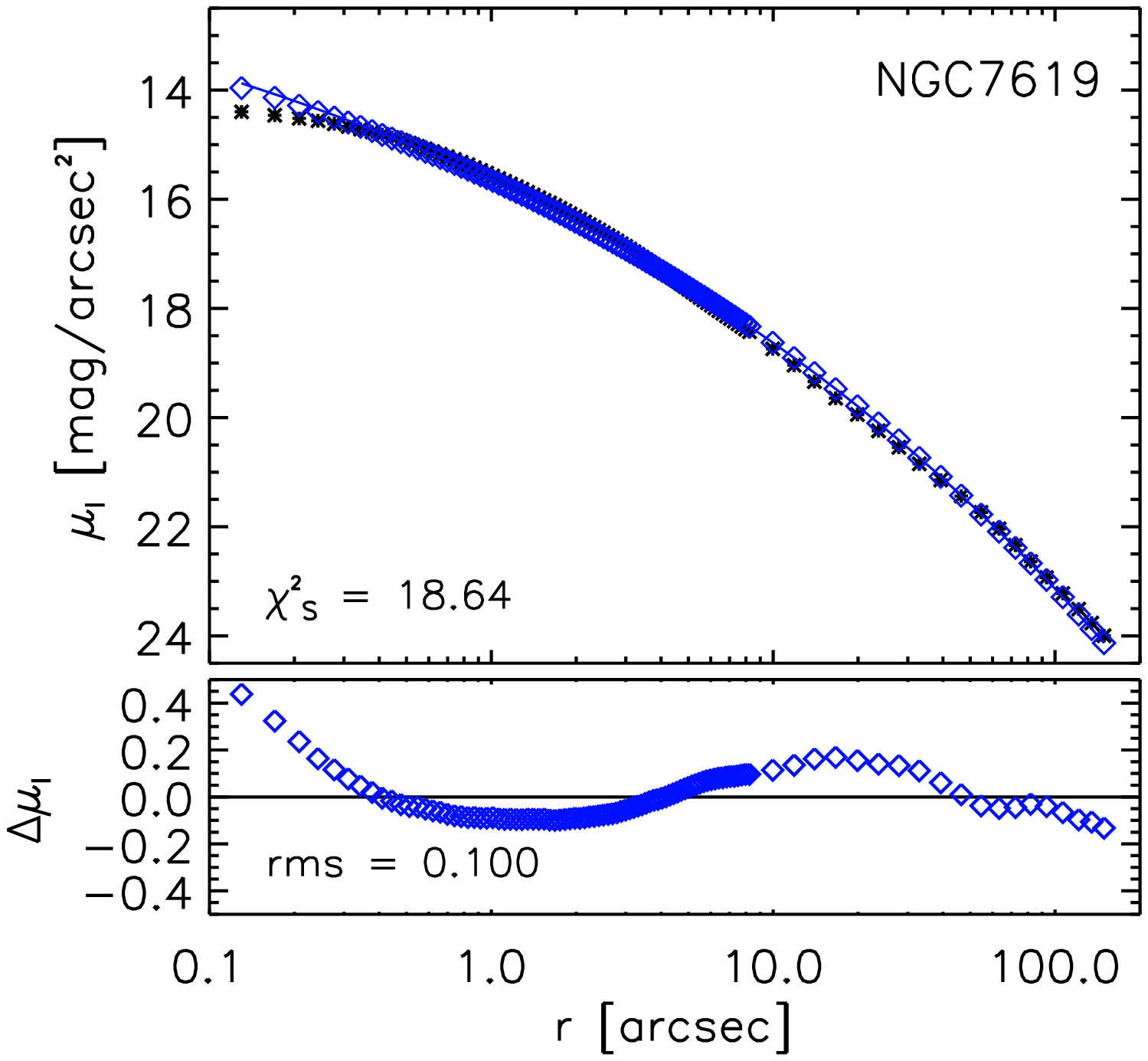}\includegraphics[scale=0.50]{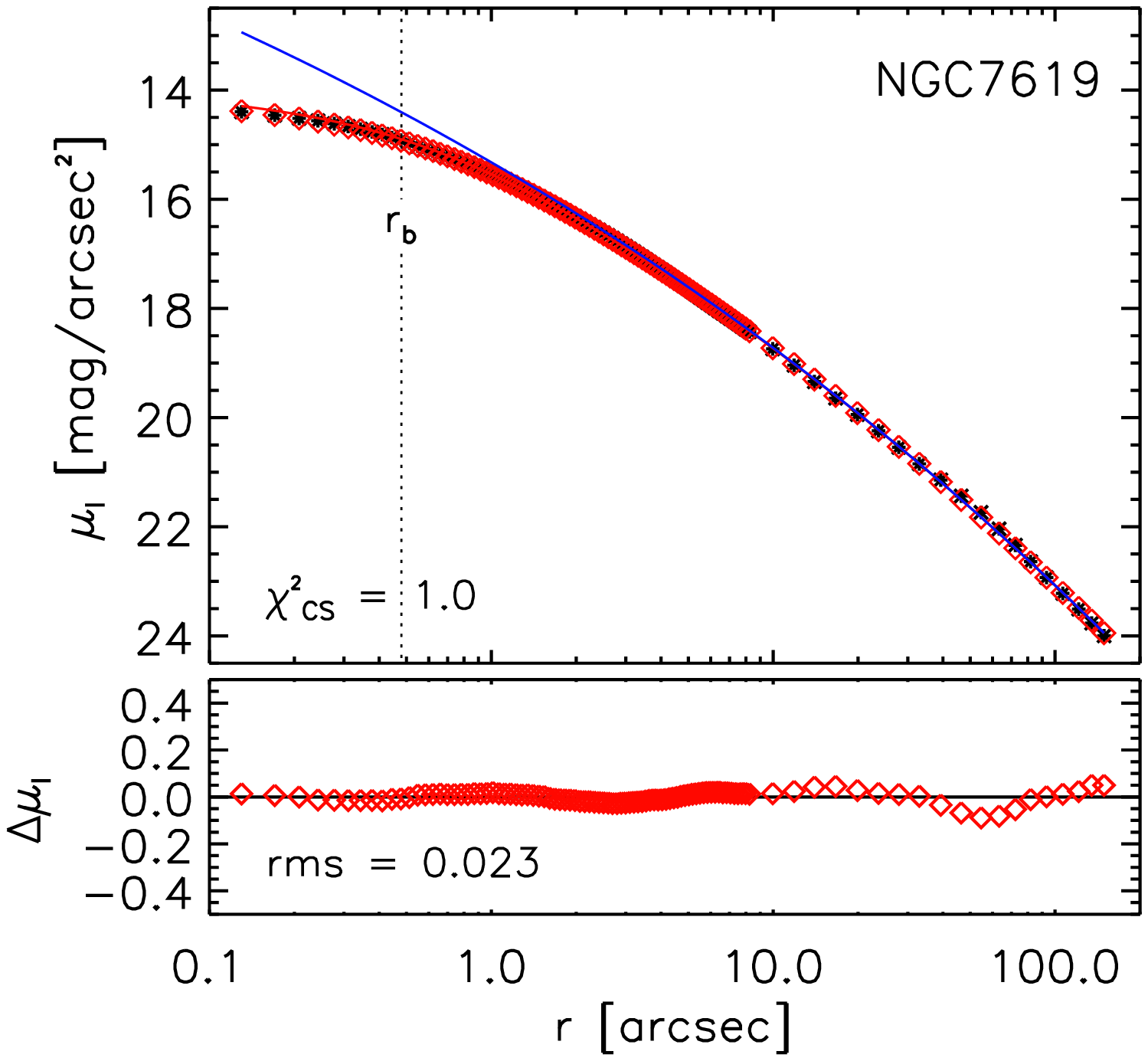}
  \includegraphics[scale=0.50]{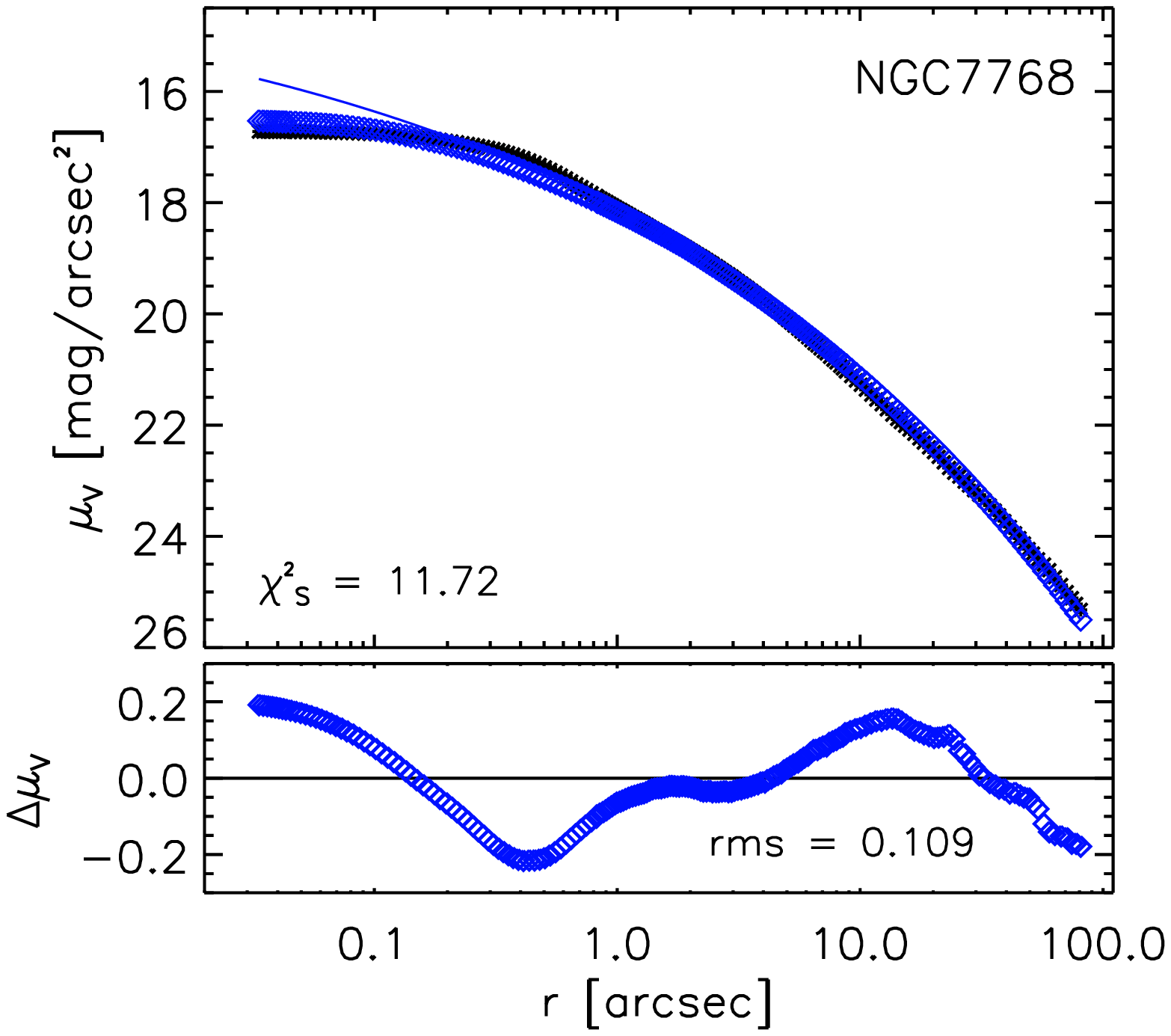}\includegraphics[scale=0.50]{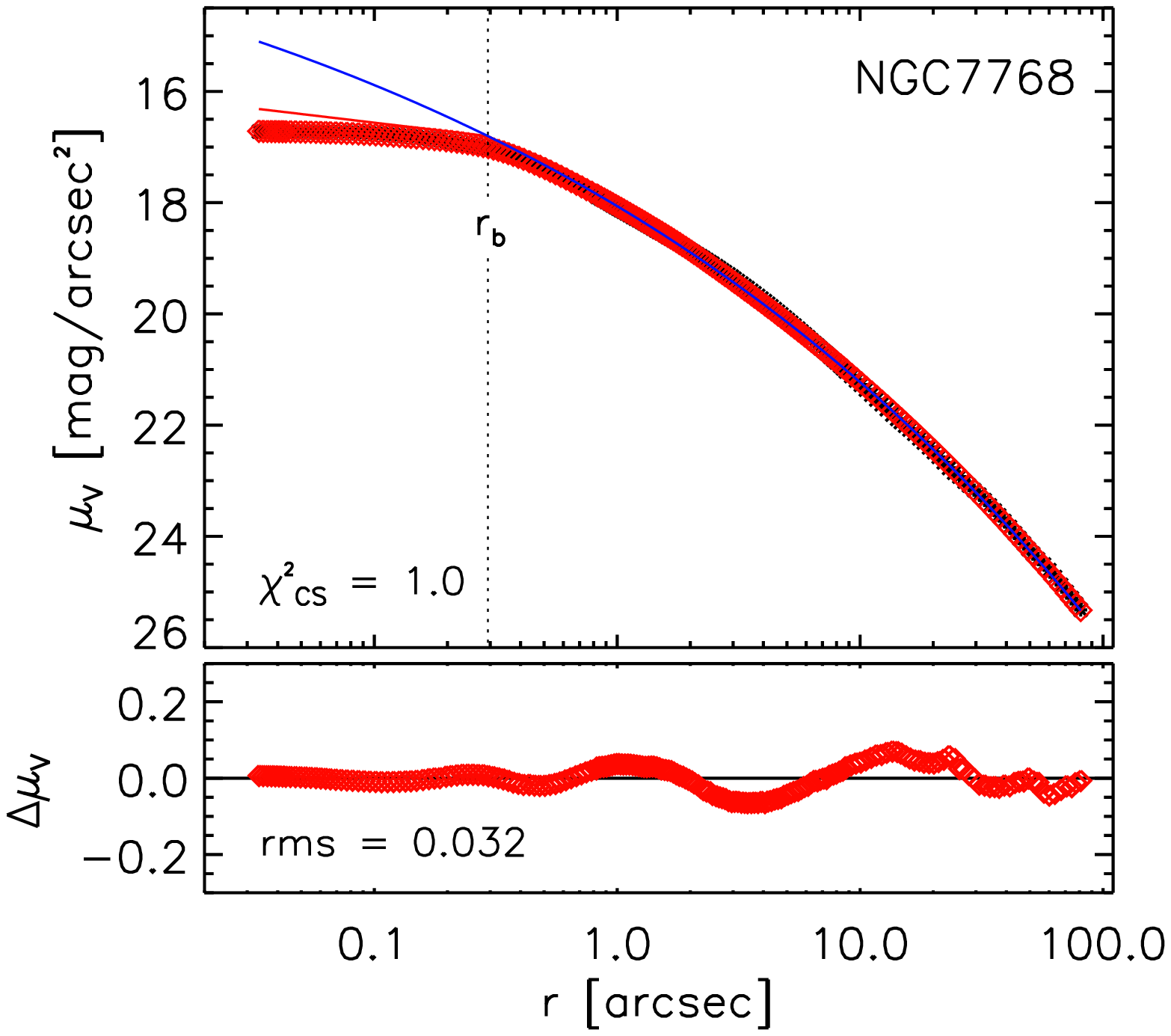}
\caption[]{continued.}
\end{figure*}

\subsection{NGC\,1374}
\label{n1374}

NGC\,1374 is the one galaxy in the sample of \citetalias{Rusli-13} with
conflicting arguments about the existence of a core.
\citetalias{Lauer-05} classify NGC\,1374 as a core galaxy, based on a
Nuker profile fit to their \textit{HST} imaging data. \citet{Dullo-12},
however, do not confirm the core in this galaxy using the same data,
and \citet{Dullo-13} come to the same no-core conclusion using a more extended
ACS-WFC--based profile from \citet{Turner-12}. In contrast, \citet{Lauer-12} argues for the presence of the core.

\begin{figure*}
\centering
  \includegraphics[scale=0.50]{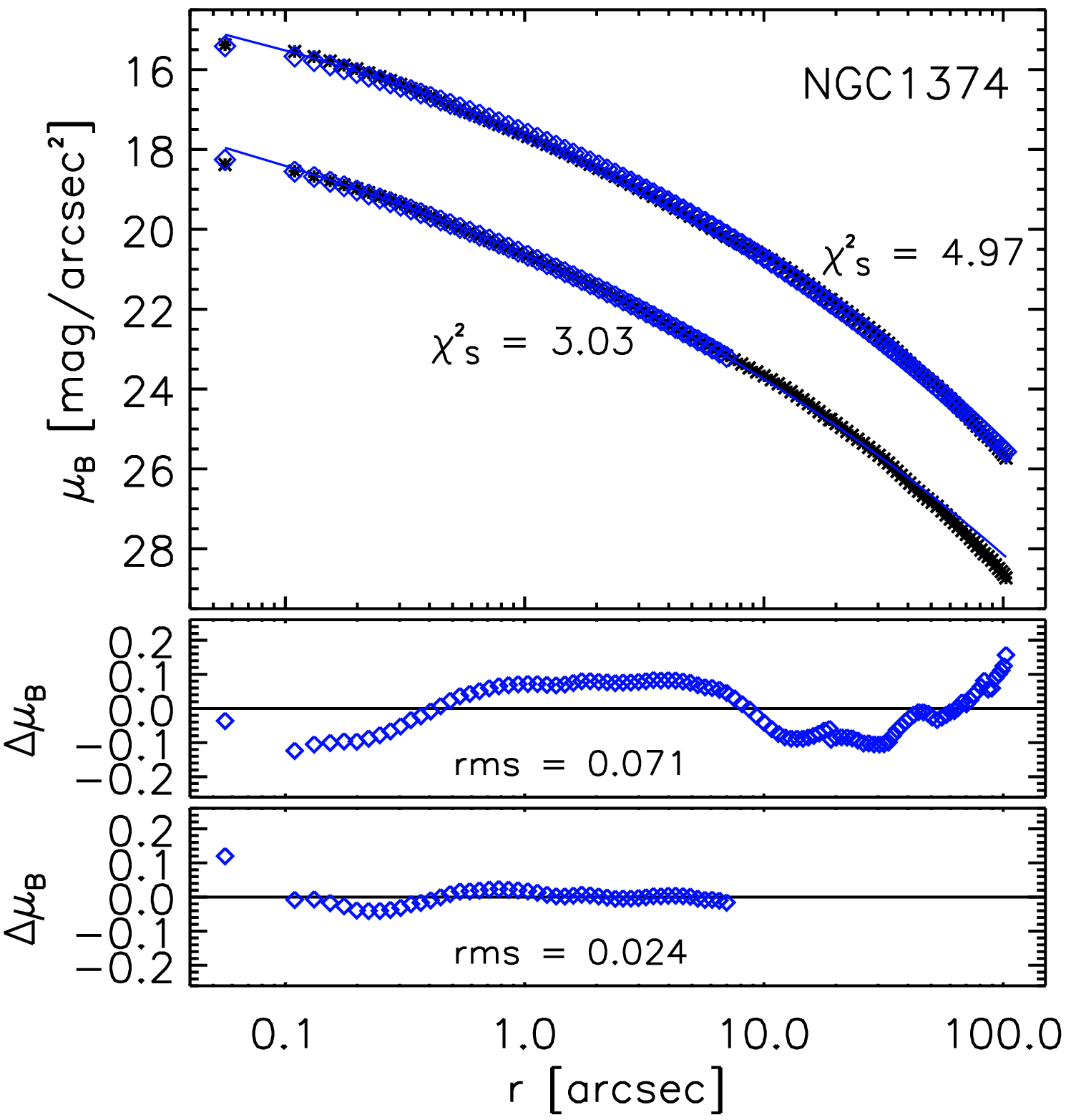}\includegraphics[scale=0.50]{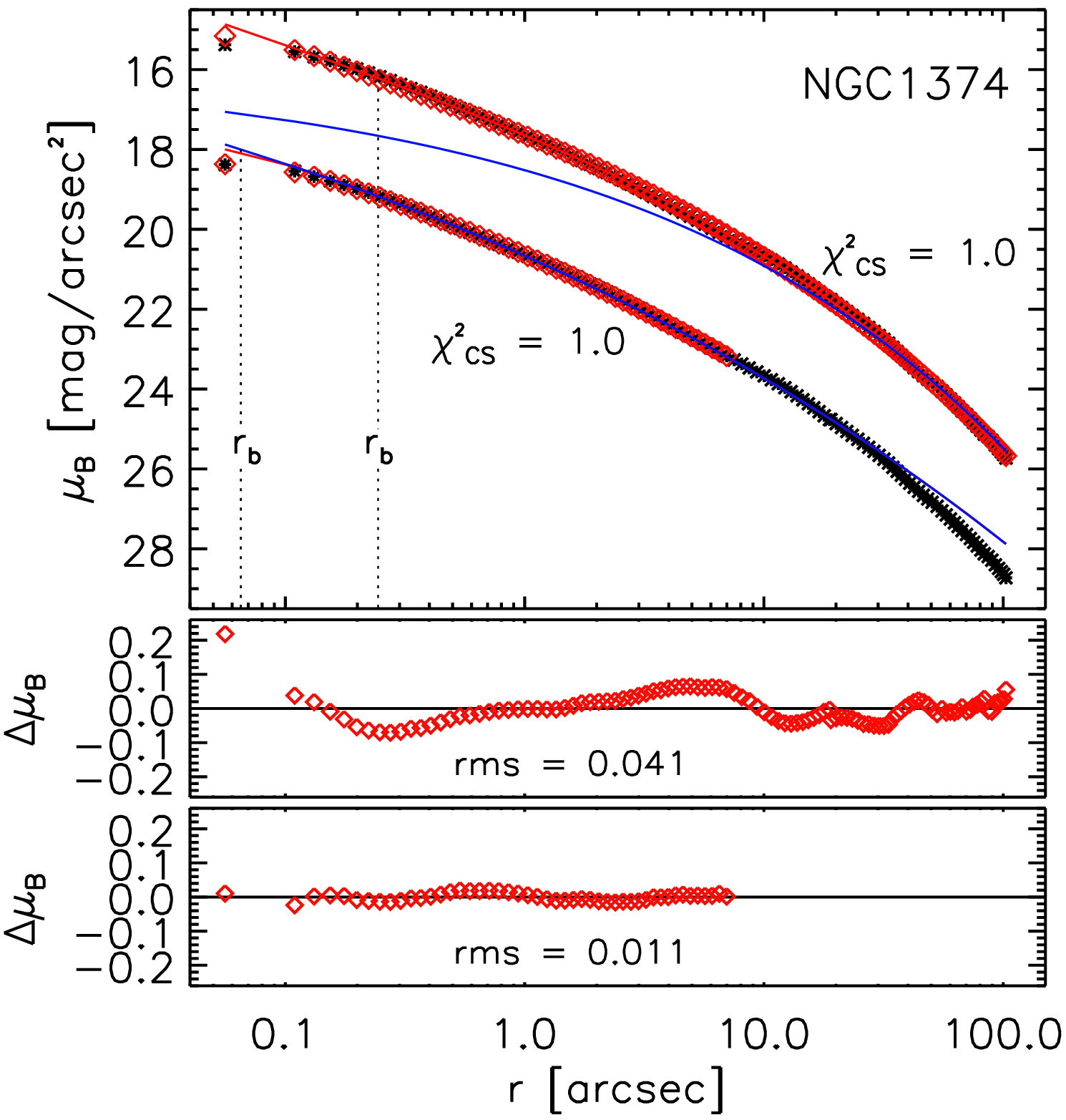}
  \includegraphics[scale=0.50]{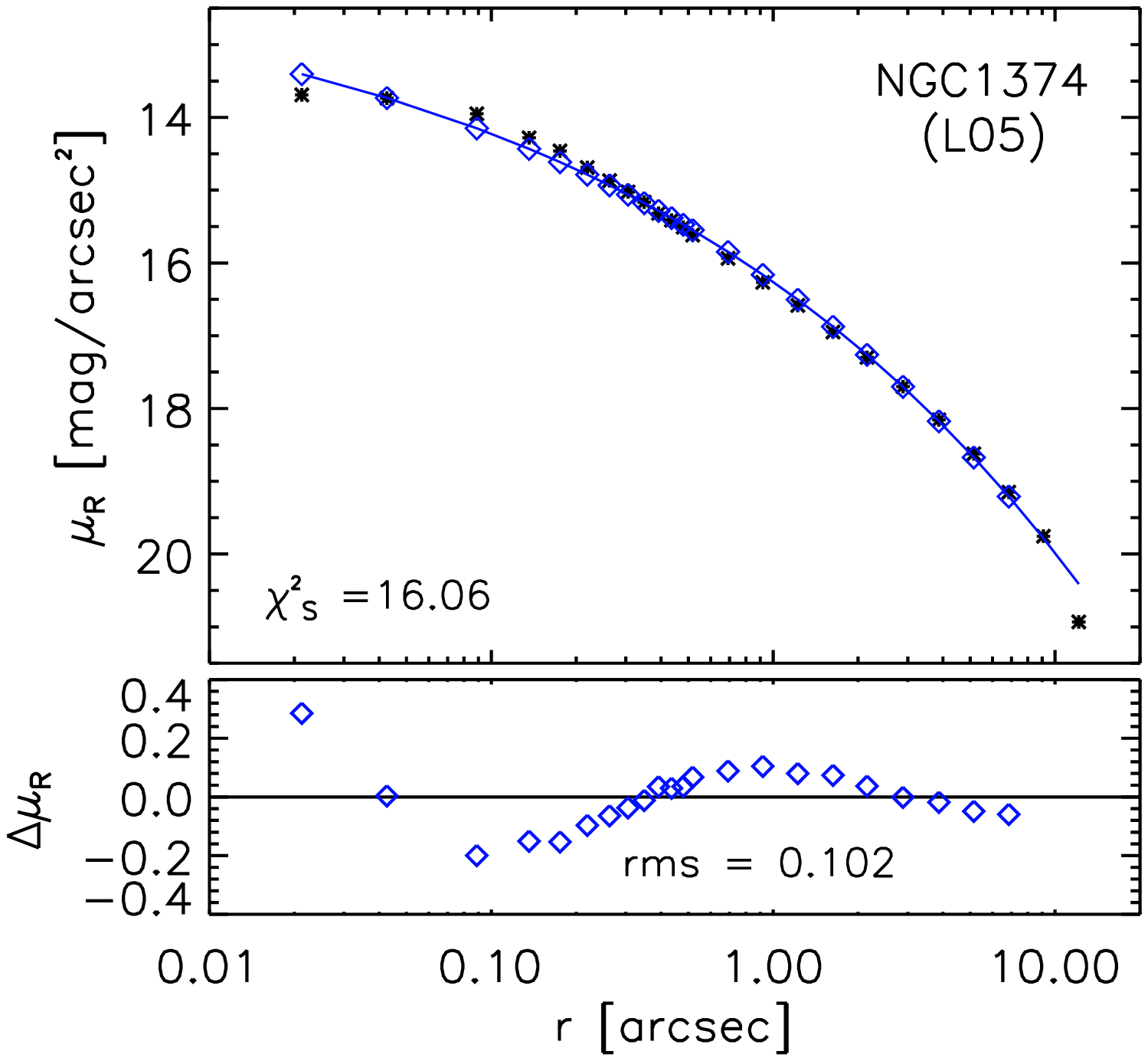}\includegraphics[scale=0.50]{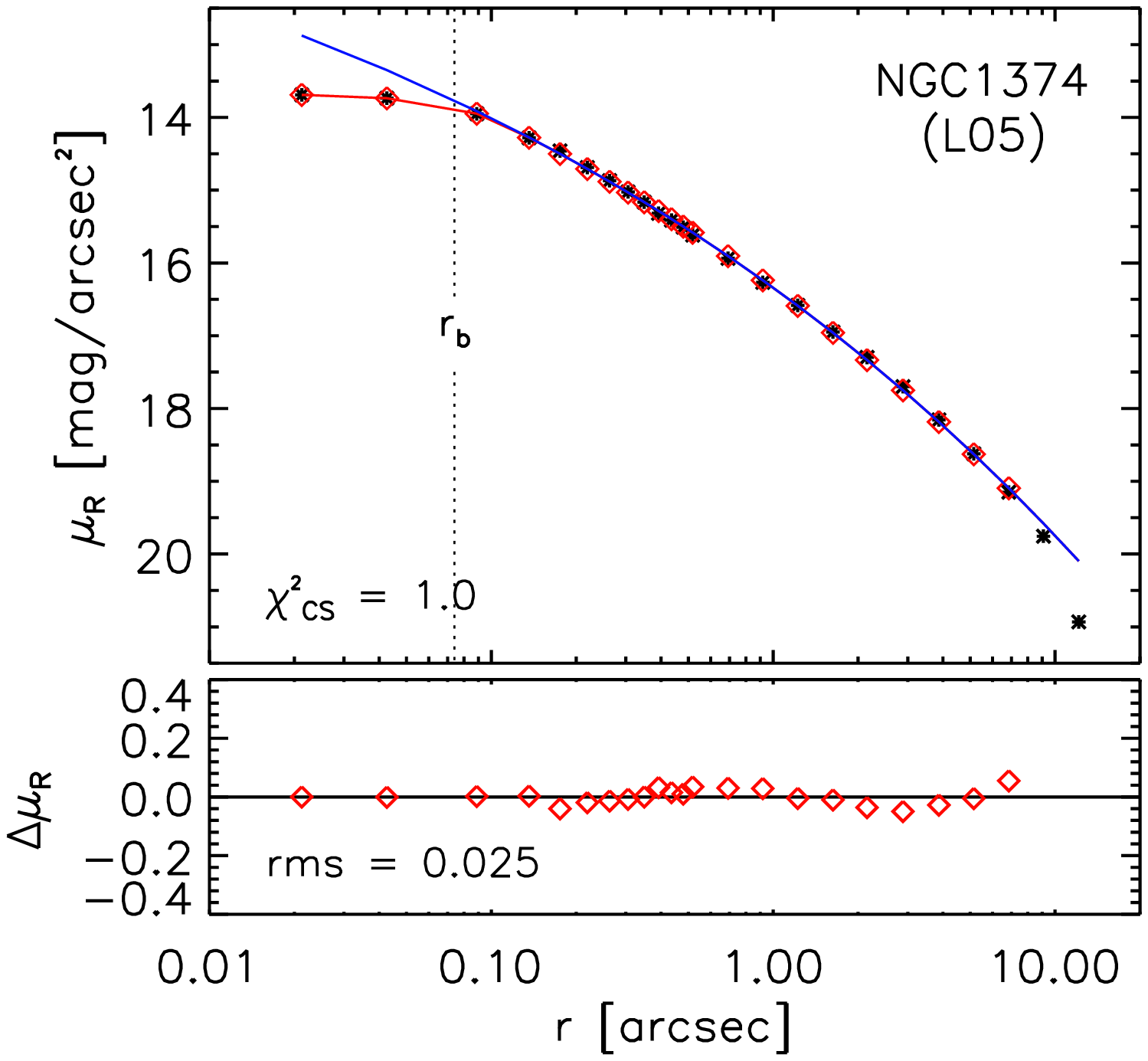}
  \caption[]{\small The model fits to the surface-brightness profiles of
  NGC\,1374. {\bf Top row:} model fits to the surface brightness profile
  derived in \citetalias{Rusli-13}. On the left side, we show the
  \sersic\ fit to all the datapoints (upper profile) and the \sersic\
  fit only to datapoints with $r < 7\arcsec$ (shifted 3 magnitudes down
  for clarity). The black asterisks show the observed profile, the blue
  diamonds are the convolved best-fit \sersic\ model and the blue line
  marks the intrinsic \sersic\ model (before PSF convolution). The upper
  black asterisks show the actual values of surface brightness.  The
  middle and bottom panels show the residuals for the fit to all
  datapoints and for the fit to $r < 7\arcsec$, respectively. The
  right-hand set of panels shows the same thing for the \cs\ fits. The
  red diamonds are the convolved best-fit \cs\ model, the red line shows
  the intrinsic \cs\ model (before PSF convolution) and the blue lines
  are the \sersic\ model that fits only the outer part of the galaxies.
  {\bf Bottom row:} model fits to the surface brightness profile
  reported in \citetalias{Lauer-05}. The fit is done for $r < 7\arcsec$.
  The vertical dotted and dashed lines mark the break radius ($r_b$) and
  the radius at which the \cs\ model deviates from the extrapolated
  \sersic\ model by more than the rms of the residuals ($r_{\rm rms}$).
  Note that the diamonds and the line with the corresponding color
  coincide with each other, since there is no PSF convolution.}
\label{fitting1374}
\end{figure*}

From fitting \cs\ models to our data, we find no solid evidence that a
core is present in this galaxy. We tested different fitting ranges by
varying the maximum radii of the fitted datapoint, but there is no
strong visibility of a core in any case. In the top row of
Fig.~\ref{fitting1374}, we show the \sersic\ and \cs\ fits to the whole
profile and to a limited profile. For the latter, we limit the fitting
range to $r < 7\arcsec$ (for a fair comparison to the fitting using
\citetalias{Lauer-05} data, see below). Judging by eye, there is no
significant difference in fit quality between the \sersic\ and the \cs\
profiles, although the overall residual rms and $\chi^2$ of the \cs\ are
smaller and $\chi^2_{S} > 2 \chi^2_{CS}$. When fitting the whole range,
the \sersic\ profile actually provides a better fit in the center and
instead of seeing a core, the \cs\ fit indicates an extra light on top
of the \sersic\ component (this might be consistent with the extra
nuclear component used by \citealt{Dullo-12,Dullo-13} in their fits to
this galaxy). There is a slight improvement and a vague indication of a
core when we fit only up to 7\arcsec. However, if we base our core
classification on \citet{Trujillo-04}'s criteria, this fit fails at the
second criterion.

Our next attempt is to fit the data of \citetalias{Lauer-05} to confirm
or reject their finding. Their profile has been corrected for the PSF
effect, so we directly fit it with the models without convolution. As
shown in Fig.~\ref{fitting1374}, we detect a core in this galaxy, albeit
weak, by discarding the two outermost datapoints ($r > 7\arcsec$).
Extending the fitting range of data does not give the same results, i.e
the existence of the core becomes inconclusive and the core disappears
altogether when all the datapoints are included in the fit.
\citetalias{Lauer-05} fit their light profile with a Nuker model and it
appears that they also made the compromise of neglecting the 3--4 last
datapoints in the fit, in order to detect the core. Moreover, our
best-fitting break radius is smaller than 0.1\arcsec, less than the
inner resolution limit that we adopt as an additional criterion of core
detection in Section \ref{identifycore}.

One might conclude that our data lack spatial resolution and that this
situation is aggravated by the PSF convolution in the fitting process.
It is, however, important to note that the core disappears when extended
datapoints are included in either our profile or the
\citetalias{Lauer-05} profile. Graham et al.\ (2003) pointed out that
fitting a Nuker model to a limited light profile, such as the case of
NGC\,1374 in \citetalias{Lauer-05}, could lead to a false conclusion
that a core exists. Having no definite evidence of a depleted core in
NGC\,1374, we exclude it from the core sample. 

\begin{table*}
\caption{Galaxies Suspected to Have Depleted Cores \label{tab:coresample}}
{\scriptsize
\begin{tabular}{llllllll}
Galaxy        &type&  Distance & Filter  & $\sigma_e$   & $M_V$         & \mbh                             & \ml \\
              &    & (Mpc)     &         & (km/s)       & (mag)         & (\msun)                          &  \\
\hline
IC\,1459  & E3 & 30.9 [5]  &  $V$ [1]    & $315.0$ [5]  & $-$22.57 [5]  & $2.8(1.6,3.9)\times10^9$ [19]    & $4.5(4.1,5.0)$\,$^\dagger$ [20]   \\
NGC\,1374 & E  & 19.23 [2] & $B,R$ [2,8] & $166.8$ [2]  & $-$20.37 [2]  & $5.8(5.3,6.3)\times10^8$ [2]     & $5.3(4.7,5.9)$ [2]                \\
NGC\,1399 & E1 & 21.1 [5]  & $B$ [1]     & $296.0$ [5]  & $-$22.12 [5]  & $9.1(4.4,18)\times10^8$ [21,22]  & $10.2(9.7,10.5)$\,$^\dagger$ [23] \\
NGC\,1407 & E0 & 28.05 [2] &   $B$* [2]  & $276.1$ [2]  & $-$22.73 [2]  & $4.5(4.1,5.4)\times10^9$  [2]    & $6.6(5.8,7.5)$ [2]             \\
NGC\,1550 & E2 & 51.57 [2] &   $R$  [2]  & $270.1$ [2]  & $-$22.30 [2]  & $3.7(3.3,4.1)\times10^9$  [2]    & $4.0(3.4,4.5)$ [2]           \\
NGC\,3091 & E3 & 51.25 [2] &  $I$ [2]    & $297.2$ [2   & $-$22.66 [2]  & $3.6(3.4,3.7)\times10^9$  [2]    & $3.8(3.6,4.1)$ [2]             \\ 
NGC\,3379 & E1 & 10.57 [4] &   $I$* [1]  & $206.0$ [5]  & $-$20.99 [4]  & $4.1(3.1,5.1)\times10^8$  [9]    & $2.8(2.6,3.0)$\,$^\dagger$  [9]  \\
NGC\,3608 & E2 & 23.0 [18] &  $V$ [1]    & $182.0$ [5]  & $-$21.05 [5]  & $4.7(3.7,5.7)\times10^8$  [18]   & $3.1(2.8,3.4)$\,$^\dagger$  [18] \\ 
NGC\,3842 & E  & 98.40 [6] &   $V$* [1]  & $270.0$ [5]  & $-$23.17 [5]  & $9.7(7.2,12.7)\times10^9$ [6]    & $6.7(5.6,7.7)$\,$^\dagger$ [6]  \\
NGC\,4261 & E2 & 31.60 [3] &   $V$* [3]  & $315.0$ [5]  & $-$22.60 [4]  & $5.2(4.1,6.2)\times10^8$  [13]   & $9.1(7.9,10.3)$\,$^\dagger$[15] \\
NGC\,4291 & E  & 25.0 [18] &   $V$ [1]   & $242.0$ [5]  & $-$20.67 [5]  & $9.2(6.3,1.21)\times10^8$ [18]   & $5.4(4.7,6.1)$\,$^\dagger$  [18]   \\
NGC\,4374 & E1 & 18.45 [3] &   $V$* [3]  & $296.0$ [5]  & $-$22.63 [4]  & $9.2(8.4,10.2)\times10^8$ [14]   & $6.8(6.3,7.2)$ [16]  \\
NGC\,4472 & E2 & 17.14 [3] &   $V$* [3]  & $300.2$ [2]  & $-$22.86 [2]  & $2.5(2.4,2.8)\times10^9$  [2]    & $4.9(4.5,5.3)$ [2]  \\
NGC\,4486 & E0 & 17.22 [3] &   $V$* [3]  & $324.0$ [5]  & $-$22.95 [4]  & $6.2(5.7,6.6)\times10^9$  [10]   & $6.2(5.3,6.9)$\,$^\dagger$ [10]  \\
NGC\,4552 & E0 & 15.85 [3] &   $V$* [3]  & $252.0$ [24] & $-$21.37 [25] & $5.0(4.5,5.5)\times10^8$ ** [24] & $7.1(6.7,7.6)$ [16] \\
NGC\,4649 & E2 & 17.30 [3] &   $V$* [3]  & $341.0$ [5]  & $-$22.75 [4]  & $5.0(3.9,6.1)\times10^9$  [11]   & $7.3(6.5,8.1)$\,$^\dagger$ [11]  \\
NGC\,4889 & E4 & 103.2 [6] &   $R$ [12]  & $347.0$ [5]  & $-$23.72 [5]  & $2.1(0.5,3.7)\times10^{10}$[6]   & $5.8(4.1,7.4)$\,$^\dagger$ [6]  \\
NGC\,5328 & E1 & 64.1 [2]  &  $V$ [2]    & $332.9$ [2]  & $-$22.80 [2]  & $4.7(2.8,5.6)\times10^9$   [2]   & $4.9(4.3,5.5)$ [2] \\
NGC\,5516 & E3 & 58.44 [2] &   $R$* [2]  & $328.2$ [2]  & $-$22.87 [2]  & $3.3(3.0,3.5)\times10^9$ [2]     & $5.2(5.1,5.5)$ [2]  \\
NGC\,5813 & E1 & 32.2 [24] & $i$ [1]     & $230.0$ [24] & $-$22.06 [25] & $7.0(6.3,7.7)\times10^8$ ** [24] & $4.7(4.4,5.0)$ [16] \\  
NGC\,5846 & E0 & 24.9 [24] & $i$ [1]     & $238.0$ [24] & $-$22.03 [25] & $1.1(1.0,1.2)\times10^9$ ** [24] & $5.2(4.9,5.5)$ [16] \\
NGC\,6086 & E  & 133.0 [17]&   $R$* [1]  & $318.0$ [17] & $-$23.12 [5]  & $3.6(2.5,5.3)\times10^{9}$ [17]  & $4.2(3.6,4.5)$\,$^\dagger$ [17]\\
NGC\,7619 & E  & 51.52 [2] &   $I$  [7]  & $292.2$ [2]  & $-$22.86 [2]  & $2.5(2.2,3.3)\times10^9$ [2]     & $3.0(2.6,3.3)$ [2]  \\
NGC\,7768 & E  & 112.8 [6] &   $V$  [1]  & $257.0$ [6]  & $-$22.92 [6]  & $1.3(0.9,1.8)\times10^9$ [6]     & $7.8(6.3,9.3)$\,$^\dagger$ [6]  \\
\hline
\end{tabular}
}
\tablecomments{The number in square brackets refers to the literature
  source: [1] this paper (Appendix), [2] \citetalias{Rusli-13}, [3] \citetalias{Kormendy-09a}, [4] \citetalias{Kormendy-09b},
  [5] \citet{McConnell-11b}, [6] \citet{McConnell-12}, [7] \citet{Pu-10},
  [8] \citet{Lauer-05}, [9] \citet{vandenBosch-10}, [10]
  \citet{Gebhardt-09}, [11] \citet{Shen-10}, [12] \citet{Thomas-07}, [13]
  \citet{Ferrarese-96}, [14] \citet{Walsh-10}, [15] \citet{Haering-04},
  [16] \citet{Cappellari-06}, [17] \citet{McConnell-11a} [18],
  \citet{Schulze-11}, [19] \citet{Cappellari-02}, [20] \citet{Haering-04},
  [21] \citet{Houghton-06}, [22] \citet{Gebhardt-07}, [23]
  \citet{Kronawitter-00}, [24] \citet{Hu-08} from \citet{Cappellari-08},
  [25] $V_T^0$ from \citet{deVaucouleurs-91}.  For the galaxies
    IC\,1459, NGC\,3379, NGC\,4261, and NGC\,4374 and the 
    three objects with dubious black hole masses (see below) the \ml\ values come from single-component dynamical 
    modeling (i.e., without dark matter; see text). Whenever the literature source of \mbh\ or the
  mass-to-light ratio \ml\ is different from that for the distance, we
  write the values of \mbh\ or \ml\ after distance correction.
  Furthermore, we apply an extinction correction following
  \citet{Schlegel-98}, as reported in NED, to the marked ($\dagger$) \ml\
  values, assuming they were not yet corrected. Column 4 is the
  photometric band of \ml\ and the light profile used in this paper; the
  asterisks in this column mark the galaxies whose innermost isophotes come
  from a deconvolved image, thus eliminating the need of PSF convolution
  during the fitting. The derivation of light profiles of NGC\,3379,
  NGC\,3842, NGC\,6086 and NGC\,7768 is described in the Appendix. For
  NGC1399, we use the average \mbh\ from the two papers and take the
  extreme lower and upper limits of both measurements to define to error
  bar. For NGC\,4552, NGC\,5813 and NGC\,5846, the quoted values are based
  on a single figure in a non-refereed publication. We do not consider
  these to be well-determined measurements, and mark them with asterisks;
  they are included in our plots but not used in our fits. We classify
  NGC~1550 and NGC~5516 as ellipticals, though literature classifications
  vary. Our extended kinematics and photometry show no signs of rotation
  or obvious (face-on) disks in these two galaxies.\\}
\end{table*}

\begin{table*}
\caption{The Core Sample: The \cs\ Function Parameters \label{tab:csparams}}
\begin{tabular}{lrrrrrr}
Galaxy          & $n$                & $\mu_b$             & $r_b$              & $r_e$            & $\alpha$      & $\gamma$ \\
                &                    & (mag arcsec$^{-2}$) & (\arcsec)          & (\arcsec)        &               &  \\
\hline
IC\,1459        &  $7.62\pm0.89$     & $15.53\pm0.08$      & $0.69\pm0.07$      & $45.43\pm4.87$   & $1.39\pm0.12$ & $0.13\pm0.04$   \\
NGC\,1399*      &  $7.38\pm0.41$     & $17.92\pm0.02$      & $2.47\pm0.05$      & $36.20\pm0.67$   & $1.93\pm0.09$ & $0.12\pm0.01$ \\
NGC\,1407*      & $2.17\pm0.18$      & $18.37\pm0.01$      & $2.01\pm0.06$      & $9.35\pm0.69$    & $3.80\pm0.17$ & $0.16\pm0.00$ \\
NGC\,1550       &  $7.96\pm0.24$     & $16.48\pm0.07$      & $1.19\pm0.09$      & $39.08\pm1.24$   & $5.24\pm1.78$ & $0.52\pm0.05$   \\
NGC\,3091       &  $9.29\pm0.57$     & $15.13\pm0.05$      & $0.62\pm0.04$      & $91.01\pm11.07$  & $1.90\pm0.19$ & $0.13\pm0.05$   \\ 
NGC\,3379       &  $5.80\pm0.10$     & $14.40\pm0.02$      & $1.09\pm0.04$      & $55.14\pm0.87$   & $2.52\pm0.16$ & $0.22\pm0.01$   \\
NGC\,3608       &  $6.29\pm0.07$     & $15.11\pm0.07$      & $0.21\pm0.02$      & $56.92\pm1.36$   & $4.14\pm1.32$ & $0.28\pm0.05$   \\
NGC\,3842       &  $6.27\pm0.18$     & $17.42\pm0.03$      & $0.70\pm0.03$      & $58.75\pm2.89$   & $5.08\pm2.21$ & $0.19\pm0.01$   \\
NGC\,4261       &  $6.33\pm0.16$     & $16.65\pm0.03$      & $1.18\pm0.05$      & $77.10\pm2.56$   & $3.78\pm0.58$ & $0.07\pm0.03$   \\
NGC\,4291       &  $5.63\pm0.06$     & $15.17\pm0.02$      & $0.33\pm0.01$      & $15.35\pm0.14$   & $3.60\pm0.28$ & $0.10\pm0.02$    \\
NGC\,4374       &  $7.06\pm0.23$     & $16.12\pm0.04$      & $1.47\pm0.07$      & $126.17\pm5.30$  & $3.41\pm0.51$ & $0.14\pm0.02$   \\
NGC\,4472       &  $5.60\pm0.09$     & $16.48\pm0.03$      & $1.82\pm0.09$      & $199.13\pm3.84$  & $3.05\pm0.38$ & $0.06\pm0.02$   \\
NGC\,4486       &  $8.91\pm0.44$     & $18.06\pm0.04$      & $8.14\pm0.29$      & $180.85\pm4.13$  & $2.13\pm0.14$ & $0.23\pm0.01$  \\
NGC\,4552*      &  $3.82\pm0.40$     & $15.03\pm0.06$      & $0.34\pm0.02$      & $19.80\pm1.92$   & $7.00\pm3.16$ & $0.03\pm0.04$ \\
NGC\,4649       &  $5.90\pm0.13$     & $16.84\pm0.02$      & $2.86\pm0.09$      & $124.54\pm2.70$  & $3.34\pm0.43$ & $0.19\pm0.01$   \\
NGC\,4889       &  $9.78\pm0.36$     & $17.19\pm0.03$      & $1.58\pm0.05$      & $169.24\pm14.76$ & $4.82\pm0.42$ & $0.11\pm0.01$   \\
NGC\,5328       & $11.13\pm0.45$     & $17.07\pm0.04$      & $0.85\pm0.04$      & $76.84\pm5.57$   & $2.54\pm0.22$ & $0.07\pm0.05$ \\
NGC\,5516*      &  $6.34\pm0.90$     & $16.39\pm0.04$      & $0.61\pm0.03$      & $24.22\pm7.02$   & $2.04\pm0.15$ & $0.12\pm0.02$   \\
NGC\,5813*      &  $2.07\pm0.04$     & $15.25\pm0.02$      & $0.35\pm0.01$      & $3.49\pm0.03$    & $2.81\pm0.26$ & $0.0\pm0.0$   \\
NGC\,5846       &  $5.32\pm0.10$     & $16.36\pm0.03$      & $1.24\pm0.04$      & $113.15\pm2.82$  & $3.25\pm0.30$ & $0.0\pm0.0$ \\
NGC\,6086*      &  $3.13\pm0.36$     & $17.12\pm0.04$      & $0.55\pm0.02$      & $3.19\pm0.58$    & $3.44\pm0.27$ & $0.12\pm0.01$ \\
NGC\,7619       &  $9.32\pm0.36$     & $14.90\pm0.09$      & $0.48\pm0.06$      & $100.06\pm6.12$  & $1.51\pm0.12$ & $0.14\pm0.07$   \\
NGC\,7768       &  $6.15\pm0.07$     & $16.80\pm0.03$      & $0.29\pm0.01$      & $46.07\pm1.12$   &    100.0        & $0.21\pm0.02$   \\
\hline
\end{tabular}
\tablecomments{These parameters correspond to the best-fitting \cs\
models plotted in Fig.~\ref{fittingplots}.  Asterisks indicate
galaxies which were fit using multiple components (see
Section~\ref{multi-component-fits}). For these galaxies, we list
parameters for the central (\cs) component; see
Table~\ref{tab:outer-params} for the parameters of the outer
components. The surface brightness profiles are calibrated to different
bands, indicated in the y-axis title of the corresponding plots (see
also Table~\ref{tab:coresample}); $\mu_b$ values are not corrected for
extinction. The uncertainties are calculated from 100 Monte Carlo
realizations, described in Section \ref{coredeficit}. The $\alpha$
parameter in NGC\,7768 is fixed to 100.0 as an approximation to
$\alpha=\infty$ (see text).}
\end{table*}

\section{Fitting Galaxy Profiles with Multiple Components}
\label{multi-component-fits}

While the current paradigm for elliptical galaxies is that their
surface-brightness profiles (outside of the nuclear region) are best fit
with the \sersic\ function, there are also long-standing arguments that
at least \textit{some} elliptical galaxies may have light at large radii
attributable to an extra component. In particular, several studies have
shown that brightest cluster galaxies (BCGs), including but not limited
to cD galaxies, are better fit with a combination of two profiles: an
inner \sersic\ (or $r^{\frac{1}{4}}$) function plus an outer envelope,
the latter modeled with exponential, $r^{\frac{1}{4}}$, or \sersic\
functions
\citep[e.g.,][]{Gonzalez-03,Gonzalez-05,Seigar-07,Donzelli-11}.
\citet{Hopkins-09} have shown that a plausible outcome of mergers is an
outer \sersic\ profile plus an inner, exponential-like excess due to
central, post-merger starbursts; such multi-component profiles could
persist even after ``a moderate number'' of subsequent dry mergers. They
suggested out that modeling core-galaxy profiles with single components
(such as the \cs\ profile) could result in biased estimates of the core
light and mass deficits; one possible signature of this is \sersic\ or
\cs\ fits with large values of $n$. \citet{Donzelli-11} noted that those
BCGs which were best fit with two-component profiles had, when fit with
just a single \sersic\ function, large values of $n$ and $r_e$ ($> 8$
and $> 300$ kpc, respectively). Most recently, \citet{Huang-13} have
argued, using 2D fits to ground-based images, that many ellipticals may
be better modeled as the sum of \textit{three} (or in some cases even
\textit{four}) \sersic\ functions.

\begin{figure}
\centering
  \includegraphics[scale=0.5]{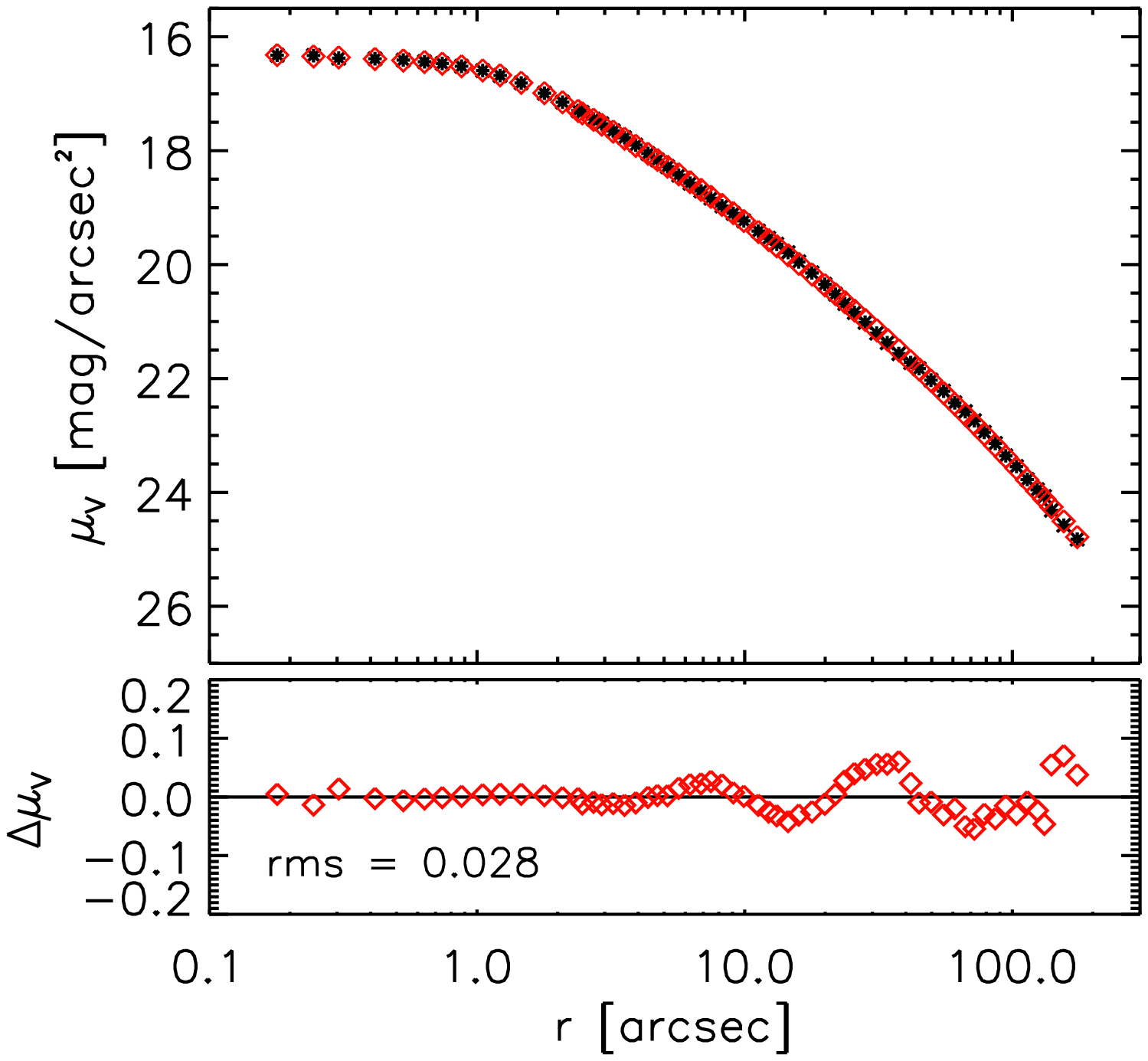}\includegraphics[scale=0.5]{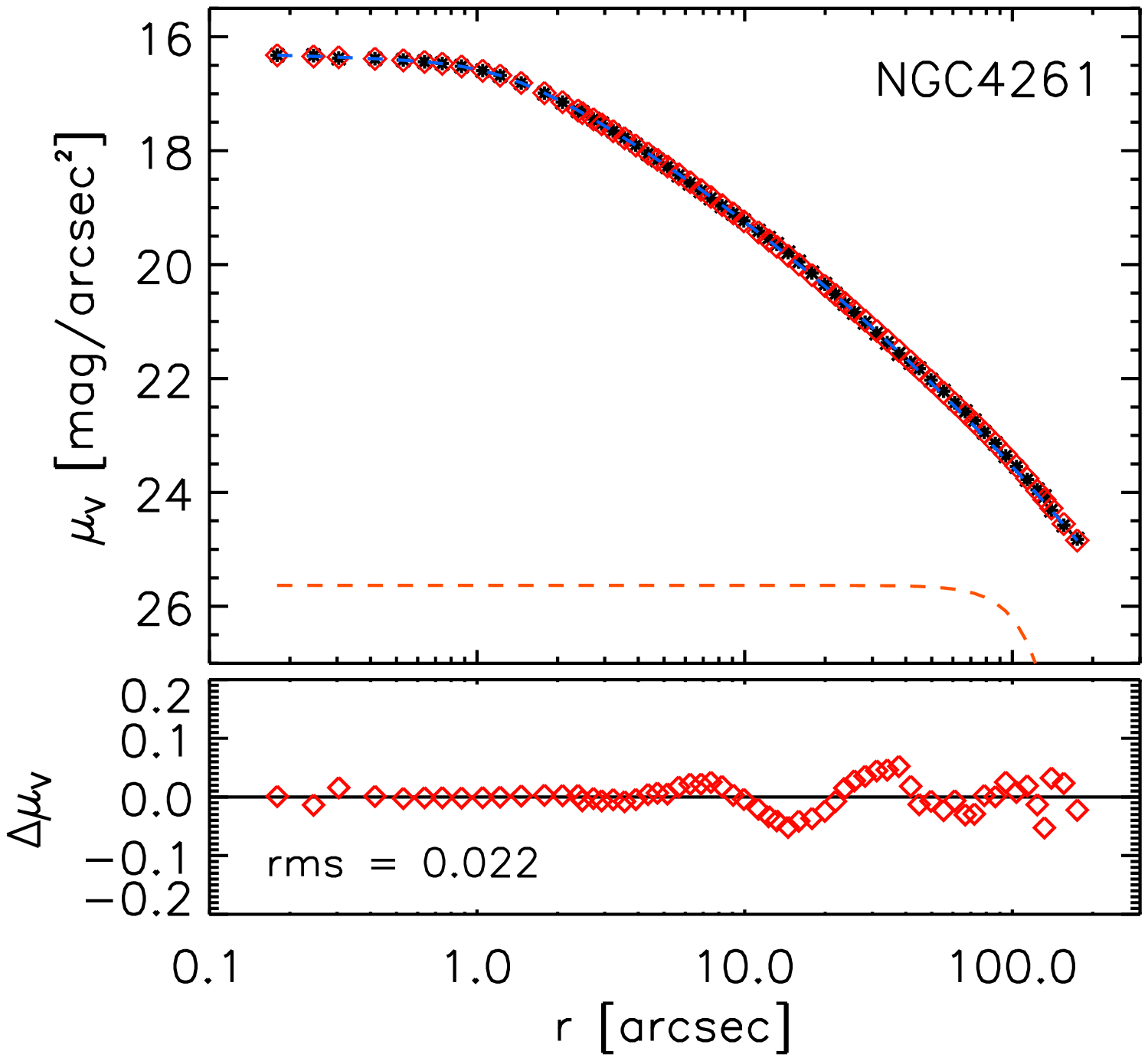}

\caption{Example of a galaxy (NGC\,4261) where the preferred fit is a
simple \cs. Left: Best-fitting \cs\ model (red diamonds) compared to the
data (black asterisks). Right: Same profile, but now with the
best-fitting sum of a \cs\ (dashed blue line) and a \sersic\ profile.
The \sersic\ profile (dashed red line) is everywhere much fainter than
the \cs\ profile, so that the latter is almost identical to the result
of fitting a single \cs\ function by itself. The majority of galaxies in
our sample are similar to this, with little or no evidence for a
significant outer envelope in addition to the \cs\ profile.}
\label{fig:multifit-minimal}
\end{figure}

\begin{figure}
\centering
  \includegraphics[scale=0.5]{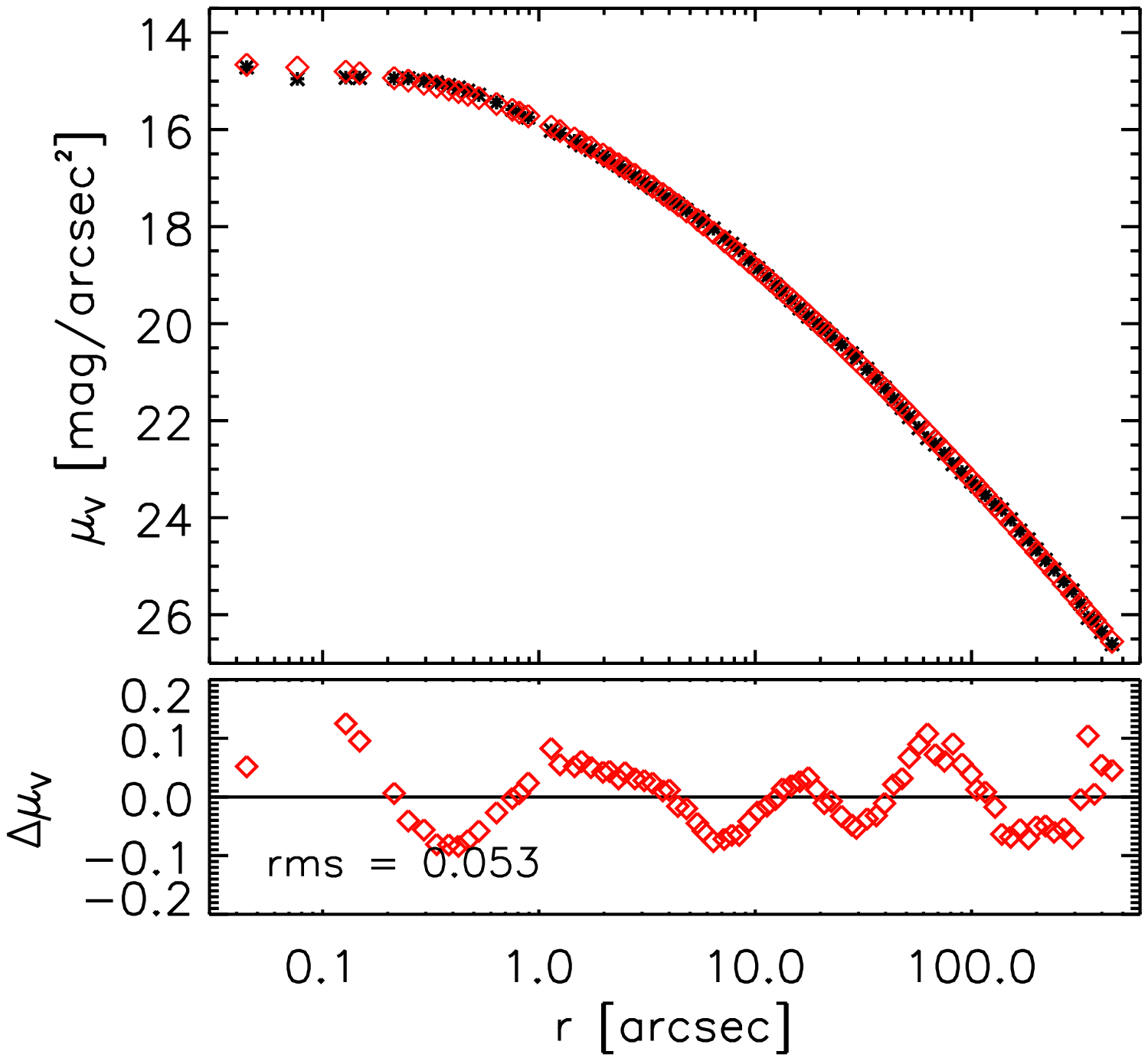}\includegraphics[scale=0.5]{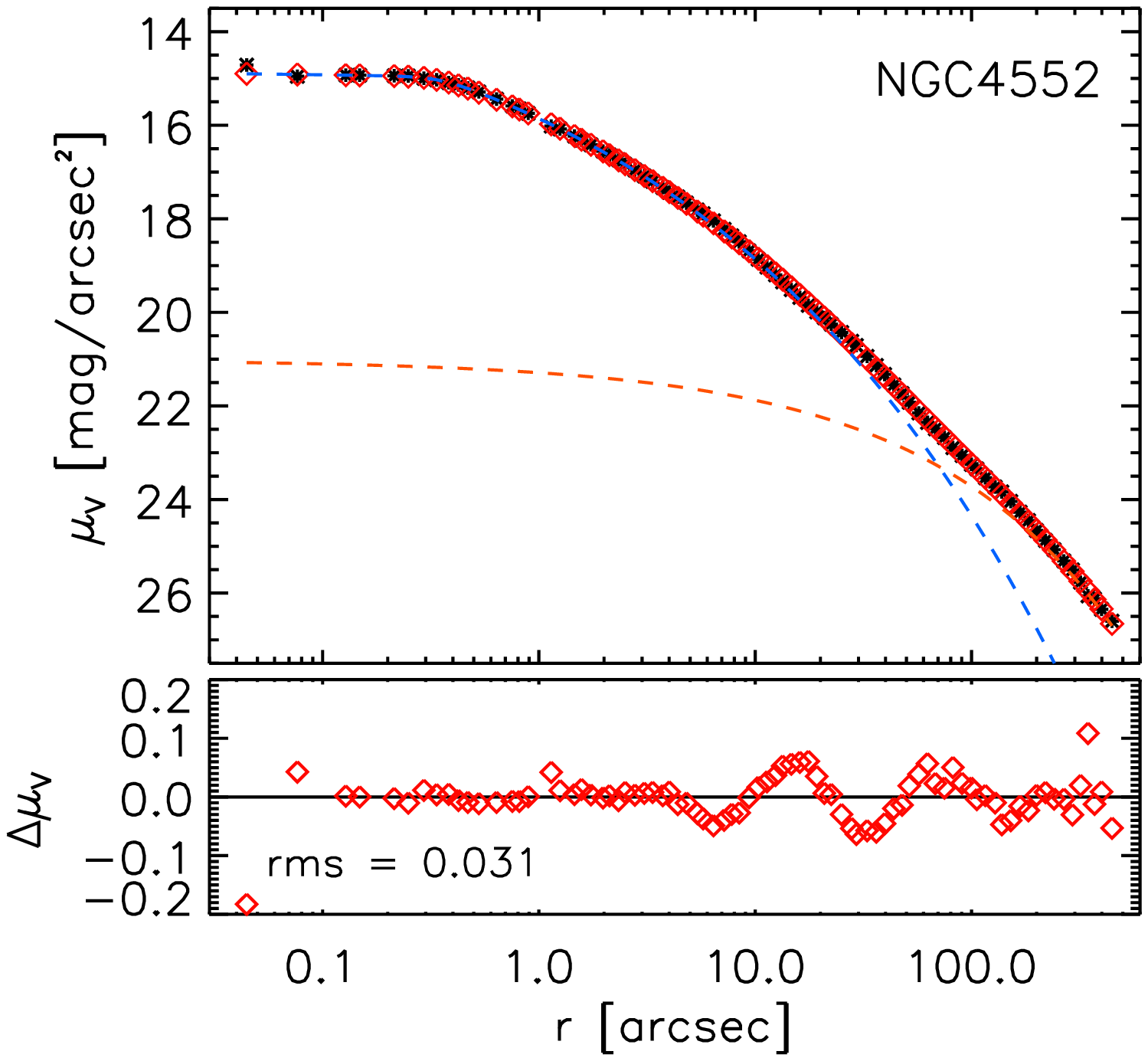}

\caption{As for Fig.~\ref{fig:multifit-minimal}, but now showing a
galaxy (NGC\,4552) where we adopt a multi-component fit as the best fit
to the surface brightness profile. Left: Best-fitting \cs\ fit (red
diamonds) and data (black asterisks). This \cs\ fit has an unusually
large value of $n$ (13.9). Right: Same data, but now showing the
best-fitting sum of a \cs\ and a \sersic\ profile. 	The \cs\
component (dashed blue line) now has $n = 3.8$; the outer \sersic\
component (dashed red line) has $n = 2.0$. The residuals are smaller as
well, including those in the core region.}
\label{fig:multifit-compare}
\end{figure}

Some of the galaxies in our sample are BCGs (e.g., NGC\,4472 and
NGC\,6086), which suggests that we should consider the possibility of
outer envelopes. In addition, some of our \cs\ fits to the
surface-brightness profiles (including both BCG and non-BCG galaxies)
resulted in very large values of the index $n$ (e.g., $n > 10$ or even
20) and/or the effective radius $r_{e}$ (e.g., $r_e \gg$ radius of
outermost data point). To investigate the possibility of outer
envelopes, we re-fit \textit{all} the galaxies in our sample (excepting
NGC~1374, which we judged to not be a core galaxy) using the sum of a \cs{}
and an extra S\'ersic profile. In the majority of cases, the result was
essentially identical to fitting the profile with just a single-\cs{}
component, with the extra S\'ersic component producing minimal
contribution at all radii, and the parameters of the \cs{} component
changing very little (e.g., mean change in $r_b$ $\approx 11$\%); see
Figure~\ref{fig:multifit-minimal} for an example. Galaxies for which the
fit was notably improved -- with the outer component contributing the
majority of light at the largest radii -- were almost always cases where
the single \cs{} fit had excessively large values of $n$ and/or $r_e$
(see Figure~\ref{fig:multifit-compare}).

In order to decide which galaxies were most plausibly better fit
by a multi-component profile, we adopted the following set of conditions for using a multi-component fit in
place of a single \cs\ fit. If at least one of the following is true, then we use
the multi-component fit instead of the \cs\ fit:
\begin{enumerate}
\item The index $n$ of the single-\cs\ fit is $> 10$, \textit{and} the multi-component fit has \cs\
and \sersic\ indices $< 10$. (See Figure~\ref{fig:multifit-compare} for an example.)

\item The effective radius $r_e$ of the single-\cs\ fit is $>$ the radius of
the outermost fitted data point, \textit{and} the reverse is true for the multi-component fit.

\item The rms of the multi-component fit is at least four times smaller than the rms
of the single-\cs\ fit.
\end{enumerate}

In the end we found that approximately one fourth of our sample
galaxies were best fit using multiple components: NGC\,1399, NGC\,1407,
NGC\,4552, NGC\,5516, NGC\,5813, and NGC\,6086. The first two galaxies
were special cases where fitting the envelope required \textit{two}
additional components. For NGC\,1399, we used the sum of a \sersic\ and
an exponential, while for NGC\,1407 we used the sum of two \sersic\
functions. In at least two cases -- NGC\,5813 and NGC\,6086 -- we can plausibly
associate the outer \sersic\ component with outer isophotes which are
significantly more elliptical than the inner part of the galaxy (where the \cs\
component dominates the light).

What is the effect of using multiple-component fits on the core
properties? On average, the derived core sizes ($r_b$ of the \cs\
component) are $\sim 10$--15\% smaller when multiple components are
used. Using multiple-component fits has a stronger effect on the
estimated light and mass deficits (see Section~\ref{coredeficit}). The
most common effect is to lower the estimated deficit (the median deficit
is 0.44 times what it is when a single--\cs\ fit is used), with an
extreme in the case of NGC\,4552, for which the deficit is only 0.9\% of
the value for the single--\cs\ fit. For two galaxies (NGC\,1399 and
NGC\,5516), on the other hand, the deficit roughly doubles in size.

\begin{table*}
\caption{Parameters of Outer \sersic\ Components in Multi-Component Galaxies \label{tab:outer-params}}
\begin{tabular}{llllr}
Galaxy & band & $n$ & $\mu_e$          & $r_e$  \\
       &      &     & (mag arcsec$^{-2}$) & (\arcsec) \\
\hline
NGC\,1399 & $B$ & 1.33 & 25.08 & 189.5 \\
NGC\,1407 & $B$ & 0.44 & 22.96 & 27.0 \\
NGC\,1407 & $B$ & 1.25 & 23.95 & 112.7 \\
NGC\,4552 & $V$ & 2.02 & 25.04 & 222.0 \\
NGC\,5516 & $R$ & 1.33 & 25.08 & 189.5 \\
NGC\,5813 & $i$ & 0.77 & 23.73 & 57.2 \\
NGC\,6086 & $R$ & 2.06 & 24.25 & 50.8 \\
\hline
\end{tabular}
\tablecomments{Parameters of the outer envelope (\sersic) component for the six galaxies
which we fit with \cs{} plus an outer envelope.  NGC\,1407 is fit with \textit{two} outer
\sersic{} components; the NGC\,1399 fit also includes an outer exponential
with $\mu_0 = 25.93$ and scale length $h = 1012.4\arcsec$.}
\end{table*}

\section{Comparison with Previous Core-S\'ersic Fits}
\label{previousfits}

Several previous studies have performed \cs\ fits to the
surface-brightness profiles of some of our galaxies. The main relevant
studies are those of \citet{Ferrarese-06}, \citet{Richings-11}, and
\citet{Dullo-12}. We concentrate on values of the break radius $r_{b}$
and the index $n$ describing the curvature of the profile outside of the
core, since these are reported by all three
studies.\footnote{\citet{Dullo-12} do not report $r_{e}$ values for
their fits.}

\begin{figure}
\centering
\includegraphics[scale=0.9]{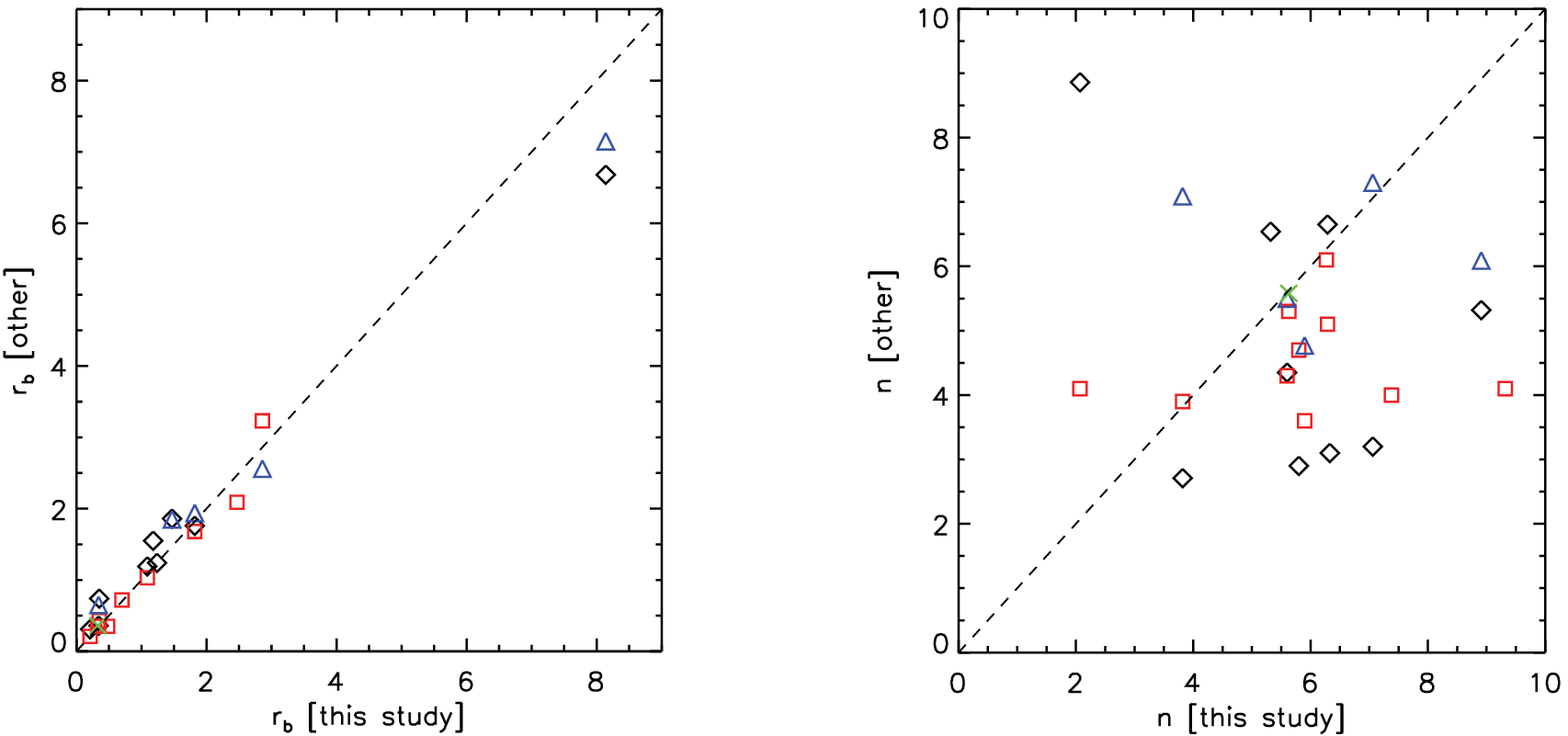}  
\caption{Left panel: Comparison of literature values of $r_{b}$ from \cs\ fits to
galaxies in our sample, using results from \citet[][blue
triangles]{Ferrarese-06}, \citet[][open black diamonds]{Richings-11},
and \citet[][red squares]{Dullo-12}; the green X is NGC\,4291 from \citet{Trujillo-04}. 
Right panel: Same, but now comparing values of the \cs\ index $n$.}
\label{fig:rb-n-vs-others}
\end{figure}

The left panel of Figure~\ref{fig:rb-n-vs-others} compares $r_{b}$
values from the various studies with those from our fits. The
agreement
 is generally excellent; a correlation analysis shows that
the $r_{b}$
 values from all three studies are well correlated with
our values for
 galaxies in common: Spearman $r_{s} = 1.0$, 0.97, and
0.96 for the
 \citet{Ferrarese-06}, \citet{Richings-11}, and
\citet{Dullo-12} values,
 respectively,
 compared to ours (the
corresponding probabilities are 0.0, $2.2 \times
 10^{-5}$, and $7.3
\times 10^{-6}$). The \citet{Ferrarese-06} break
 radii are on
average $1.20 \pm 0.43$ times ours (median = 1.07), the
\citet{Richings-11} values are $1.23 \pm 0.39$ times ours (median =
1.09), and the \citet{Dullo-12} values are $0.99 \pm 0.14$ times ours
(median = 1.01), so there is no clear evidence for systematic offsets
or biases between the different studies when it comes to the break
radius.

The right panel of Figure~\ref{fig:rb-n-vs-others} compares the $n$
values of the various \cs\ fits with ours. Here, the agreement is
clearly much worse, and there are hints of possible systematic bias: the
previous measurements tend to be \textit{lower} than our values of $n$.
The \citet{Ferrarese-06} values are a median of 0.98 times ours (mean $=
1.07 \pm 0.46$), while the values for \citet{Richings-11} and
\citet{Dullo-12} are 0.71 (mean $= 1.12 \pm 1.21$) and 0.81 (mean $=
0.89 \pm 0.43$), respectively.  6 of the 9 Richings et al.\
and 8 of the 10 Dullo \& Graham $n$ values are lower than ours, though
this is not a statistically significant bias.  More generally, we can
note that a correlation analysis shows almost no agreement between the
various values of $n$: $r_s = 0.10$, $-0.13$ and 0.09 ($P = 0.87$, 0.73,
and 0.81) for the Ferrarese et al., Richings et al., and Dullo \& Graham
measurements compared to ours. In addition, a Kolmogorov-Smirnov test
suggests that the Dullo \& Graham $n$ values, and possibly the Ferrarese
et al.\ values as well, are drawn from different distributions than
ours ($P = 0.0069$ and 0.031). 

While it is true that some of our profiles are fit with a different
model -- \cs\ + outer envelope(s) instead of just a \cs\ function -- this
is true for only 1/4 of our total sample and cannot be the cause of the
discrepancy. If we exclude those galaxies in common with the other
studies for which we use multi-component fits (NGC~1399, NGC\,4552, and
NGC\,5813), the correlations and K-S test results do not improve: the
correlation coefficients become $r_s = 0.60$, $-0.14$, and $-0.071$ for
the three studies, with $P = 0.40$, 0.76, and 0.88, while the K-S test
probabilities become 0.53, 0.13, and 0.0042. This might seem to suggest that
the Ferrarese et al.\ values have the best agreement with ours, but
there are only four galaxies in common when we exclude our multi-component
fits, so we really cannot make that conclusion.

We suspect that the main reason for the poor agreement between
previous measurements of \cs\ $n$ and ours for these galaxies is the
much smaller radial range of the surface-brightness profiles used in
the
 previous studies. Both \citet{Richings-11} and \citet{Dullo-12}
use
 profiles that extend to only $\sim 10$--15\arcsec, which means
that the
 outer \sersic\ part of the \cs\ profile is poorly
constrained.  Even
 though the galaxies from \citet{Ferrarese-06}
have profiles which extend
 to $r \sim 100\arcsec$, our profiles for
the same galaxies extend to
 4--7 times further out in
radius. Note that for 6 galaxies we used profiles from
  \citet{Kormendy-09a}. Their Sersic n measurements agree well with our
  values (at the 1.4 sigma level on average, having excluded NGC 4552,
  where we adopt a two-component fit).

An additional factor might be sky-subtraction errors. For example,
\citet{Dullo-12} used profiles originally published by
\citetalias{Lauer-05}. These profiles were extracted from the PC chip of
the \textit{HST} WFPC2 array, and a sky subtraction was applied using
measurements made in the far corners of the WF3 chip. Since almost all
of these galaxies are larger than the WFPC2 array -- sometimes much
larger (e.g., NGC\,1399) -- this ``sky'' measurement will necessarily
include some galaxy light, and thus the resulting profile will be
over-subtracted. Over-subtracted profiles decline more steeply at
large radii, and are better fit by S\'ersic profiles with lower values
of $n$.\footnote{For the five \citet{Ferrarese-06} galaxies in common
with our sample, sky subtraction was estimated based on the pointing of
\textit{HST}. \citet{Richings-11} provide no information as to what, if
any, sky subtraction was done for their data.}

It is clear from this comparison that the outer part of the \cs\ fit is
sensitive to how far out the surface brightness profile extends; fits to
profiles which only extend to $\sim 10\arcsec$ are likely to be poorly
constrained when it comes to the values of $n$. It is important to note
that the \textit{core} parameters are much more robust: the size of the
break radius is relatively insensitive to how much of the outer profile
is included in the fit.

\section{The light and mass deficit in the core}
\label{coredeficit}

We define the light deficit in the inner part of the galaxy as the
difference between the luminosity of the \sersic\ component of the
best-fitting \cs\ model and the luminosity of the \cs\ model itself,
both are integrated out to the radius of the last datapoint. (For
galaxies with multiple-component fits, we consider only the \cs\ component
and ignore the outer envelope.) Since this radius is much larger than
the break radius, the difference represents the total light deficit and
is independent of the choice of the maximum radius.

We first calculate intensity in the units of solar luminosity per ${\rm
pc}^2$ from the surface brightness $\mu$ (mag arcsec$^{-2}$) :
\begin{eqnarray}
  I(\lsun/{\rm pc}^2) & = & (206265)^2\times 10^{0.4({\rm M_{sun}}-\mu-5)} \\
               & = & 4.255\times10^8\times 10^{0.4({\rm M_{sun}}-\mu)}
\end{eqnarray}
where ${\rm M_{sun}}$ is the absolute magnitude of the sun in the same
band as $\mu$. We then obtain the luminosity $L$ interior to the last
datapoint $r_{\rm last}$ by numerically integrating the intensity over a
projected area $\pi\,r_{\rm last}^2$:
\begin{equation}
  L(r\le r_{\rm last}) = \int^{r_{\rm last}}_0{I(r) \, 2 \pi \, rdr}
\end{equation}
This is calculated for both the \cs\ model and the extrapolated \sersic\
component of that model. The difference between the two is the
luminosity deficit $\Delta L$ (in \lsun). For comparison with the work
of \citetalias{Kormendy-09b}, the light deficit is computed in
magnitudes according to $\mdef = -2.5\times {\rm log_{10}}(\Delta L) +
{\rm M_{sun}}$, where $\lambda$ indicates that the conversion is
done in the same band.  In order to convert the light deficit into a
stellar mass deficit \massdef, we use the stellar
mass-to-light-ratio \ml\ derived from dynamical models (see
Table~\ref{tab:coresample}). 

For 16 of the 20 galaxies with reliable black hole mass
determinations, these \ml\ values stem from two-components models that
fit simultaneously for the \ml\ which scales the light-density profile
and for the parameterized dark-matter halo density distribution (in
addition to the black hole mass, of course). For the remaining four
galaxies -- IC\,1459, NGC\,3379, NGC\,4261, and NGC\,4374 (along with
the three objects with dubious black hole masses, which we do not use in
our analyses) -- the \ml\ values come from single-component dynamical
modeling (i.e., without dark matter), so the true \ml\ could be
overestimated by $\approx 20$\% \citep[see][]{Thomas-07}. We prefer this
approach to the use of stellar-population $\Upsilon_{\star}$
determinations, because of the uncertainty which affects the latter
concerning the proper stellar initial mass function (IMF) to be used.
The recent results of \citet{Thomas-11}, \citet{Wegner-12},
\citet{Cappellari-12,Cappellari-13}, and \citet{Conroy-12} all show that
the IMF may change as one goes from low- to high-velocity-dispersion ellipticals, from
a Kroupa IMF to an IMF even ``heavier'' than Salpeter, but the scatter in
this trend is too large to predict the IMF on the basis of the velocity
dispersion; moreover, some high-dispersion ellipticals may not have a
``heavy'' IMF \citep{vandenBosch-12, Rusli-13,Smith-13}. Possible spectroscopic
information constraining the IMF \citep{Conroy-12} is not available for
most core galaxies.
We apply a simple multiplication of $\ml
\times \Delta L$ (given that both are measured in the same band and
assuming that \ml\ is constant within the galaxy) to obtain the mass
deficit.

The uncertainties are calculated through a Monte Carlo simulation. Using
the rms of the best fit as an estimate of the measurement error,
Gaussian noise is added to the best-fit model profile to create 100
realizations of the observed surface brightness profile. For each
realization, a best-fit \cs\ (or multiple-component) profile was derived
and the light deficit was calculated in the same way as done for the
real profile. The rms of those 100 values of light deficit becomes the
uncertainty. The error of the mass deficit is derived from the
uncertainties of both the light deficit and \ml. From these 100
realizations, we also derive the rms of each \cs\ parameter and adopt
this as the error of the parameter (given in Table~\ref{tab:csparams}).
This Monte Carlo approach provides estimates of random errors, not
systematic errors. The latter could include errors in describing the
PSF, errors in the sky subtraction, failure to account for possible
substructure in the galaxies, and the sampling of the surface brightness
profiles. One should also keep in mind that the circularized profiles
that we use can be systematically different -- depending on the
ellipticity profiles -- than the semi-major axis profiles commonly used
in other papers.

To enable a statistical comparison of the light deficit \mdef\
among the galaxies, this quantity should be presented in a single
photometric filter. For this purpose, we convert \mdef\ in the
given band to the $V$-band ($\mvdef$) by applying a color
correction based on a value obtained from the literature, specific for
each galaxy. In the case where multiple aperture measurements are
available, we take the color corresponding to the smallest aperture
since we are mostly interested in the inner part of the galaxy where the
core is present. We assume a typical uncertainty of 0.05 mag in color,
taken into account in the error calculation for $\mvdef$.
Table~\ref{tab:massdef} lists the color correction, the light and mass
deficit for each galaxy.

\begin{table*}
\caption{The light and mass deficits in the core \label{tab:massdef}}
\begin{tabular}{lllllll}
Galaxy         & $\mdef$              & Color    & Value    & $\mvdef$                & \massdef{}                 &  $f(M_{\rm BH})$  \\
               & (mag)                &          &          & (mag)                   & (\msun)                    &  \\
\hline
IC\,1459       & $-19.15\pm0.35$      & \ldots{} & \ldots{} & $-19.15\pm0.35$    & $(1.78\pm0.75)\times10^9$  &  0.64 \\
NGC\,1399      & $-18.37\pm0.13$      & $B-V$    & 0.68     & $-19.05\pm0.14$    & $(3.52\pm0.46)\times10^{10}$  &  38.7 \\
NGC\,1407      & $-16.56\pm0.17$      & $B-V$    & 0.98     & $-17.54\pm0.18$    & $(4.31\pm0.92)\times10^9$    &  0.96 \\
NGC\,1550      & $-19.17\pm0.11$      & $V-R$    & 1.08     & $-18.09\pm0.12$    & $(1.09\pm0.19)\times10^{10}$     &  2.95 \\
NGC\,3091      & $-19.90\pm0.20$      & $V-I$    & 1.39     & $-18.51\pm0.21$    & $(1.57\pm0.34)\times10^{10}$  &  4.37 \\ 
NGC\,3379      & $-17.56\pm0.11$      & $V-I$    & 1.29     & $-16.27\pm0.12$    & $(1.26\pm0.15)\times10^9$   &  3.07 \\
NGC\,3608      & $-13.95\pm0.10$      & \ldots{} & \ldots{} & $-13.95\pm0.10$    & $(1.01\pm0.13)\times10^{8}$   &  0.21 \\
NGC\,3842      & $-17.88\pm0.11$      & \ldots{} & \ldots{} & $-17.88\pm0.11$    & $(8.04\pm1.50)\times10^9$  &  0.83 \\
NGC\,4261      & $-17.67\pm0.12$      & \ldots{} & \ldots{} & $-17.67\pm0.12$    & $(9.11\pm1.58)\times10^9$   &  17.5 \\
NGC\,4291      & $-15.92\pm0.06$      & \ldots{} & \ldots{} & $-15.92\pm0.07$    & $(1.07\pm0.16)\times10^{9}$    &  1.16 \\
NGC\,4374      & $-17.68\pm0.15$      & \ldots{} & \ldots{} & $-17.68\pm0.15$    & $(6.85\pm1.10)\times10^{9}$  &  7.45 \\
NGC\,4472      & $-17.35\pm0.14$      & \ldots{} & \ldots{} & $-17.35\pm0.14$    & $(3.66\pm0.58)\times10^9$  &  1.46 \\
NGC\,4486      & $-20.27\pm0.16$      & \ldots{} & \ldots{} & $-20.27\pm0.16$    & $(6.76\pm1.34)\times10^{10}$  &  10.9 \\
NGC\,4552      & $-14.12\pm0.15$      & \ldots{} & \ldots{} & $-14.12\pm0.15$    & $(2.70\pm0.40)\times10^{8}$  &  0.54 \\
NGC\,4649      & $-18.12\pm0.12$      & \ldots{} & \ldots{} & $-18.12\pm0.12$    & $(1.11\pm0.17)\times10^{10}$  &  2.22 \\
NGC\,4889      & $-20.62\pm0.07$      & $V-R$    & 0.65     & $-19.97\pm0.09$    & $(5.97\pm1.76)\times10^{10}$  &  2.84 \\
NGC\,5328      & $-19.52\pm0.09$      & \ldots{} & \ldots{} & $-19.52\pm0.09$    & $(2.70\pm0.41)\times10^{10}$  &  6.23 \\
NGC\,5516      & $-19.12\pm0.29$      & $V-R$    & 0.60     & $-18.52\pm0.30$    & $(1.35\pm0.31)\times10^{10}$  &  4.09 \\
NGC\,5813      & $-16.08\pm0.22$      & $V-i$    & 0.71     & $-15.37\pm0.24$    & $(5.80\pm1.24)\times10^8$  &  0.83 \\
NGC\,5846      & $-17.48\pm0.15$      & $V-i$    & 0.79     & $-16.69\pm0.16$    & $(2.32\pm0.33)\times10^{9}$  &  2.11 \\
NGC\,6086      & $-18.69\pm0.20$      & $V-R$    & 0.60     & $-18.10\pm0.21$    & $(7.36\pm1.70)\times10^{9}$  &  2.05 \\
NGC\,7619      & $-20.05\pm0.18$      & $V-I$    & 1.23     & $-18.82\pm0.19$    & $(1.34\pm0.28)\times10^{10}$  &  5.36 \\  
NGC\,7768      & $-16.41\pm0.06$      & \ldots{} & \ldots{} & $-16.41\pm0.06$    & $(2.46\pm0.50)\times10^9$  &  1.89 \\
\hline
\end{tabular}
\tablecomments{\mdef\ is the absolute magnitude of the light deficit in the given filter 
(see Table~\ref{tab:coresample}). This quantity is converted to the $V$-band ($\mvdef$) 
using the color and its value in col. 4 and 5. The value is specific for each galaxy and in general 
taken from \citet{Prugniel-98} for the smallest aperture available and then corrected for extinction 
\citep{Schlegel-98}; we assume a typical uncertainty of 0.05 mag for each color. For NGC\,3091, 
$I-$F814W is 0.03 as derived in \citetalias{Rusli-13} and for the extinction correction we adopt the value in 
the Landolt $I$-band. 
For NGC 5813 and NGC 5846, the conversion from SDSS $i$-band to Cousins $I$
uses central $r-i$ colors measured from the SDSS images and the equations of \citet{Jordi-06}, 
yielding $i-I = 0.50$ for NGC\,5813 and $i-I = 0.51$ for NGC\,5846.
The uncertainty in the mass deficit \massdef{} takes into account the uncertainties in both the light 
deficit and \ml.\\}
\end{table*}

\section{Black hole--core correlation}
\label{bhcore}

\begin{table*}
\caption{Relations Between Core Size and Galaxy Parameters \label{tab:core-relations}}
\begin{tabular}{lcccrrl}
Fit type & $a$             & $b$             & $\epsilon$     & rms   & $r_{s}$ &  $P$  \\
\hline
\multicolumn{5}{c}{$r_b$ vs.\ ($\sigma_e/200$ \kms)}                  & 0.53    & 0.017 \\
Direct   & $-1.41\pm0.17$  & $4.19\pm1.00$   & $0.27\pm0.06$  &  0.24 & & \\
Reverse  & $-1.84\pm0.28$  & $6.95\pm1.71$   & $0.34\pm0.11$  &  0.32 & & \\
Bisector & $-1.57\pm0.15$  & $5.23\pm0.96$   & $0.30\pm0.09$  &  0.25 & & \\
\hline
\multicolumn{5}{c}{$r_b$ vs.\ $M_{V}$}                                & $-0.65$ & 0.0018 \\
Direct   & $-0.55\pm0.06$  & $-0.41\pm0.07$  & $0.24\pm0.05$  &  0.20 & & \\
Reverse  & $-0.45\pm0.09$  & $-0.61\pm0.13$  & $0.29\pm0.09$  &  0.30 & & \\
Bisector & $-0.50\pm0.05$  & $-0.51\pm0.07$  & $0.26\pm0.07$  &  0.22 & & \\
\hline
\multicolumn{5}{c}{$r_b$ vs.\ $(\mbh/3\times10^{9} \msun)$}           & 0.77   & $7.3\times10^{-5}$ \\
Direct   & $-0.68\pm0.06$  & $0.73\pm0.16$   & $0.25\pm0.05$  &  0.23 & & \\
Reverse  & $-0.64\pm0.10$  & $1.15\pm0.37$   & $0.32\pm0.15$  &  0.32 & & \\
Bisector & $-0.66\pm0.05$  & $0.92\pm0.17$   & $0.28\pm0.11$  &  0.26 & & \\
\hline
\end{tabular}

\tablecomments{Coefficients of fits between core size $r_b$ (in kpc) and galaxy
parameters (velocity dispersion, luminosity, or black hole mass) for the
20 core galaxies with well-determined black hole masses. For each
relation we list intercept $a$, slope $b$, and intrinsic Gaussian
scatter $\epsilon$ for direct, reverse, and bisector fits, along with
the rms of residuals (in $\log r_b$) from the fit; see Section~\ref{bhcore}. The
Spearman correlation coefficient $r_s$ and its associated probability
$P$ are listed for each overall relation in the 6th and 7th columns.}

\end{table*}

\begin{table*}
\caption{Relations Between Light/Mass Deficits and Galaxy Parameters \label{tab:mdef-relations}}
\begin{tabular}{lcccrrl}
Fit type & $a$             & $b$             & $\epsilon$     & rms   & $r_{s}$  &  $P$  \\
\hline
\multicolumn{5}{c}{\mvdef{} vs.\ ($\sigma_e/200$ \kms)}               & $-0.75$  & 0.00013 \\
Direct   & $-14.93\pm0.48$ & $-19.53\pm2.82$ & $0.68\pm0.18$  &  0.70 & & \\
Reverse  & $-14.37\pm0.54$ & $-23.09\pm3.10$ & $0.70\pm0.21$  &  0.78 & & \\
Bisector & $-14.67\pm0.36$ & $-21.16\pm2.11$ & $0.69\pm0.20$  &  0.73 & & \\
\hline
\multicolumn{5}{c}{\mvdef{} vs.\ $(\mbh/3\times10^{9} \msun)$}        & $-0.59$ & 0.0059 \\
Direct   & $-18.25\pm0.28$ & $-2.67\pm0.73$  & $1.14\pm0.24$  &  1.07 & & \\
Reverse  & $-18.45\pm0.41$ & $-4.86\pm1.32$  & $1.53\pm0.56$  &  1.57 & & \\
Bisector & $-18.32\pm0.23$ & $-3.47\pm0.71$  & $1.34\pm0.43$  &  1.20 & & \\
\hline
\multicolumn{5}{c}{$(\massdef/3\times10^{9} \msun)$ vs.\ ($\sigma_e/200$ \kms)}   & 0.68 & 0.00098 \\
Direct   & $8.55\pm0.23$   & $8.38\pm1.35$   & $0.34\pm0.09$  &  0.36 & & \\
Reverse  & $8.24\pm0.27$   & $10.36\pm1.62$  & $0.35\pm0.12$  &  0.41 & & \\
Bisector & $8.41\pm0.17$   & $9.27\pm1.10$   & $0.34\pm0.11$  &  0.38 & & \\
\hline
\multicolumn{5}{c}{$(\massdef/3\times10^{9} \msun)$ vs.\ $(\mbh/3\times10^{9} \msun)$}   & 0.57 & 0.0086 \\
Direct   & $9.97\pm0.13$   & $1.10\pm0.34$   & $0.54\pm0.11$  &  0.51 & & \\
Reverse  & $10.09\pm0.21$  & $2.27\pm0.74$   & $0.76\pm0.32$  &  0.79 & & \\
Bisector & $10.01\pm0.12$  & $1.54\pm0.33$   & $0.65\pm0.24$  &  0.58 & & \\
\hline
\end{tabular}
\tablecomments{As for Table~\ref{tab:core-relations}, but now showing coefficients of fits
(and overall correlations) between light deficit \mvdef\ or mass deficit \massdef\ of the
core and either $\sigma_e$ or \mbh\ of the galaxy. The rms residuals are for \mvdef{} in
the first two relations and for $\log \massdef$ in the second two.}
\end{table*}

\begin{figure*}[h!]
\centering
   \includegraphics[scale=0.75,trim=8mm 0mm 5mm 0mm,clip]{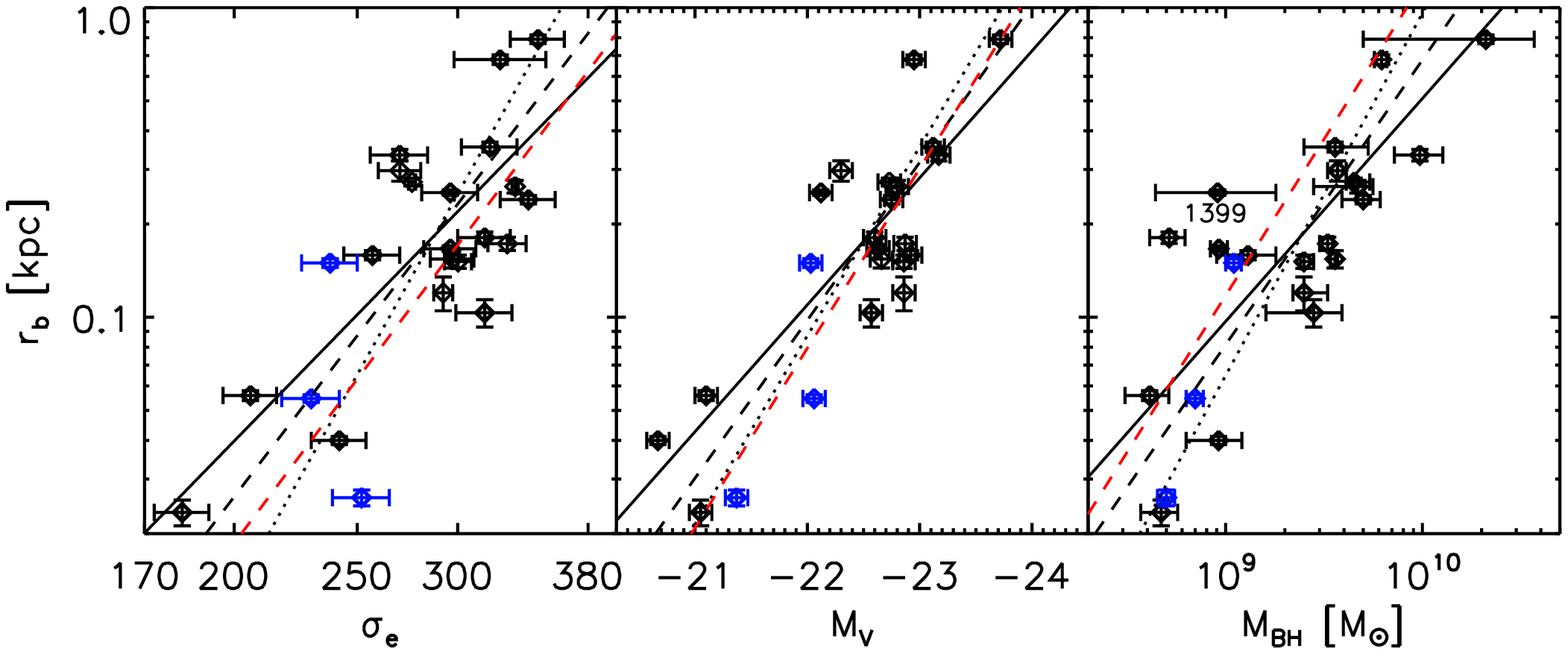}     
  \caption[]{From left to right: the break radius is plotted as a
  function of velocity dispersion within the effective radius
  $\sigma_e$, luminosity of the host galaxy and black hole mass. The 20
  galaxies with reliable \mbh\ are represented by the black diamonds.
  The additional three galaxies without reliable \mbh\ (NGC\,4552,
  NGC\,5813, NGC\,5846) are shown in blue. The black lines show our fits
  to all black datapoints. The solid lines show the fits when $x$-axis
  variables are treated as independent variables/predictors, the dotted 
  lines are fits when these parameters are the response, and the
  dashed lines represent the symmetrical bisector regression; see
  Table~\ref{tab:core-relations}. The red dashed lines are the bisector
  fits from \citet{Dullo-12}, i.e. their equations 5, 6, and 12,
  respectively.}
\label{rbplots}
\end{figure*}

Taking the break radius ($r_b$) as a measure of the core size, we
examine its relationships with galaxy velocity dispersion, luminosity
and the black hole mass. These are displayed in Fig.~\ref{rbplots}. The
fit to the datapoints (also for all the other fits in this Section) is
based on a Bayesian method with Gaussian errors and intrinsic scatter
$\epsilon$, described in \citet{Kelly-07}, and the results are shown by
the black lines. For each diagram we perform non-symmetrical regression
using either of the parameter to predict the other. We further compute
the bisector of the two fitting results (with flipped $x$ and $y$-axes) to
give a symmetrical representation of the correlation. 
The coefficients of the fits involving $r_b$ are shown in Table~\ref{tab:core-relations}.
All the fits are linear fits to the logarithm of the relevant quantities expressed
in the following form:
\begin{equation}
y \; = \; a \, + \, b x,
\label{eqn:basic}
\end{equation}
where $y$ is $\log r_b$ (in kpc) and $x$ is $\log \sigma_e / (200 \:
\kms)$, $M_V$, or $\log \mbh / (3\times10^{9} \msun)$. For the ``direct''
fits $\sigma_e$, $M_V$, and $\mbh$ are the independent variables in the
fit, while in the ``reverse'' fits $r_b$ is the independent variable.
The errors on the coefficient values are taken as the rms of the
a-posteriori parameter probability distribution. We also list the
Spearman correlation coefficients and the corresponding probabilities
for each relation in Table~\ref{tab:core-relations}.

\citet{Dullo-12} analysed the same relationships using bisector fits. In
the $r_b$--$\sigma_e$ diagram, our fit predicts a slightly higher
$r_b$ for a given velocity dispersion, though the fits are
compatible within the errors.  In the $r_b$--${\rm M_V}$ diagram, our
fit coincides with their fit for high-luminosity galaxies (${\rm M_V}
\gtrsim 23$).  In the case of the $r_b$--$M_{\rm BH}$ relation, Graham
\& Dullo performed fits using both measured black hole masses (their
Equation~12) and proxy black hole masses estimated from the $M_{\rm
BH}$--$\sigma$ and $M_{\rm BH}$--$L_{V}$ relations (their Equation~13
and 14); only the former is actually plotted (their Fig.~20). Since our
fits use only measured black hole masses, we compare our results with
their Equation~12 (lower right panel of Figure~6). For this relation,
our fit suggests larger $M_{\rm BH}$ for the same value of break radius
(or, conversely, smaller break radius for a given $M_{\rm BH}$). We note
that the fit in Dullo \& Graham's Equation~12 is based on only 7
galaxies, all but one of which have black hole masses $< 10^{9} \msun$,
versus the 20 galaxies used in our fit, with 14 having $M_{\rm BH} >
10^{9} \msun$.

The measured rms scatter is almost identical for the $r_b$--$\sigma_e$
and $r_b$--\mbh{} fits (0.24 dex versus 0.23 dex). The tightest
relationship turns out to be between $r_b$ and ${\rm M_V}$, where the
rms is 0.20 dex.  The correlation coefficients indicate that all three
relations are strong and statistically significant, with the strongest
being that between $r_{b}$ and $M_{\rm BH}$ ($r_s = 0.77$, $P =
7.3\times10^{-5}$). Both \citet{Lauer-07} and \citet{Dullo-12} use a
larger number of datapoints than ours in examining the
$r_{b}$--$\sigma_{e}$ and $r_{b}$--$M_{V}$ relationships, and they find
opposing results as to which parameter ($\sigma_{e}$ or $M_{V}$) is
better correlated with $r_{b}$; our results suggest that $M_{V}$ is a
better predictor of $r_{b}$ than $\sigma_{e}$. (Note that the
$r_{b}$--$M_{V}$ correlation uses the \textit{total} galaxy luminosity;
if for the multi-component galaxies we use the luminosity of just the
\cs\ component instead, the correlation is weaker and only marginally
significant, with $r_{s} = -0.41$ and $P = 0.070$.) Despite the
moderately large scatter, the $r_{b}$--$M_{\rm BH}$ relation is
apparently the strongest one among the three, suggesting that $M_{\rm
BH}$ is the most relevant parameter for determining $r_{b}$.

\begin{figure*}[b!]
\centering
  \includegraphics[scale=0.85,trim=8mm 0mm 5mm 0mm,clip]{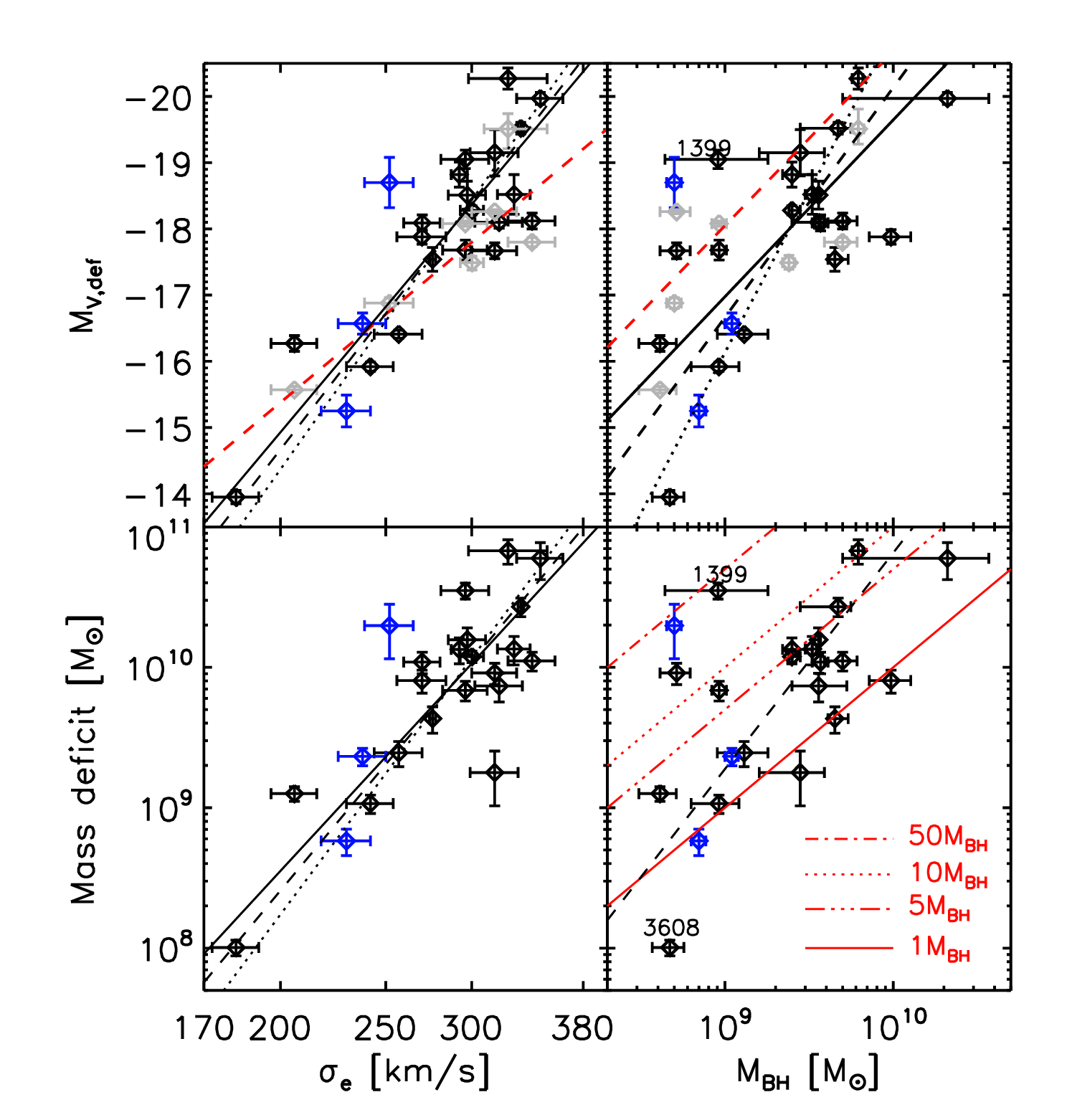}
  \caption[]{{\bf Upper panels:} the plot of $\mvdef$ vs
      $\sigma_e$ and \mbh. $\mvdef$ is the magnitude of the
      light deficit in the center, measured in the $V$-band. The black
      diamonds show light deficit measurements for all 20
      galaxies; the additional three galaxies (NGC\,4552, NGC\,5813, and
      NGC\,5846) are shown in blue. The gray points, for the
      seven galaxies we have in common with \citetalias{Kormendy-09b}, mark the light
      deficit values from that paper. The dashed red lines are
      the relation reported in \citetalias{Kormendy-09b}. The black lines are our fits
      to the black diamonds: solid lines for direct fits ($x$-axis variables
      as predictors), dotted for reverse fits, and dashed lines for bisector
      fits; see Table~\ref{tab:mdef-relations}. {\bf Lower panels:} mass deficits \massdef\
      plotted against the velocity dispersions (left) and black hole
      masses (right). The symbols here are the same as in the upper
      panels.}
\label{deficitplots}
\end{figure*}

Fig.~\ref{deficitplots} explores relationships between the light or mass
deficit in the core and the black hole mass or the stellar velocity
dispersion. The upper left panel shows the plot of the light deficit in
the $V$-band versus the stellar velocity dispersion calculated within
the effective radius $\sigma_e$. The correlation between stellar
velocity dispersion of host galaxies and $\mvdef$ determined in
\citetalias{Kormendy-09b} is overplotted in red. Note that the velocity
dispersion used to construct the latter correlation is defined
differently: it is based on the velocity dispersion averaged within a
slit aperture with a length of two half-light radii. Although there is
probably no significant difference between these two methods of
determining $\sigma_e$, this means that the individual datapoints cannot
be strictly compared to the \citetalias{Kormendy-09b} relation. For
seven galaxies that overlap with \citetalias{Kormendy-09b}'s sample, we
indicate their values of light deficit using gray diamonds. On average,
our values are more negative (i.e. more luminous) than those of
\citetalias{Kormendy-09b} by 0.30 mag, and the rms of the difference is
1.15 mag.  In the upper right panel of Fig.~\ref{deficitplots}, the plot
between $\mvdef$ and the black hole masses is shown. Similarly, we
overplot the correlation between these two quantities, as derived in
\citetalias{Kormendy-09b}, in red.

Coefficients of the fits plotted in Fig.~\ref{deficitplots} are listed in
Table~\ref{tab:mdef-relations}. The meaning of the coefficients is the same as
in Equation~\ref{eqn:basic}, except now $y$ corresponds to either \mvdef\ or
$\log \massdef$. In the direct fits, the independent variables are either $\log \sigma_e$ or
$\log \mbh$.

Compared to \citetalias{Kormendy-09b}, our $\mvdef$--$\sigma_e$ correlation is
steeper and our $\mvdef$--\mbh\ exhibits a shift towards a
larger \mbh\ at a given value of light deficit. This highlights the
difference between estimates based on the \msig\ relation and the more
reliable dynamical estimates of \mbh\ that we use. The scatter is
notably higher in $\mvdef$--\mbh\ diagram, opposite to what is
found in \citetalias{Kormendy-09b}.

The most interesting relation from a theoretical point of view is that
between the \textit{mass} deficit and the black hole mass, since this is
the quantity which simulations predict (see references in the
Introduction). The lower panels of Fig.~\ref{deficitplots} show this
relation, and also the relation between mass deficit and velocity
dispersion; the coefficients of the fits are in the bottom half of
Table~\ref{tab:mdef-relations}.  The large intrinsic scatter (and
the large rms values) that we derive indicate that the linear fit is not
representing the data well.

Of the seven correlations we have examined (Figs. \ref{rbplots} and
\ref{deficitplots}), those between $r_b$ and $M_{\rm BH}$ and between
$\mvdef$ and $\sigma_e$ appear to be the strongest ($r_{s} = 0.77$ and
$-0.75$, respectively). Moreover, the rms scatter in $\log M_{\rm BH}$
for the fit between $M_{\rm BH}$ and $r_b$ is only 0.30 dex (it is 0.35
dex for the correlation with \massdef{} and 0.33 dex for that with the
light deficit). This is comparable to or even lower than the typical rms
scatter for the $M_{\rm BH}$--$\sigma$ relation
\citep[e.g.,][]{Park-12,Graham-13,McConnell-13}. This suggests that
measuring $r_{b}$ can allow one to estimate $M_{\rm BH}$ (in core
galaxies) with a precision similar to that obtained using the $M_{\rm
BH}$--$\sigma$ relation, without the need for spectroscopy. We give here
below the corresponding equation (derived from the fits in
Table~\ref{tab:core-relations}):
\begin{equation}
\log \left(\frac{M_{\rm BH}}{3\times10^9 \msun}\right) = (0.59\pm0.16) + (0.92\pm0.20) \log \left(\frac{r_b}{\rm kpc}\right); \: \epsilon=0.28\pm0.06~\mathrm{dex}.   
\label{rbmbh2rev}
\end{equation}

The ratio between \massdef\ and \mbh\ seems to vary strongly, but is
biased toward the low end of the range. Most galaxies have a mass
deficit between 1 and 5 \mbh, and the median value is 2.2 \mbh. In three
galaxies (IC\,1459, NGC\,3842 and NGC\,5813), the mass deficit is just
below the black hole mass. NGC\,3608 has a remarkably low mass deficit:
only 0.21 \mbh. The largest mass deficit that we find is just below 40
\mbh, occurring in NGC\,1399. This unusually large value should perhaps
be taken with a grain of salt, however, given the large uncertainty in
the black hole mass of NGC\,1399. The two publications which have
presented dynamical measurements of the black hole mass
\citep{Houghton-06, Gebhardt-07} provide values which differ by almost a
factor of 3, and neither measurement included the dark matter halo in
the modelling, which could likely result in underestimating \mbh\ and
thereby overestimating the ratio between the mass deficit and \mbh\
\citepalias[see][]{Rusli-13}.  The next-largest mass deficit is 18 times
\mbh, in NGC\,4261, with the one other galaxy having a mass deficit
larger than 10\mbh\ being NGC\,4486. The broad range of $\massdef/\mbh$
(between 0.2 and 18 if we exclude NGC\,1399) is in better agreement with
\citet{Dullo-12} than with \citetalias{Kormendy-09a}. The former also
shows a number of galaxies having mass deficits of 1--4 \mbh\ while the
latter has no galaxies in their sample with mass deficits below 5 \mbh.
Considering that \citetalias{Kormendy-09a} and \citet{Kormendy-09b} use mostly outdated \mbh\
values and an old \msig\ relationship for their analysis, it is likely
that their ratios of $\massdef/\mbh$ are simply too high. (For the seven
of our galaxies in common with those in \citet{Kormendy-09b}, we find that their
mass deficits are somewhat higher than ours -- median ratio $\sim 1.2$ --
while their black hole masses are only $\sim 55$\% of ours.) The
simulations of \citet{Gualandris-08} indicate that mass deficits up to
$5 \, M_{BH}$ can be generated from single dry mergers. Therefore, our
results can probably be explained with a reasonably low number of major
dry mergers. Low $\massdef/\mbh$ ratios could also be obtained if gas to
refill the core via new star formation is available (i.e., the merger is
``wet'').

Why does $r_b$ correlate so well with $M_{\rm BH}$, while the light and
the mass deficit do not? We note that our estimate of $M_{\rm V,def}$,
from which the mass deficit is in turn derived, is based on the
assumption that the inward extrapolation of the outer S\'ersic fit is
the correct representation of the inner projected density profile of the
galaxy before the binary black hole scouring takes place. This might not
be the case for real galaxies, and the large scatter about the relations
of Fig.~\ref{deficitplots} could simply reflect this \citep[see,
e.g.][]{Hopkins-10}.  The size of the core may be a more robust quantity
in this respect. It will be interesting to investigate whether the
internal kinematics of the galaxies change at this radius.

\section{Summary}
\label{coresummary}

We have presented measurements for cores in 23 elliptical galaxies,
where the cores are clearly detected based on the criteria suggested in
\citet{Trujillo-04}. These detections and measurements were made using
core-S\'ersic fits to extended surface-brightness profiles, the majority
of which go out significantly beyond the half-light radius; for a
quarter of the sample, we found that the best fits were with the sum of
\cs\ and an outer envelope, the latter represented by an exponential or
\sersic\ component. For 20 of these galaxies, reliable $M_{\rm BH}$
measurements are available from the literature or from our own work
\citepalias{Rusli-13}; the majority of these are derived from recent
stellar dynamical modelling.  We computed the light deficit in the core
(assumed to be the result of supermassive black hole mergers which
follow galaxy mergers) assuming that the original profile, before the
core was formed, followed a S\'ersic model matching the outer part of
the galaxy, based on the core-S\'ersic fit (or the outer part of
the \cs\ component in the case of multi-component fits). We then derived
the corresponding stellar mass deficit using stellar mass-to-light ratios
from dynamical models, available for all of the galaxies in the sample.

We find that the size of the core (break radius $r_{b}$) is most
strongly correlated with the black hole mass, while the second strongest
correlation (and the linear fit with the lowest scatter) is with the
luminosity of the host galaxy; the weakest correlation is that between
the break radius and the galaxy's stellar velocity dispersion.  The low
scatter of the $r_b$--$M_{\rm BH}$ relation (0.30 dex when $r_{b}$
is the independent variable) opens up the possibility of accurately
estimating black hole masses in core galaxies without the need for
spectroscopy.

The stellar mass deficits we derive range between 0.2 and 39 times the
black hole mass; for about 75\% of the sample, the range is 1--5 $M_{\rm BH}$,
and the median value for the whole sample is 2.2 \mbh.
Given the results of recent simulations, these values can be explained
by a reasonably small number of (dry) mergers.

\section*{Acknowledgements}

We thank the anonymous referee for a number useful comments that helped us 
significantly improve the paper.

SPR was supported by the DFG Cluster of Excellence ``Origin and
Structure of the Universe''. PE was supported by the Deutsche
Forschungsgemeinschaft through the Priority Programme 1177 ``Galaxy
Evolution''.

Some of the data presented in this paper were obtained using SINFONI at
the Very Large Telescope VLT and from the Mikulski Archive for Space
Telescopes (MAST). The VLT is operated by the European Southern
Observatory on Cerro Paranal in the Atacama Desert of northern Chile.
STScI is operated by the Association of Universities for Research in
Astronomy, Inc., under NASA contract NAS5-26555.

This research is based in part on data from the Isaac Newton Group
Archive, which is maintained as part of the CASU Astronomical Data
Centre at the Institute of Astronomy, Cambridge.

Funding for the creation and distribution of the SDSS Archive has been
provided by the Alfred P. Sloan Foundation, the Participating
Institutions, the National Aeronautics and Space Administration, the
National Science Foundation, the U.S. Department of Energy, the Japanese
Monbukagakusho, and the Max Planck Society.  The SDSS Web site is
http://www.sdss.org/.

The SDSS is managed by the Astrophysical Research Consortium (ARC) for
the Participating Institutions.  The Participating Institutions are The
University of Chicago, Fermilab, the Institute for Advanced Study, the
Japan Participation Group, The Johns Hopkins University, the Korean
Scientist Group, Los Alamos National Laboratory, the
Max-Planck-Institute for Astronomy (MPIA), the Max-Planck-Institute for
Astrophysics (MPA), New Mexico State University, University of
Pittsburgh, University of Portsmouth, Princeton University, the United
States Naval Observatory, and the University of Washington.

Finally, this work has made use of the NASA/IPAC Extragalactic Database
(NED) which is operated by the Jet Propulsion Laboratory, California
Institute of Technology, under contract with the National Aeronautics
and Space Administration.

\appendix

\section{Photometry of IC\,1459}

We constructed the circularized $V$-band photometry of IC\,1459 by
matching \textit{HST} and ground based data. Inside 12\arcsec{} we used
an \textit{HST} WFPC2 F814W profile (Proposal ID: 5454, PI: Marijn Franx)
derived using the software of \citet{Bender-87} while masking the central
dusty features. At larger radii we used CCD data from
\citet{Bender-89} (see also \citealt{Scorza-98}), calibrated with
photoelectric photometry available in Hyperleda.

\section{Photometry of NGC\,1399}

The circularized $B$-band photometry of NGC\,1399 was constructed by
matching the extended profile considered in \citet{Saglia-00} to the
profile derived using the IRAF software from an \textit{HST} ACS F606W
image (Proposal ID: 10129, PI: Thomas Puzia) of the galaxy in the inner
12\arcsec.

\section{Photometry of NGC\,1407}

The light profile for this galaxy is based on that used in \citetalias{Rusli-13}
for the BH mass measurement. To facilitate the exploration of possible outer-envelope
models (Section~\ref{multi-component-fits}), we replaced the inner 7.5\arcsec{} of the
previous profile with data from a deconvolved version of the \textit{HST}-ACS F435W image (20
iterations of Lucy-Richardson deconvolution). This allowed us to fit the profile without
PSF convolution.

\section{Photometry of NGC\,3379}

The light profile was constructed using WFPC2 F814W images from the
\textit{HST} archive (Proposal ID: 5512, PI: Sandra Faber) and SDSS
$i$-band images. The PC chip image of \textit{HST} was deconvolved using
20 rounds of Lucy-Richardson deconvolution and then the profile was
measured through ellipse fitting. That profile was used for $r < 10\arcsec$. 
At larger radii, we used an ellipse-fit profile measured using a
mosaic of SDSS $i$-band images, constructed using SWarp
\citep{Bertin-02}. The SDSS $i$-band magnitudes were converted to
Cousins $I$ using an $r-i$ color of 0.44 (measured from ellipse fits to
single SDSS $r$ and $i$ images) and the transformation of
\citet{Jordi-06}.

\section{Photometry of NGC\,3608}

The surface-brightness profile for NGC\,3608 is a combination of ellipse
fits to an F555W WFPC2 image (Proposal ID: 5454, PI: Marijn Franx) for
$r_{\rm eq} < 7\arcsec$ and ellipse fits to an SDSS $g$-band image for
$r_{\rm eq} > 7\arcsec$. The SDSS image was calibrated to $V$-band using
the SDSS $g$ and $r$-band images and the color transformations of
Lupton.\footnote{http://www.sdss.org/dr7/algorithms/sdssUBVRITransform.html\#Lupton2005} 
We matched the two datasets using fixed-ellipse-fit
profiles to the SDSS image and the WFPC2 mosaic, using the overlap
between 6--85\arcsec\ to determine the best scaling and background
subtraction to apply to the WFPC2 data.

\section{Photometry of NGC\,3842}

The surface brightness profile for NGC\,3842 is calibrated to the
$V$-band and is assembled from three pieces. The innermost part ($a <
1.2\arcsec$) comes from the published F555 WFPC2 PC1 profile of \citetalias{Lauer-05}. The
profile in the intermediate radii is derived from an ellipse fit to a
F555W WFPC2 mosaic image (galaxy center on WF3 chip; Proposal ID: 265,
PI: Douglas Geiser). This is matched to an ellipse fit to an SDSS
g-band image, including determining the level of sky background of the
WFPC2 image. The combination then consists of WFPC2 data at $1.2 < a <
38\arcsec$ and SDSS data at $a > 38\arcsec$. The final surface brightness
profile is constructed by matching the combined profile and the \citetalias{Lauer-05}
profile.

\section{Photometry of NGC\,4291}

For NGC\,4291, the profile is a combination of ellipse fits to an WFPC2
F702W image [Proposal ID: 6357, PI: Walter Jaffe; see \citealt{Rest-01}]
and ellipse fits to a Sloan $r$-band Isaac Newton Telescope Wide Field
Camera (INT-WFC) image retrieved from the Isaac Newton Group (ING)
Archive. The latter was a 120s exposure originally obtained on 2004
January 24. The INT-WFC image was calibrated to $V$-band using aperture
photometry from \citet{Prugniel-98}. We matched the two datasets using
fixed-ellipse-fit profiles to the INT-WFC image and the WFPC2 mosaic,
using the overlap between 10--80\arcsec{} to determine the best scaling
and background subtraction to apply to the WFPC2 data. The final profile
used data from the WFPC2 PC chip for $r < 7\arcsec$ and data from the
INT-WFC image for $r > 7\arcsec$.

\section{Photometry of NGC\,5516}

The profile for this galaxy was constructed by combining data from
WFPC2 F814W images from the \textit{HST} archive (Proposal ID: 6579, PI:
John Tonry) and an $R$-band image from the Wide Field Imager of the
ESO-MPI 2.2m Telescope. The sky background for the WFPC2 data was
determined by matching ellipse-fit profiles of the full WFPC2 mosaic
with a profile from the WFI image.  The PC chip image of \textit{HST}
was deconvolved using 20 rounds of Lucy-Richardson deconvolution and
then the profile was measured through ellipse fitting; the resulting
profile was used for $r < 9\arcsec$.  At larger radii, we used an
ellipse-fit profile from the WFI image.

\section{Photometry of NGC5813}

The profile for this galaxy is a combination of ellipse fits to an F814W
WFPC2 image (Proposal ID: 5454, PI: Marijn Franx) and an SDSS $i$-band
image. Since patchy dust was present in the nucleus, we combined the
F814W image with the corresponding F555W image to perform a simple dust
extinction correction on the F814W image. Sky backgrounds for the two
filters were determined by matching ellipse-fit profiles of the full
WFCP2 mosaic image with ellipse-fit profiles from the sky-subtracted
SDSS $g$ and $i$ images. The F555W and F814W images were also used,
along with the prescriptions in \citet{Holtzman-95}, to generate
$V$-band and $I$-band calibrations for each, iterating from an intial
guess for the central ($r < 10\arcsec$) $V - I$ color until the measured
color converged. We then formed a $V - I$ PC-chip image and used this to
color-correct the F814W PC image, assuming a simple screen model for the
dust. The final profile used the (extinction-corrected) WFPC2 PC-chip
data for $r < 2.8\arcsec$ and SDSS $i$-band data at larger radii.

\section{Photometry of NGC5846}

Since NGC\,5846 had available the same kind of data as NGC\,5813
\textit{and} similar patchy dust extinction in the circumnuclear region,
we followed an essentially identical process for this galaxy. The final
profile used the (extinction-corrected) WFPC2 PC-chip data for $r <
15\arcsec$ (Proposal ID: 5920, PI: Jean Brodie) and SDSS $i$-band data
at larger radii (the difference in transition radius compared with
NGC\,5813 is due to the larger extent of circumnuclear dust in
NGC\,5846).

\section{Photometry of NGC\,6086}

We used a WFPC2 F814W image (proposal ID: 7281, PI: Roberto Fanti) and
an SDSS $i$-band image to construct the surface brightness profile based
on combined ellipse fits. The part of the profile inside 11\arcsec{} comes
from the PC chip of WFPC2; and the innermost 1.5\arcsec{}  is from the PC
chip of WFPC2 after performing 20 rounds of Lucy-Richardson
deconvolution. The light profile is calibrated to the $R$-band. This was
done by first converting the SDSS $i$-band profile to SDSS $r$ by
matching with a profile from the SDSS $r$-band image. The SDSS $r$-band
magnitudes were then converted to Cousins $R$ using a $g-r$ color of
0.95 (measured from the SDSS g and r images) and the conversion of
\citet{Jordi-06}.

\section{Photometry of NGC\,7768}

The light profile of NGC\,7768 is mostly based on WFPC2 F814W images
(Proposal ID: 8184, PI: P. C{\^o}t{\'e}), which were combined to make
both a PC image and a mosaic. The PC image was corrected for dust
extinction using the F814W image and F450W WFPC2 images, but there was
still considerable residual extinction inside 1\arcsec, so an
\textit{HST}-NICMOS2 F160W image was used for the inner region. The
final profile is obtained through ellipse fits to the following:
(1)NICMOS F160W image for $a < 0.93\arcsec$, (2)WFPC2 F814W PC image for
$a$ between 0.93 and 10\arcsec{} and (3)WFPC2 F814W mosaic for $a > 10\arcsec$.

The F814W image was corrected for sky background by matching a
fixed-ellipse-fit profile with a similar fixed-ellipse-fit profile from
an $r$-band image with the Isaac Newton Telescope Wide Field Camera
(100s, 3 Dec 2002), taken from the Isaac Newton Group archive. The
profiles were calibrated to $V$-band using aperture photometry from
\citet{Prugniel-98} (specifically, the INT-WFC $r$-band image was
calibrated to $V$, and then the WFPC2 F814W profile was matched to the
INT-WFC $r$-band profile.)

\section{Comparison with the \sersic\ fitting of KFCB09}
\label{comparisonkfcb09}

Of all the galaxies presented in Fig.~\ref{rbplots}, there are five
galaxies which are also part of \citetalias{Kormendy-09b} sample, i.e.
NGC\,4472, NGC\,4486 (M87) and NGC\,4649, NGC\,4261 and NGC\,4374. Using
the photometric data available in \citetalias{Kormendy-09a}, we derive
the missing light using the \cs\ function as described in Sections
\ref{themethod} and \ref{coredeficit}. From these five measurements, we
find that some of our $M_{\rm V,def}$ fall outside the uncertainties
provided by \citetalias{Kormendy-09b}. The difference is largest for
NGC\,4486 and NGC\,4261 (by $\sim$0.7 magnitude or a factor of two in
luminosity) and it is not systematic. The disagreement probably reflects
the different details that were involved in the fitting and the
derivation of the missing light:

1. We use the circularized radius; \citetalias{Kormendy-09a} uses
semi-major axis radius. Galaxies are not perfectly round or circular in
projection. Non-zero ellipticity anywhere in the galaxy, not just in the
center, will induce a deviation in $M_{\rm V,def}$ since the light
deficit in both \citetalias{Kormendy-09a} and this work depends also on
the \sersic\ fit to the outer part of the galaxy. The magnitude of the
deviation is difficult to estimate, since this depends very much on the
characteristics of each galaxy. 

2. A different fitting range introduces more discrepancies. Because we
use circularized radius, we can use only the datapoints for which the
ellipticity is given. In the photometric profiles of
\citetalias{Kormendy-09a}, the ellipticity is often not specified for a
few innermost or outermost radii; therefore we cannot use those points
in our fits. In addition, in the particular case of NGC\,4486,
\citetalias{Kormendy-09b} used a S\'ersic fit (from
\citetalias{Kormendy-09a}, using semi-major axis) which excluded data
outside 400\arcsec, because it apparently gave rise to a S\'ersic index
which was too large (almost 12) due to the cD halo. However, we found no
difficulties in fitting a core-S\'ersic function to the entire profile
(out to $\sim1000\arcsec$), using the circularized radius; we obtain a
S\'ersic index of 8.9 (almost identical to \citetalias{Kormendy-09a}'s
value), albeit with different values of $r_e$ and $\mu_e$.

3. We fit the galaxy profile with a \cs\ profile,
\citetalias{Kormendy-09a} fit the outer part of the galaxy with a
\sersic\ profile. The critical point is the transition region between
the \sersic\ and the core. In the way \citetalias{Kormendy-09a} defines
the core, there is no transition region: the core immediately starts
when the \sersic\ model no longer fits the datapoints in the inner part.
We use a \cs\ parametrization and since we fit the $\alpha$ parameter,
instead of fixing it to infinity, we allow some part of the galaxy to be
partly \sersic, partly power-law. This leads to a different inner limit
of the radial range that is fitted only by the \sersic\ model, which in
the end results in a different best-fitting \sersic\ profile. 

4. The light deficit in \citetalias{Kormendy-09b} is obtained by
calculating the difference of the integrated intensity between the
extrapolated \sersic\ function and the actual (observed) profile after
PSF deconvolution. The ellipticity of the \sersic\ model is fixed to the
ellipticity at the inner limit of the fitted radial range. Since some of
our light profiles are not PSF-deconvolved, we instead compute the
difference between the best-fitting \cs\ function and the extrapolated
\sersic\ component. We also implicitly assume an ellipticity of zero
when integrating the extrapolated \sersic\ function, since we have used
a circularized profile in the first place.

\bibliographystyle{aj}

\end{document}